%% file: main.tex
\documentclass{IEEEtran}
\usepackage{cite}
\usepackage{amsmath,amssymb,amsfonts}
\usepackage{graphicx,color}
\usepackage{textcomp}
\usepackage[most]{tcolorbox}
\usepackage{xcolor}
\definecolor{myblue}{RGB}{200,220,255}   
\usepackage{bm}
\usepackage{subfigure}
\usepackage{url}
\usepackage{algorithm}
\usepackage{algpseudocode}
\usepackage{booktabs}
\usepackage{makecell}
\usepackage{tikz}
\usetikzlibrary{arrows.meta,positioning}

\def\BibTeX{{\rm B\kern-.05em{\sc i\kern-.025em b}\kern-.08em
    T\kern-.1667em\lower.7ex\hbox{E}\kern-.125emX}}
\AtBeginDocument{\definecolor{tmlcncolor}{cmyk}{0.93,0.59,0.15,0.02}\definecolor{NavyBlue}{RGB}{0,86,125}}

\usepackage[acronym,shortcuts]{glossaries}
\usepackage{tcolorbox}
\definecolor{mycolor}{RGB}{242,242,242}

\newacronym{2D}{2D}{two-dimensional}
\newacronym{3D}{3D}{three-dimensional}
\newacronym{ADoA}{ADoA}{angle difference of arrival}
\newacronym{AN}{AN}{anchor node}
\newacronym{AoA}{AoA}{angle of arrival}
\newacronym{DoF}{DoF}{degrees of freedom}
\newacronym{EDM}{EDM}{Euclidean distance matrix}
\newacronym{GPS}{GPS}{Global Positioning System}
\newacronym{IoT}{IoT}{internet of things}
\newacronym{MDS}{MDS}{multidimensional scaling}
\newacronym{GEK}{GEK}{Gram edge kernel}
\newacronym{SIMO}{SIMO}{single-input single-output}
\newacronym{MSE}{MSE}{mean square error}
\newacronym{MRC}{MRC}{maximal ratio combining}
\newacronym{MRC-SMDS}{MRC-SMDS}{maximal ratio combining super multidimensional scaling}
\newacronym{QD-MRC-SMDS}{QD-MRC-SMDS}{quaternion-domain maximal ratio combining super multidimensional scaling}
\newacronym{LAN}{LAN}{Local Are Network}
\newacronym{LOS}{LOS}{line-of-sight}
\newacronym{NLOS}{NLOS}{non-line-of-sight}
\newacronym{PDF}{PDF}{probability density function}
\newacronym{RSSI}{RSSI}{received signal strength indicator}
\newacronym{SMDS}{SMDS}{super multi-dimensional scaling}
\newacronym{CD-SMDS}{CD-SMDS}{complex-domain SMDS}
\newacronym{QD-SMDS}{QD-SMDS}{quaternion-domain super multi-dimensional scaling}
\newacronym{SVD}{SVD}{singular value decomposition}
\newacronym{EVD}{EVD}{eigen value decomposition}
\newacronym{QSVD}{QSVD}{quaternion singular value decomposition}
\newacronym{TN}{TN}{target node}
\newacronym{ToA}{ToA}{time of arrival}
\newacronym{TDoA}{TDoA}{time difference of arrival}
\newacronym{WSN}{WSN}{wireless sensor network}
\newacronym{CDF}{CDF}{cumulative distribution function}
\newacronym{UWB}{UWB}{ultra-wideband}
\newacronym{DoA}{DoA}{direction of arrival}
\newacronym{SDP}{SDP}{semidefinite programming}
\newacronym{CSI}{CSI}{channel state information}
\newacronym{FLOP}{FLOP}{floating-point operation}
\newacronym{IZMA}{IZMA}{interest zone matrix approximation}
\newacronym{SDR}{SDR}{semidefinite relaxation}
\newacronym{SLAM}{SLAM}{simultaneous localization and mapping}

\def\OJlogo{\vspace{-4pt}$<$Society logo(s) and publication title will appear here.$>$}
\def\seclogo{\vspace{10pt}$<$Society logo(s) and publication title will appear here.$>$}

\def\authorrefmark#1{\ensuremath{^{\textbf{#1}}}}

\begin{document}

\markboth{}{Author {et al.}}

\title{Quaternion-Domain Super MDS\\ for Robust 3D Localization}

\author{
Alessio~Lukaj\authorrefmark{1},
Keigo~Masuoka\authorrefmark{2},
Takumi~Takahashi\authorrefmark{2},
Giuseppe~Thadeu~Freitas~de~Abreu\authorrefmark{3},
and Hideki~Ochiai\authorrefmark{2}
\thanks{\authorrefmark{1}A. Lukaj is with the Department of Information Technology and Electrical Engineering, ETH Zurich, 8092 Zurich, Switzerland.}%
\thanks{\authorrefmark{2}K. Masuoka, T. Takahashi, and H. Ochiai are with the Graduate School of Engineering, The University of Osaka, 2-1 Yamada-oka, Suita 565-0871, Japan.}%
\thanks{\authorrefmark{3}G. T. F. de Abreu is with the School of Computer Science and Engineering, Constructor University, 28759 Bremen, Germany.}%
}

\maketitle

\begin{abstract}
\input{TXT/0_abstraction}
\end{abstract}


\glsresetall

\maketitle

\section{INTRODUCTION}
\label{sec:intro}
\input{TXT/1_Introduction}

\vspace{-1ex}
\section{Preliminaries}
\label{sec:NaP}
\input{TXT/2_Notation_and_Preliminaries}

\section{Problem Formulation and the SMDS Algorithm}
\input{TXT/3_Problem}

\section{Quaternion-Domain Super MDS}
\input{TXT/4_QD-SMDS}

\section{Quaternion-Domain MRC-SMDS}
\label{sec:MRC}
\input{TXT/5_QD-MRC-SMDS}

{\section{Computation Effort Comparisons and Discussions}}
\label{sec:wall_clock}

\input{TXT/6_Discussions}

\section{Conclusion}
\label{sec:conclusion}
\input{TXT/7_Conclusion}

\appendix
\input{TXT/appendix}

\vfill\pagebreak

\end{document}

%% file: TXT/0_abstraction.tex
%
This paper proposes a novel low-complexity \ac{3D} localization algorithm for wireless sensor networks, termed \ac{QD-SMDS}.
The algorithm is based on a reformulation of the \acs{SMDS}, originally developed in the real domain and for \ac{2D} problems, using quaternion algebra. 
By representing \ac{3D} coordinates as quaternions, the method constructs a rank-$1$ \ac{GEK} matrix that integrates both relative distance and angular information between nodes, which enhances the noise reduction effect achieved through low-rank truncation employing \ac{SVD}, thereby improving robustness against information loss.
To further reduce computational complexity, we also propose a variant of \ac{QD-SMDS} that eliminates the need for the computationally expensive \ac{SVD} by leveraging the inherent structure of the quaternion-domain \ac{GEK} matrix. 
This alternative directly estimates node coordinates using only matrix multiplications within the quaternion domain.
Simulation results demonstrate that the proposed method significantly improves localization accuracy compared to the original \acs{SMDS} algorithm, especially in scenarios with substantial measurement errors.
The proposed method also achieves comparable localization accuracy without requiring \ac{SVD}.

%% file: TXT/1_Introduction.tex
%
\IEEEPARstart{W}{ith} recent advancements in sensor technology, \acp{WSN} have increasingly emerged as fundamental information infrastructures across a wide range of industrial domains~\cite{ChenCST22, Praveen2022, TrifunAccess2022}, including precision agriculture~\cite{Mowla2023}, smart factories~\cite{Xiang2024}, and medical sensing~\cite{Javaid2022}.
Among the various types of sensor data, location information plays a particularly vital role, as it not only enhances the value of the sensed data itself but also contributes significantly to the efficient operation of \acp{WSN}.
In fact, its importance is often regarded as comparable to that of payload data in conventional wireless communications.
Accordingly, many services that utilize such location information are designed under the assumption of large-scale networks composed of numerous small, low-power sensor terminals (hereafter referred to as \textit{nodes})~\cite{Chaloupka2017,Yassin2017}.

In these scenarios, there is a strong demand for algorithms capable of simultaneously estimating the positions of multiple nodes with high accuracy and low computational complexity, based on the aggregated multidimensional information available from \acp{WSN}~\cite{Ahmad,Xiao2016,Wymeersch2009}.

Localization algorithms can be broadly categorized into three main approaches based on their underlying mathematical frameworks: Bayesian inference~\cite{Xiong,Naseri2019,Ihler2005,Li2022}, convex optimization~\cite {Shi2017}, and isometric embedding methods.
When selecting an appropriate algorithm, it is crucial to consider the trade-off between computational complexity and estimation accuracy.
This paper focuses on the isometric embedding approach, commonly known as \ac{MDS}~\cite{Torgerson1952}.
Unlike Bayesian inference and convex optimization, \ac{MDS} offers the notable advantage of fixed computational complexity.
While the convergence speed of Bayesian and optimization-based methods can vary significantly due to measurement noise or missing data, \ac{MDS} maintains consistent complexity regardless of such factors.
This makes \ac{MDS} particularly well-suited for systems with limited computational resources and strict time constraints.

Among various \ac{MDS}-based localization methods, \ac{SMDS}~\cite{Abreu2007, Macagnano2011} stands out as a hybrid algorithm that integrates both distance and angle information.
It has demonstrated superior performance over classical \ac{MDS}—which relies solely on distance information—even in the presence of angular uncertainties of approximately $\pm35^\circ$ ~\cite{Macagnano2013}.
\Ac{CD-SMDS} is a technique that reduces computational complexity and improves precision by tailoring \ac{SMDS} for \ac{2D} localization~\cite{Ghods2018TWC}.
Unlike conventional \ac{SMDS}, which represents \ac{2D} node coordinates as real-valued vectors, \ac{CD-SMDS} uses complex-valued scalars to express node positions.
By constructing a \ac{GEK} matrix in the complex domain that consolidates all measurement data, the method enables rank reduction to one and enhances accuracy through noise suppression via low-rank truncation using \ac{SVD}~\cite{Nishi2023}.
This demonstrates that translating real-valued vectors into single complex scalars not only simplifies representation but also provides a foundation for extending \ac{SMDS} to a \ac{3D} localization framework. 
In other words, if \ac{3D} node coordinates can be similarly represented as scalars, it may be possible to develop an \ac{SMDS} algorithm optimized for \ac{3D} localization that {reduces the rank of the \ac{GEK} matrix to one, favoring matrix completion techniques, and thereby improves localization accuracy, as demonstrated by \ac{CD-SMDS} for \ac{2D} coordinates}.

As an intuitive candidate for representing \ac{3D} coordinates with a single scalar, one might initially consider \textit{bicomplex} numbers~\cite{Luna2016}.
%
%
However, to the best of our knowledge, no existing literature addresses \ac{SVD} or matrix completion for bicomplex matrices, and their algebraic structure renders their integration into \ac{SMDS} impractical.
Accordingly, in this study, we turn our attention to \textit{quaternions} as a mathematical tool for representing \ac{3D} coordinates.

{ Quaternions constitute the smallest and most widely established extension of complex numbers preserving an analogous algebraic structure, and have been extensively adopted across a broad range of spatial estimation algorithms.
Unit quaternions are the preferred parameterization of orientation in image processing~\cite{Barthélemy2015}, computer graphics~\cite{Ken1985}, pose estimation~\cite{Horn1987}, and \ac{SLAM} systems~\cite{Cadena2016}, where the rotational state of a robot or camera platform is encoded as a unit quaternion within extended Kalman filters.
In these applications, however, quaternions are used exclusively to represent \textit{rotations}; the translational coordinates of a node continue to be expressed as ordinary real-valued vectors.
To the best of our knowledge, no existing work has applied quaternions to encode node \textit{positions} themselves in range- or angle-based wireless sensor network localization.
In particular, the idea of representing \ac{3D} coordinates as quaternions to construct a rank-1 kernel matrix for 3D position estimation, thus  maximizing noise suppression via low-rank approximation, has not been previously explored, and constitutes a core novelty of the proposed \ac{QD-SMDS}.}

A quaternion consists of one real component and three imaginary components~\cite{Baek2017}.
By utilizing three of its four \ac{DoF}, it becomes possible to represent \ac{3D} node coordinates and construct a rank-$1$ \ac{GEK} matrix in the quaternion domain.
{Furthermore, a mature matrix toolbox for quaternions is available, providing well-defined conjugation and norm operations, as well as frameworks for performing algebraic computations such as \ac{SVD}~\cite{Miao2019}}.
This enables the exploitation of \ac{SVD}-based low-rank truncation for noise suppression, thereby improving localization accuracy and leading to the development of \ac{QD-SMDS}.

However, the computational efficiency gained by \ac{QD-SMDS} is quite limited and may even result in increased computational cost in some systems.
This is because performing \ac{SVD} in the quaternion domain requires converting the target quaternion matrix into a complex-valued matrix of double the size, followed by applying \ac{SVD} to the resulting complex-equivalent matrix.
Given that the size of the \ac{GEK} matrix increases proportionally with the number of node pairs, this process can become impractical—particularly in systems with constrained computational resources.
To eliminate the need for \ac{SVD} in \ac{SMDS}, an iterative approach based on \ac{MRC} was proposed in~\cite{Ghods2018TWC}.
However, due to the non-commutative nature of quaternion multiplication, directly extending this method to \ac{QD-SMDS} is not straightforward.
In this paper, we address this challenge by explicitly considering quaternion non-commutativity in the combining operation and propose a novel, low-complexity \ac{3D} localization algorithm—termed \ac{QD-MRC-SMDS}—which is constructed solely from simple quaternion multiplications.
%
%
%
%
The contributions of this paper are summarized as follows\footnote{\setlength{\baselineskip}{10pt}The conference paper~\cite{Keigo2025SPAWC} is a {shorter and} earlier version of this work and was presented at the IEEE SPAWC 2025, introducing the basic concept of \ac{QD-SMDS} and providing a brief performance evaluation by comparing it with the preceding \ac{SMDS}. 
In contrast, the present paper offers a detailed derivation of \ac{QD-SMDS} and introduces a new lower-complexity variant, \ac{QD-MRC-SMDS}, together with a comprehensive performance assessment.}:

\begin{itemize}
    \item A novel \ac{3D} localization algorithm, termed \ac{QD-SMDS} is proposed. This algorithm is derived by reformulating the conventional \ac{SMDS}, originally developed in the real domain, into the quaternion domain. By representing \ac{3D} coordinates using quaternions, a rank-1 \ac{GEK} matrix can be constructed, enabling enhanced noise suppression through low-rank approximation. As a result, improved localization accuracy can be achieved, especially under large measurement errors.
    
    \item To reduce the computational complexity, we propose a novel low-complexity variant, termed \ac{QD-MRC-SMDS}. While \ac{QD-SMDS} relies on \ac{SVD}-based low-rank truncation, \ac{QD-MRC-SMDS} performs localization using only simple multiplication operations derived from a closed-form expression.
    Furthermore, we extend this method into an iterative scheme that leverages the internal structure of the quaternion-domain \ac{GEK} matrix, significantly improving estimation accuracy with only a marginal increase in computational cost.
    \item To validate the efficacy of the proposed methods, we conducted computer simulations. As \ac{QD-SMDS} requires additional measurement parameters compared to conventional \ac{SMDS}, its performance was evaluated under various scenarios, including cases where these parameters are available, unavailable, or partially missing. The results demonstrate that \ac{QD-SMDS} outperforms conventional \ac{SMDS}, particularly under large angle measurement errors. Moreover, the computationally efficient variant achieves performance asymptotically close to that of \ac{QD-SMDS} across nearly all scenarios.
\end{itemize}

\textit{Notation}:
The following notation is used throughout unless otherwise specified. 
Sets of real and complex numbers are denoted by $\mathbb{R}$ and $\mathbb{C}$, respectively.
Vectors and matrices are denoted by lower- and upper-case bold-face letters, respectively.
The conjugate, transpose, and conjugate transpose operators are denoted by $(\cdot)^*$, $(\cdot)^\mathsf{T}$, and $(\cdot)^\mathsf{H}$, respectively.
The $a\times a$ square identity matrix is denoted by $\bm{I}_{a}$.
The $a\times b$ all-zeros matrix and all-ones matrix are denoted by $\bm{0}_{a\times b}$ and $\bm{1}_{a\times b}$, respectively.
The diagonal matrix constructed by placing the elements of a vector $\bm{a}$ on its main diagonal is denoted by $\mathrm{diag}\left[\bm{a}\right]$.
The Euclidean norm and Frobenius norm are denoted by $\|\cdot\|$ and $\|\cdot\|_{\mathrm{F}}$, respectively.
The inner product and the outer product are denoted by $\left<\cdot,\cdot\right>$ and $\left|\cdot\times\cdot\right|$, respectively.
The determinant of a matrix $\bm{A}$ is denoted by $\mathrm{det}\left[\bm{A}\right]$.
%
The matrix formed by stacking the odd-numbered rows of a matrix $\bm{A}$ is denoted by $\mathrm{r}_{\mathrm{odd}}\left[\bm{A}\right]$, and the matrix formed by collecting the odd-numbered columns is denoted by $\mathrm{c}_{\mathrm{odd}}\left[\bm{A}\right]$.
The element-wise $k$-th power of the entries in a matrix $\bm{A}$ is denoted by $\bm{A}^{\odot k}$.

%% file: TXT/2_Notation_and_Preliminaries.tex

In this section, we first provide a brief overview of quaternion algebra, and then introduce the \ac{QSVD}.

\subsection{Basics of Quaternion Algebra}

The quaternion space was first introduced by W. Hamilton ~\cite{Miron2023} as a natural extension of the complex space.
A quaternion consists of one real component and three imaginary components, as follows:
\begin{equation}
    \mathbb{H}
    \triangleq
    \left\{
    a + \mathbf{i}b + \mathbf{j}c + \mathbf{k}d
    : a,b,c,d
    \right\},
\end{equation}
where $a,b,c,d\in\mathbb{R}$ are real numbers, and $\mathbf{i,j,k}$ are the imaginary units, which obey the following rules:

\begin{equation}
    \mathbf{i}^2 = \mathbf{j}^2 = \mathbf{k}^2 = -1,
\end{equation}
with
\begin{subequations}
      \begin{eqnarray}
        \mathbf{i}\cdot \mathbf{j}
        \!\!&\!\!=\!\!&\!\!
        -\mathbf{j}\cdot \mathbf{i}
        =
        \mathbf{k}, \\
        \mathbf{j}\cdot \mathbf{k}
        \!\!&\!\!=\!\!&\!\!
        -\mathbf{k}\cdot \mathbf{j}
        =
        \mathbf{i}, \\
        \mathbf{k}\cdot \mathbf{i}
        \!\!&\!\!=\!\!&\!\!
        -\mathbf{i}\cdot \mathbf{k}
        =
        \mathbf{j}.
        \end{eqnarray}
\end{subequations}

If the real part of a quaternion is zero (\textit{i.e.}, $a=0$), the quaternion is called a \textit{pure quaternion}. 
The square of a pure quaternion is the negative sum of the squares of its imaginary components:
\begin{equation}
    (\mathbf{i}b + \mathbf{j}c + \mathbf{k}d)^2
    =
    -b^2 -c^2 -d^2,
\end{equation}
while multiplications in $\mathbb{H}$ is generally \textit{noncommutative}, multiplication in $\mathbb{H}$ by real numbers is commutative, thus,
\begin{equation}
    \begin{array}{l}
        aq \neq qa \quad \mathrm{if} \;a \in \mathbb{H},\quad q \in \mathbb{H},\\
        aq = qa \quad \mathrm{if} \;a \in \mathbb{R},\quad q \in \mathbb{H}.
    \end{array}
\end{equation}

For a quaternion $q = a + \mathbf{i}b + \mathbf{j}c + \mathbf{k}d$, its conjugate is defined as
\begin{equation}
    q^* = a - \mathbf{i}b - \mathbf{j}c - \mathbf{k}d.
\end{equation}

Conjugation in $\mathbb{H}$ shares similar properties with conjugation in $\mathbb{C}$.
In $\mathbb{H}$, we have the following identities (assuming $q_1$ and $q_2$ are two arbitrary quaternions)
\begin{subequations}
      \begin{eqnarray}
    \left(
    q_1 +q_2
    \right)^*
    &=&
    q_1^* + q_2^*,\\
      \left(
    q_1q_2
    \right)^*
    &=&
    q_2^* q_1^*, \\
    \left(
    q_1^*
    \right)^*
    &=&
    q_1.
        \end{eqnarray}
\end{subequations}
The norm of a quaternion $q$ is defined as
\begin{equation}
    \|q\| = \sqrt{qq^*} = \sqrt{q^*q} = \sqrt{a^2 + b^2 + c^2+ d^2},
\end{equation}
while the reciprocal of a quaternion $q$ is defined as
\begin{equation}
    q^{-1} = \frac{q^*}{\|q\|^2}.
\end{equation}

\subsection{Quaternion Singular Value Decomposition (QSVD)}
\label{subsec:QSVD}

In this subsection, we introduce the \ac{QSVD} as an algebraic tool to be used in the subsequent discussion.
A quaternion matrix $\dot{\bm{Q}} \in \mathbb{H}^{M\times N}$ is written as $\dot{\bm{Q}} = \bm{Q}_0 + \mathbf{i}\bm{Q}_1 + \mathbf{j}\bm{Q}_2 + \mathbf{k}\bm{Q}_3$, where $\bm{Q}_l\in\mathbb{R}^{M\times N}$ for $l = 0,1,2,3$.
Using the Cayley-Dickson notation~\cite{Miao2019}, $\dot{\bm{Q}}$ can be expressed as
\begin{eqnarray}
\label{eq:Cayley-Dickson}
    \dot{\bm{Q}} = \bm{Q}_\mathrm{a} + \mathbf{j}\bm{Q}_\mathrm{b},
\end{eqnarray}
where $\bm{Q}_\mathrm{a} = \bm{Q}_0 + \mathbf{i}\bm{Q}_1 \in \mathbb{C}^{M\times N}, \bm{Q}_\mathrm{b} = \bm{Q}_2 + \mathbf{i}\bm{Q}_3 \in \mathbb{C}^{M\times N}$.

Accordingly, the equivalent complex-valued representation of the quaternion matrix $\dot{\bm{Q}}$ is given by
\begin{equation}
\label{eq:Q_c}
    \bm{Q}_\mathrm{c}
    =
    \begin{bmatrix}
        \bm{Q}_\mathrm{a}&\bm{Q}_\mathrm{b}\\
        -\bm{Q}_\mathrm{b}^*&\bm{Q}_\mathrm{a}^*
    \end{bmatrix}
    \in
    \mathbb{C}^{2M\times 2N}.
\end{equation}

The \ac{SVD} of the quaternion matrix $\dot{\bm{Q}}$ can be derived from the \ac{SVD} of its equivalent complex matrix $\bm{Q}_\mathrm{c}$.
Given $\bm{\dot{Q}} = \bm{U}_{\mathrm{q}}\bm{D}_{\mathrm{q}}\bm{V}_{\mathrm{q}}^\mathsf{H}$ and $\dot{Q}_\mathrm{c} = \bm{U}_{\mathrm{c}}\bm{D}_{\mathrm{c}}\bm{V}_{\mathrm{c}}^\mathsf{H}$, we have
\begin{subequations}
\begin{eqnarray}
\bm{D}_\mathrm{q}
&=&
\mathrm{r}_{\mathrm{odd}}\left[\mathrm{c}_{\mathrm{odd}}\left[\bm{D}_\mathrm{c}\right]\right], \\
\bm{U}_\mathrm{q}
&=&
\mathrm{c}_{\mathrm{odd}}\left[\bm{U}_1\right] + \mathbf{j}\mathrm{c}_{\mathrm{odd}}\left[-\bm{U}_2^{*}\right],\\
\bm{V}_\mathrm{q}
&=&
\mathrm{c}_{\mathrm{odd}}\left[\bm{V}_1\right] + 
\mathbf{j}\mathrm{c}_{\mathrm{odd}}\left[-\bm{V}_2^{*}\right],
\end{eqnarray}
\end{subequations}
where
\begin{equation}
\bm{U}_\mathrm{c}=
\begin{bmatrix}
\bm{U}_1\\
\bm{U}_2
\end{bmatrix},\quad
\bm{V}_\mathrm{c}=
\begin{bmatrix}
\bm{V}_1\\
\bm{V}_2
\end{bmatrix},
\end{equation}
with $\bm{U}_1,\bm{U}_2\in\mathbb{C}^{M\times 2M}$ and $\bm{V}_1,\bm{V}_2\in\mathbb{C}^{N\times 2N}$.

%% file: TXT/3_Problem.tex
\subsection{Problem Formulation}
\label{sec:SMDS}

Consider a network embedded in \ac{3D} Euclidean space consisting of $N$ nodes, among which $N_{\mathrm{A}}$ nodes are designated as \acp{AN} with known, error-free locations.
The remaining $N_{\mathrm{T}} \triangleq N - N_{\mathrm{A}}$ nodes, referred to as \acp{TN}, have unknown locations to be estimated.
It is assumed that relative distances and angles are measurable between any pair of \acp{AN} and between each \ac{AN}--\ac{TN} pair, while such measurements are unavailable among \acp{TN}.
The objective of this paper is to estimate the coordinates of \acp{TN} based on the measured (and noisy) relative distances and angles between nodes, as well as the known coordinates of \acp{AN}.

Let the coordinates of the $n$-th node in the network be denoted by the column vector $\bm{x} \triangleq \left[a_n,b_n,c_n\right]^\mathsf{T}\in\mathbb{R}^{3\times 1}$, which represents the \ac{3D} coordinates of the node in the Cartesian coordinate system.
We define the coordinate matrix $\bm{X}_\mathrm{A}$ as the matrix that stacks the coordinate vectors of \acp{AN}:
\begin{equation}
\bm{X}_{\mathrm{A}}
\triangleq
\left[\bm{x}_1,\ldots,\bm{x}_{N_{\mathrm{A}}}\right]^\mathsf{T}
\in\mathbb{R}^{N_{\mathrm{A}}\times 3},
\label{eq:r_Xa}
\end{equation}
and similarly define the coordinate matrix that stacks the coordinate vectors of \acp{TN} as 
\begin{equation}
\bm{X}_{\mathrm{T}}
\triangleq
\left[\bm{x}_1,\ldots,\bm{x}_{N_{\mathrm{T}}}\right]^\mathsf{T}
\in\mathbb{R}^{N_{\mathrm{T}}\times 3}.
\label{eq:r_Xt}
\end{equation}

The real-valued matrix carrying the coordinate vectors of all nodes in the network can then be expressed as
\begin{equation}
\label{eq:r_X}
    \bm{X}
    \triangleq
    \left[
    \bm{X}_\mathrm{A}^\mathsf{T},\bm{X}_{\mathrm{T}}^\mathsf{T}
    \right]^{\mathsf{T}}
    \in
    \mathbb{R}^{N\times 3}.
\end{equation}

Consider the set $\mathcal{M}$ of unique index pairs $(i,j)$, arranged in ascending order, for which the pairwise distances and phases are measurable, \textit{i.e.}, any pair among \acp{AN} or between \acp{AN} and \acp{TN}: 
\begin{equation}
    \mathcal{M}\triangleq \left\{(1,2),\cdots,(1,N),(2,3),\cdots,(2,N),\cdots,(N_{\mathrm{A}},N)\right\},
\end{equation}
such that each pair $m\in\mathcal{M}$ corresponds to an edge vector $\bm{v}_m$ is defined as
%
%
\begin{equation}
    \bm{v}_m = \bm{x}_i - \bm{x}_j,\qquad j>i.
    \label{eq:r_v}
\end{equation}
 
From the above, the real-valued edge matrix consisting of the collection of all $M\triangleq|\mathcal{M}|=N_{\mathrm{A}}(N_{\mathrm{A}}-1)/2 + N_{\mathrm{A}}N_{\mathrm{T}}$ edges is defined by
\begin{equation}
    \bm{V} \triangleq    \left[\bm{v}_1,\ldots,\bm{v}_{m},\ldots,\bm{v}_{M}\right]^\mathsf{T}
    =
    \bm{C}\bm{X}\in\mathbb{R}^{M\times3},
    \label{eq:r_V}
\end{equation}
where $\bm{C}\triangleq \left[\bm{C}_{\mathrm{AA}}^\mathsf{T},\bm{C}_{\mathrm{AT}}^\mathsf{T}\right]^\mathsf{T}\in\mathbb{R}^{M\times N}$ is a structure matrix encoding the pairwise relationships between nodes and edges.

The submatrix corresponding to the edges between \acp{AN}, denoted as $\bm{C}_{\mathrm{AA}}\in \mathbb{R}^{N_\mathrm{A}(N_\mathrm{A}-1)/2 \times N}$, is expressed as
\begin{subequations}
\label{eq:C}
\begin{equation}
   \hspace{-1ex} 
   \bm{C}_{\mathrm{AA}}
\!\triangleq\!\!
\left[
\begin{array}{c|c|c|c|c|c}
\!\!\!\bm{1}_{N_{\mathrm{A}}-1\times 1}\!\! &\multicolumn{4}{c|}{-\bm{I}_{N_{\mathrm{A}}-1}} &\!\!\bm{0}_{N_{\mathrm{A}}-1 \times N_{\mathrm{T}}} \!\!\!\!\!\\\hline
\!\!\!\bm{0}_{N_{\mathrm{A}}-2\times 1}\!\!&\!\!\bm{1}_{N_{\mathrm{A}}-2\times 1} \!\! & \multicolumn{3}{c|}{-\bm{I}_{N_{\mathrm{A}}-2}} & \!\!\bm{0}_{N_{\mathrm{A}}-2 \times N_{\mathrm{T}}}\!\!\!\!\! \\\hline
\multicolumn{2}{c|}{\ddots}&\!\! \ddots \!\!&\multicolumn{2}{c|}{\ddots} &\vdots \\\hline
\multicolumn{3}{c|}{\bm{0}_{1\times N_{\mathrm{A}}-2}} &\!1  \!&\!\!\!-1 \!\!\!& \bm{0}_{1\times N_{\mathrm{T}}}
\end{array}
\right]\!\!,
\end{equation}
while the submatrix corresponding to the edges between \acp{AN} and \acp{TN}, denoted as $\bm{C}_{\mathrm{AT}}\in\mathbb{R}^{N_{\mathrm{A}}N_{\mathrm{T}}\times N}$, is expressed as
\begin{equation}
   \hspace{-1ex} 
   \bm{C}_{\mathrm{AT}}
\!\triangleq\!\!
\left[\!
\begin{array}{c|c|c|c|c|c|c}
\!\!\bm{1}_{N_\mathrm{T} \times 1}\!\!&\multicolumn{5}{c|}{\bm{0}_{N_{\mathrm{T}} \times N_{\mathrm{A}}-1}}&\!\!\!{-\bm{I}_{N_{\mathrm{T}}}}\!\!\!\!  
\\\hline
\!\!\bm{0}_{N_\mathrm{T}\times 1} \!\!& \!\!\bm{1}_{N_\mathrm{T} \times 1} \!\!& \multicolumn{4}{c|}{\bm{0}_{N_\mathrm{T} \times N_{\mathrm{A}}-2}}&\!\!\!-\bm{I}_{N_\mathrm{T}}\!\!\!\! 
\\\hline
\multicolumn{2}{c|}{\ddots}  &\!\!\ddots \!\! &\multicolumn{3}{c|}{\ddots}  & \vdots 
\\\hline
\multicolumn{3}{c|}{\bm{0}_{N_\mathrm{T}\times N_{\mathrm{A}}-2}} &\!\!\bm{1}_{N_\mathrm{T} \times 1}\!\!&\multicolumn{2}{c|}{\!\!\bm{0}_{N_\mathrm{T} \times 1}\!\!}&\!\!\! -\bm{I}_{N_\mathrm{T}}\!\!\!\! \\\hline
\multicolumn{4}{c|}{\bm{0}_{N_{\mathrm{T}}\times N_\mathrm{A}-1}} &\multicolumn{2}{c|}{\!\!\bm{1}_{N_\mathrm{T} \times 1}\!\!} &\!\!\! -\bm{I}_{N_\mathrm{T}}
\end{array}
\!\right]\!\!.
\end{equation}
\end{subequations}

{The structure matrix $\bm{C}$ has rank $N-1$, regardless of which nodes are designated as \acp{AN} or \acp{TN}. 
To see this, note that the rows of $\bm{C}$ sum to zero, so $\mathbf{1}_N$ lies in its null space and $\mathrm{rank}(\bm{C})\leq N-1$.
Conversely, the $N-1$ rows corresponding to the edges emanating from node~$1$ reduce to $[\mathbf{1}_{N-1}\,|\,-\bm{I}_{N-1}]$ up to column reordering, which is of full row rank and thus establishes $\mathrm{rank}(\bm{C})\geq N-1$.}

\vspace{-1ex}
\subsection{SMDS Algorithm Recap}

In this subsection, we briefly describe the conventional \ac{SMDS}
algorithm~\cite{Abreu2007, Macagnano2011} for estimating the coordinates of \acp{TN} based on the above formulation.

The inner product between two edge vectors $\bm{v}_m$ and $\bm{v}_p$ ($m,p\in\mathcal{M}$) can be expressed as
\begin{equation}
\label{eq:r_inner}
k_{mp}^{\mathrm{r}}
\triangleq
\left<\bm{v}_m,\bm{v}_p\right>=d_md_p\cos\alpha_{mp},
\end{equation}
where $d_m \triangleq \|\bm{v}_m\|$ denotes the Euclidean distance between the two nodes (\textit{i.e.}, $\bm{x}_i$ and $\bm{x}_j$), and $\alpha_{mp}$ denotes the \ac{ADoA} between $\bm{v}_m$ and $\bm{v}_p$.

Based on \eqref{eq:r_inner}, the real-domain \ac{GEK} matrix $\bm{K}_{\mathrm{r}}\in\mathbb{R}^{M\times M}$, which integrates both relative distance and \ac{ADoA} information, can be expressed as
\begin{eqnarray}
\label{eq:r_kernel}
\hspace{-2em}
\bm{K}_{\mathrm{r}}
\!\!\!&\triangleq&\!\!\!
\bm{V}\bm{V}^{\mathsf{T}} \nonumber \\[-2ex]
&=&\!\!\!
\mathrm{diag}\left(\bm{d}\right)
\begin{bmatrix}
\cos\alpha_{11} 
\!\!\!\!&\!\!\!\! \cdots \!\!\!\!&\!\!\!\! \cos\alpha_{1M} \\[-1ex]
\vdots & \ddots & \vdots \\
\cos\alpha_{M1}
\!\!\!\!&\!\!\!\! \cdots \!\!\!\!&\!\!\!\! \cos\alpha_{MM}
\end{bmatrix}
\mathrm{diag}\left(\bm{d}\right),
\end{eqnarray}
where $\bm{d} = \left[d_1,\ldots,d_m,\ldots,d_M\right]^\mathrm{T}\in\mathbb{R}^{M\times 1}$.

As shown in \eqref{eq:r_kernel}, it is evident that the rank of the real-domain \ac{GEK} matrix is $3$.

Assuming that all pairwise distance and \ac{ADoA} measurements required in \eqref{eq:r_kernel} are available, the real-domain \ac{GEK} matrix with measurement errors, denoted as $\tilde{\bm{K}}_{\mathrm{r}}$, can be constructed.
Given the \ac{SVD} of the real-valued \ac{GEK} matrix $\tilde{\bm{K}}_\mathrm{r}$ as $\tilde{\bm{K}}_\mathrm{r} = \bm{U}\bm{\varLambda}\bm{U}^{\mathsf{T}}$, the estimate of the edge matrix $\hat{\bm{V}}$ is then obtained as
\begin{equation}
\label{eq:r_V_estimated}
\hat{\bm{V}}
=
\bm{U}_{M\times 3}\bm{\Lambda}_{3\times 3}^{\odot\frac{1}{2}},
\end{equation}
where $\bm{U}_{\bm{M}\times 3}$ consists of the first $3$ columns of $\bm{U}$, and $\bm{\Lambda}_{3\times3}$ contains the corresponding top $3$ singular values on its diagonal.
%

%
\begin{algorithm}[!t]
\caption{\ac{SMDS} Algorithm}
\label{alg:1}
\begin{algorithmic}[1]
%
\Statex {\bf{Input:}}
\State  \textit{Measured pairwise distances and \acp{ADoA}: $\tilde{d}_m$ and $\tilde{\alpha}_{mp}$.}
\State \textit{The coordinates of at least $4$ \acp{AN}.}
\Statex {\bf{Steps:}}
\State \textit{Construct the real-domain \ac{GEK} matrix $\tilde{\bm{K}}_\mathrm{r}$ in Eq. \eqref{eq:r_kernel} using the input parameters.}
\State \textit{Perform SVD of the constructed \ac{GEK} matrix $\tilde{\bm{K}}_\mathrm{r}$.}
\State \textit{Obtain the edge matrix $\hat{\bm{V}}$ using Eq. $\eqref{eq:r_V_estimated}$}
\State \textit{Compute $\hat{\bm{X}}$ from $\hat{\bm{V}}$ in Eq. \eqref{eq:MoorePenrose}.}
\State \textit{Apply the Procrustes transform to $\hat{\bm{X}}$ if needed} \text{(\textit{e.g.}, see \cite{Fiore2001}).}
\end{algorithmic}
\end{algorithm}


Finally, the estimated coordinate matrix can be recovered from $\hat{\bm{V}}$ by inverting the relationship in \eqref{eq:r_V}, \textit{i.e.}, $\hat{\bm{X}}
=
\bm{C}^{-1}
\hat{\bm{V}}$.
%
%
However, since the rank of the structure matrix $\bm{C}$ is $N-1$, this inverse cannot be directly computed.
To circumvent this rank deficiency, we exploit the known coordinate matrix $\bm{X}_{\mathrm{A}}$ corresponding to \acp{AN} in \eqref{eq:r_Xa}.
By incorporating this prior knowledge,  the inversion can be reformulated as
\begin{equation}
\label{eq:MoorePenrose}
\left[
\begin{array}{c}
  \bm{X}_{\mathrm{A}}\\\hline
  \hat{\bm{X}}
\end{array}
\right]
=
\left[
\begin{array}{c|c}
  \bm{I}_{N_{\mathrm{A}}} & \bm{0}_{N_{\mathrm{A}}\times N_{\mathrm{T}}}\\\hline
  \multicolumn{2}{c}{\bm{C}}
\end{array}
\right]^{-1}
\left[
\begin{array}{c}

  \bm{X}_{\mathrm{A}}\\\hline
  \hat{\bm{V}}
\end{array}
\right].
\end{equation}

Moreover, since the \ac{SMDS} algorithm operates solely on the relative relationships among nodes, the inverse problem in \eqref{eq:MoorePenrose} can be characterized in multiple ways, resulting in a non-unique solution.
Therefore, a Procrustes transformation~\cite{Fiore2001} is typically required to align the estimated $\hat{\bm{X}}$ with the true coordinates $\bm{X}$ in terms of scale, orientation, and translation. 

{The computational complexity of the \ac{SMDS} algorithm is dominated by the truncated \ac{SVD} of the kernel matrix in \eqref{eq:r_kernel}, in which only the three dominant eigenpairs are computed~\cite{Calvetti1994}.
Accordingly, the asymptotic complexity of this approach is $O\left(3M^2\right)=O\left((N_{\mathrm{A}}(N_{\mathrm{A}}-1)/2+N_{\mathrm{A}} N_{\mathrm{T}})^2\right)=O\left(N_{\mathrm{A}}^2 (N_{\mathrm{A}}+N_{\mathrm{T}})^2\right)$.}

For the sake of completeness, this subsection is concluded with the pseudo-code of the full \ac{SMDS} algorithm, presented in Algorithm \ref{alg:1}.

\vspace{-2ex}
{
\subsection{Classical Localization Models and Link to \ac{SMDS} and its Quaternion Variants}
\label{subsec:classical_models}

In classical \ac{WSN} localization, unknown node positions $\bm{X}_\mathrm{T}$ are estimated using range-based models, angle-based models (\textit{e.g.}, \ac{AoA}/\ac{DoA}), and hybrid models that fuse distance and angle information~\cite{Yassin2017,Wymeersch2009,Naseri2019}.
Both \ac{SMDS} and in its quaternion-domain variants adopt such a hybrid framework, in which range estimates, \acp{ADoA} between edge vectors, and \acp{DoA} are jointly used to estimate node locations.

Given these measurements, most classical localization methods estimate $\bm{X}_\mathrm{T}$ by solving a nonlinear optimization problem, for example via maximum likelihood/least-squares formulations, Bayesian filtering or message passing techniques, or convex relaxation approaches such as semidefinite programming~\cite{Wymeersch2009,Shi2017,Biswas2006,Naseri2019}.
In contrast, \ac{SMDS} transforms the same measurement information into a structured matrix problem.
Specifically, distance and angle measurements are used to construct a low-rank \ac{GEK} matrix, from which edge-vector estimates and, subsequently, the target locations $\bm{X}_\mathrm{T}$ are recovered.
This separation of steps (measurement acquisition $\rightarrow$ \ac{GEK} construction $\rightarrow$ low-rank recovery $\rightarrow$ coordinate reconstruction) is a key feature that is exploited by the proposed quaternion-domain extensions.

The proposed quaternion-domain methods preserve the same sensing model and localization objective while modifying the \emph{algebraic representation} of \ac{3D} edges from real vectors to quaternions.
This modification alters the structure and rank of the corresponding \ac{GEK} matrix and will be shown to yield more robust position estimates, particularly in the presence of strong measurement noise and missing information.
}

%% file: TXT/4_QD-SMDS.tex
\begin{figure*}[!t]
\begin{center}
    	\subfigure[Cartesian coordinate system]{
	\includegraphics[width=0.50\columnwidth,keepaspectratio=true]{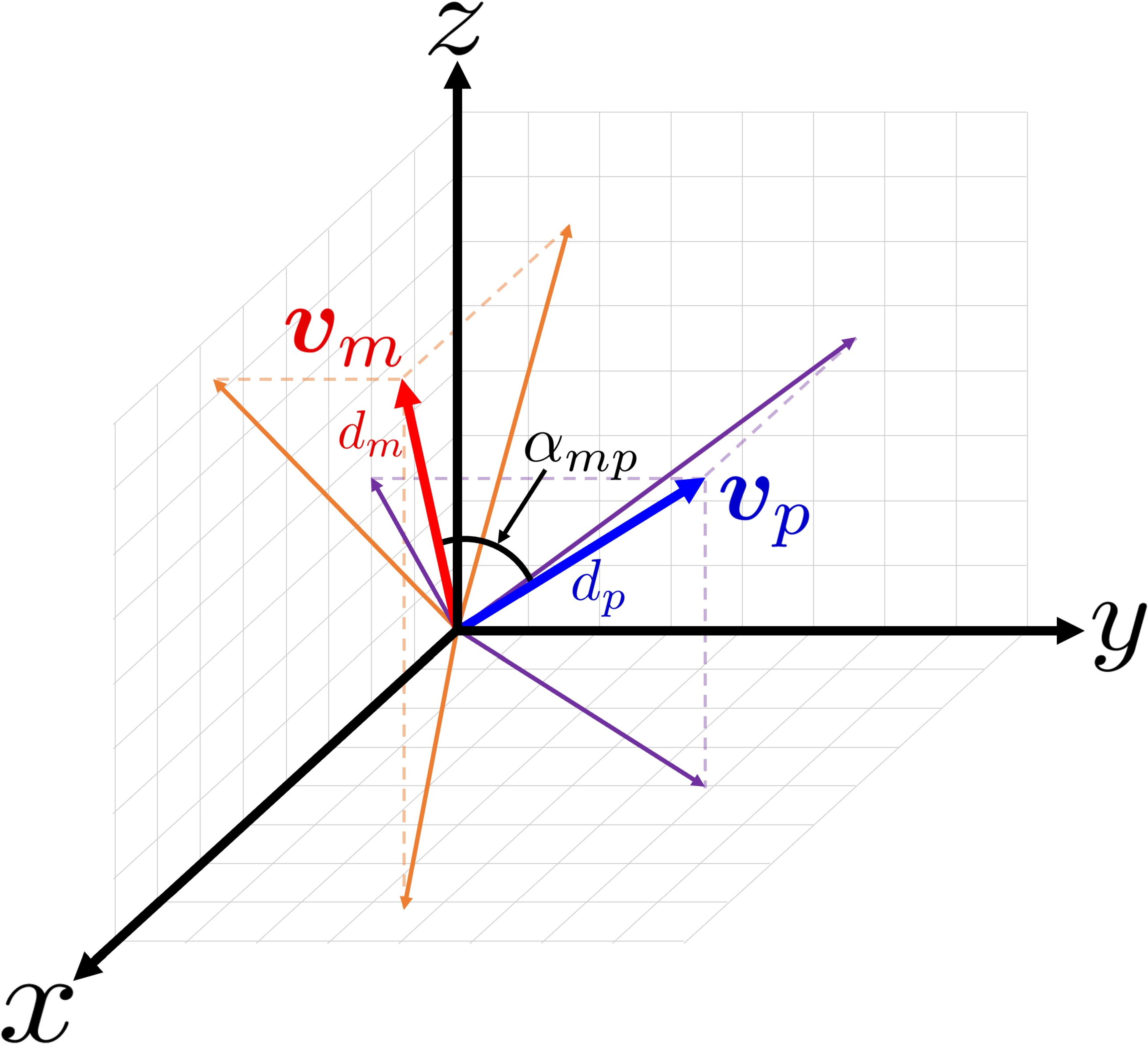}
	\label{fig:CSS}
	}
	\subfigure[$(\mathrm{x},\mathrm{y})$-plane]{
\includegraphics[width=0.42\columnwidth,keepaspectratio=true]{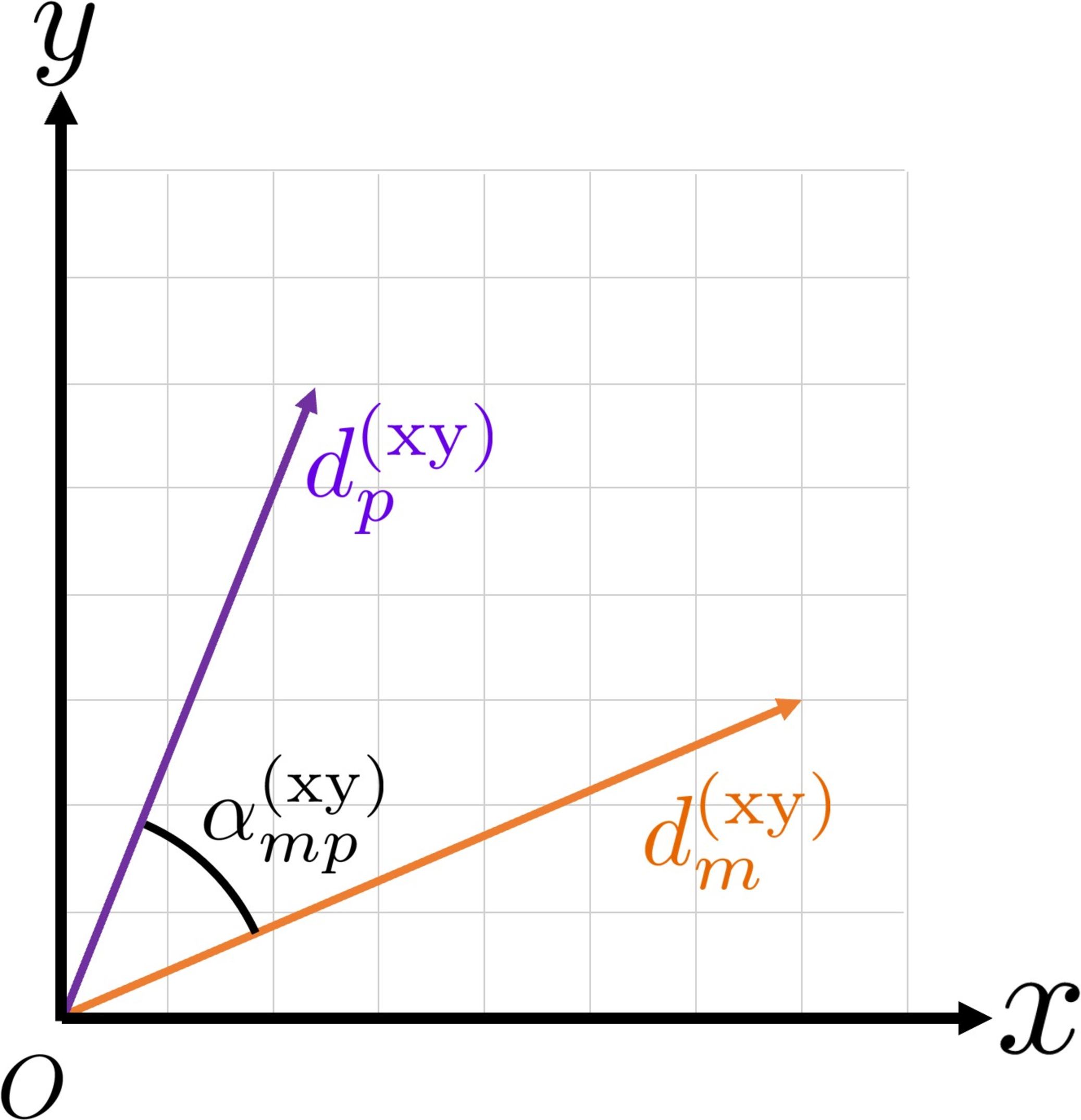}
	\label{fig:xy_plane}
	}
	\subfigure[$(\mathrm{x},\mathrm{z})$-plane]{
\includegraphics[width=0.42\columnwidth,keepaspectratio=true]{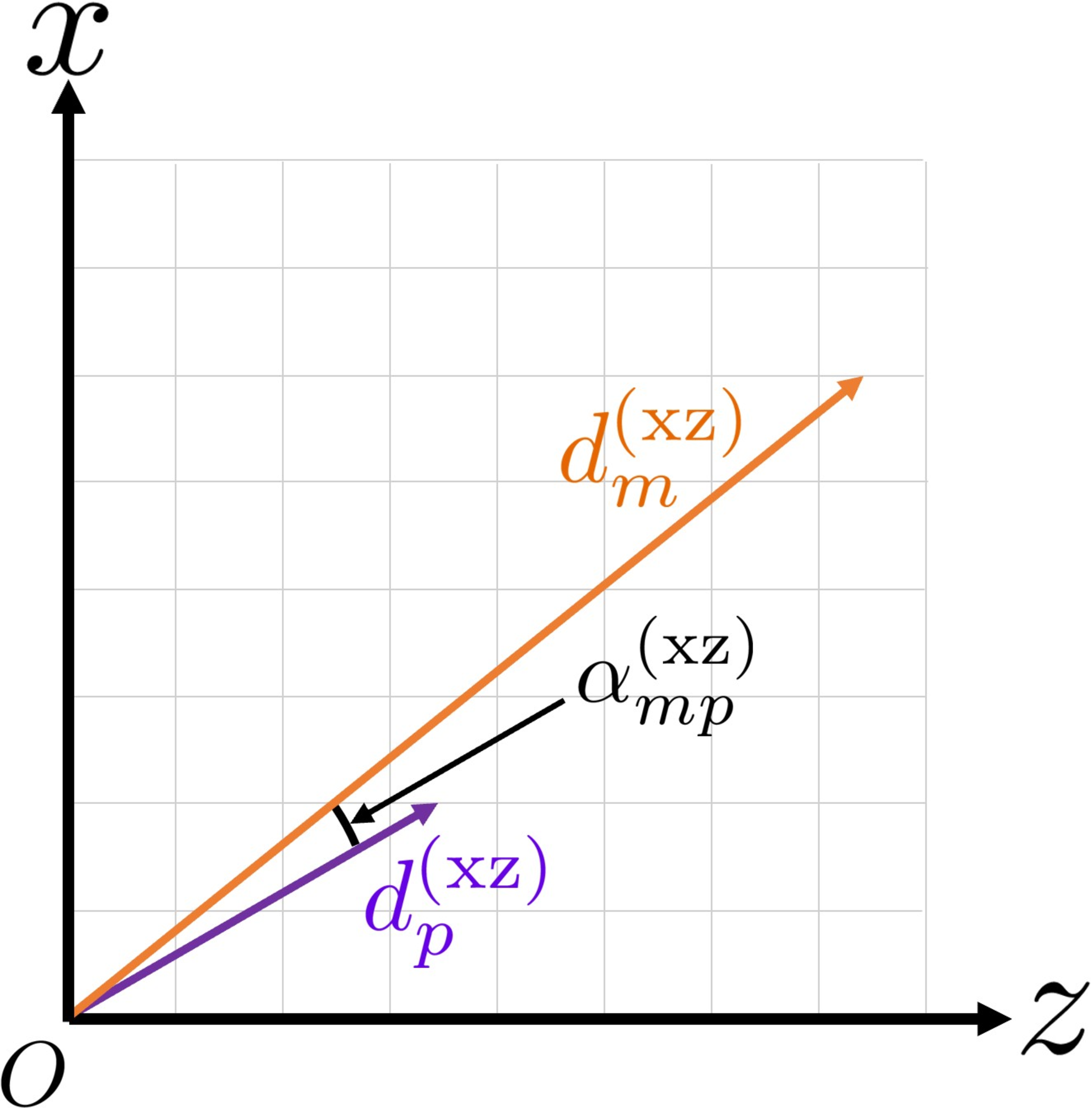}
	\label{fig:xz-plane}
	}
	\subfigure[$(\mathrm{y},\mathrm{z})$-plane]{
\includegraphics[width=0.42\columnwidth,keepaspectratio=true]{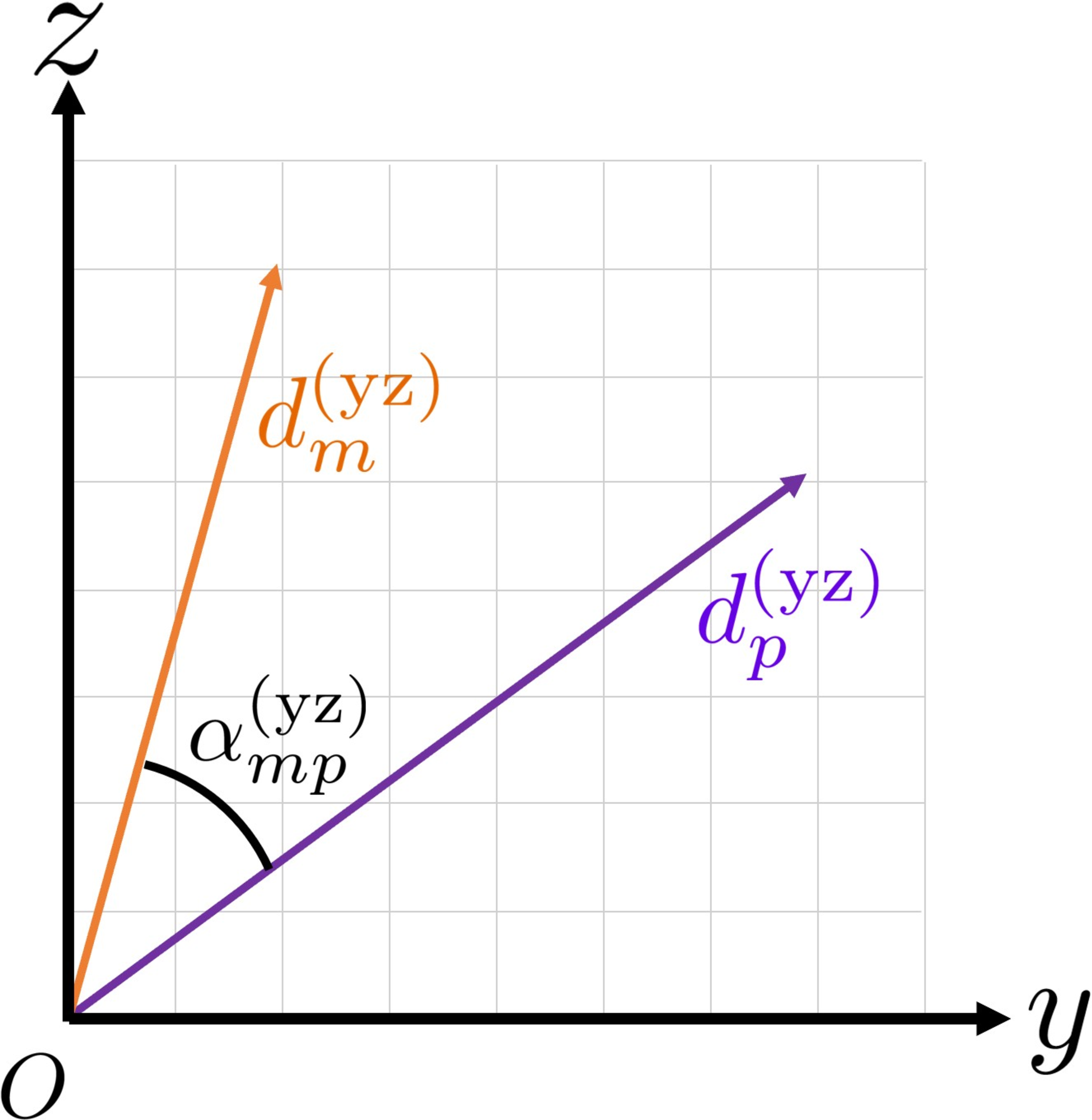}
	\label{fig:yz-plane}
	}
	\caption{
    Illustration of the parameters required to construct quaternion-domain \ac{GEK} matrix $\bm{K}_\mathrm{q}$.}
    \label{fig:param_q}
	\vspace{-2ex}
\end{center}
\end{figure*}

\subsection{\ac{QD-SMDS} Algorithm}
\label{sec:QD-SMDS}

\subsubsection{Derivation of the Proposed Algorithm}

In this subsection, we derive \ac{QD-SMDS}—an extension of the conventional \ac{SMDS} tailored for \ac{3D} localization—using the quaternion algebra introduced in Section II.
 
First, the \ac{3D} coordinates vector $\bm{x}_{n}$ of a generic node $n$ in the network can be alternatively expressed by the quaternion representation $\chi\in \mathbb{H}$ as
\begin{equation} 
\label{eq:q_x}
\bm{x}_n =\left[a_n, b_n, c_n\right]
\Longleftrightarrow
\chi_{n} = a_n+\mathbf{i}b_n+\mathbf{j}c_n+\mathbf{k}\cdot 0,
\end{equation}
where out of the four \ac{DoF}, three are used for the $(x,y,z)$ coordinates, and the remaining one is set to $0$\footnote{\setlength{\baselineskip}{10pt}The choice of embedding the \ac{3D} coordinates into these components is arbitrary.
In principle, any three of the four components can be used, and the resulting \ac{GEK} matrix retains the same structure, with the angle and distance matrices appearing with different imaginary units.
When extracting the estimated coordinates, the components corresponding to the three imaginary units selected to represent the $(x,y,z)$ coordinates must then be taken.}.
%

Accordingly, the quaternion coordinate vector corresponding to the real coordinate matrix defined in \eqref{eq:r_X} can be expressed as
\begin{equation}
    \bm{\chi} \triangleq
    \left[\chi_1,\ldots,\chi_N\right]\in\mathbb{H}^{N\times1}.
    \label{eq:q_X}
\end{equation}
 
Similarly, the edge vector $\bm{v}_m$ between any two nodes $\bm{x}_i$ and $\bm{x}_j$ in \eqref{eq:r_v} can be represented as
\begin{equation}
\label{eq:q_v}
\nu_{m}
=
(\underbrace{a_i-a_j}_{\grave{a}_m})
+\mathbf{i}
(\underbrace{b_i-b_j}_{\grave{b}_m})
+\mathbf{j}
(\underbrace{c_i-c_j}_{\grave{c}_m})
+\mathbf{k}
\cdot 0,
\end{equation}
where $\grave{a}_m\triangleq a_i-a_j,\grave{b}_m\triangleq b_i-b_j,\grave{c}_m\triangleq c_i-c_j$.
From the above, the quaternion edge vector corresponding to the real edge matrix defined in \eqref{eq:r_V} can be written as
\begin{equation}
    \bm{\nu}
    \triangleq
    \left[\nu_1,\ldots,\nu_m,\ldots,\nu_M\right]^\mathsf{T}
    =
    \bm{C}\bm{\chi}
    \in \mathbb{H}^{M\times 1},
    \label{eq:q_V}
\end{equation}
with the structure matrix $\bm{C}$ defined in \eqref{eq:C}.

Next, the \ac{QD-SMDS} algorithm is derived based on the quaternion formulation given in \eqref{eq:q_x}--\eqref{eq:q_V}.
In this context, the inner product defined in \eqref{eq:r_inner} can be rewritten as
\begin{eqnarray}
\!\!\!\!&\!\!\!\!&\!\!\!\!
\left<\bm{v}_m,\bm{v}_p\right>
\triangleq
\left[\grave{a}_m,\grave{b}_m,\grave{c}_m\right]
\begin{bmatrix}
\grave{a}_p\\
\grave{b}_p\\
\grave{c}_p
\end{bmatrix}
\nonumber \\
\!\!\!\!&\!\!\!\!&\!\!\!\!
\quad=
\grave{a}_m\grave{a}_p+\grave{b}_m\grave{b}_p
+\grave{c}_m\grave{c}_p =
d_{m}d_{p}\cos\alpha_{mp}.
\label{eq:inner_prod}
\end{eqnarray}

In turn, the outer product of the vectors obtained by projecting the edge vectors onto the $(\mathrm{x},\mathrm{y})$ plane can be expressed as
\begin{subequations}
\label{eq:outer_prod}
\begin{eqnarray}
\!\!\!\!&\!\!\!\!&\!\!\!\!
\left|\bm{v}_{m}^{(\mathrm{xy})}\times \bm{v}_{p}^{(\mathrm{xy})}\right|
\triangleq
\mathrm{det}
\left[
\begin{bmatrix}
\grave{a}_m\quad \grave{b}_m \\
\grave{a}_p\quad \grave{b}_p
\end{bmatrix}
\right]
\nonumber \\
\!\!\!\!&\!\!\!\!&\!\!\!\!
\quad=
\grave{a}_m\grave{b}_p-\grave{a}_p\grave{b}_m
=
d_{m}^{(\mathrm{xy})}d_{p}^{(\mathrm{xy})}\sin\alpha_{mp}^{(\mathrm{xy})}.
\label{eq:outer_xy}
\end{eqnarray}
 
Similarly, the outer products on the $(\mathrm{x},\mathrm{z})$ and $(\mathrm{y},\mathrm{z})$ planes, respectively, can be expressed as
\begin{equation}
\label{eq:outer_xz}
\left|\bm{v}_{m}^{(\mathrm{xz})}\times \bm{v}_{p}^{(\mathrm{xz})}\right|
=
\grave{a}_m\grave{c}_p-\grave{a}_p\grave{c}_m
=
d_{m}^{(\mathrm{xz})}d_{p}^{(\mathrm{xz})}\sin\alpha_{mp}^{(\mathrm{xz})},
\end{equation}
\begin{equation}
\label{eq:outer_yz}
\left|\bm{v}_{m}^{(\mathrm{yz})}\times \bm{v}_{p}^{(\mathrm{yz})}\right|
=
\grave{b}_m\grave{c}_p-\grave{b}_p\grave{c}_m
=
d_{m}^{(\mathrm{yz})}d_{p}^{(\mathrm{yz})}\sin\alpha_{mp}^{(\mathrm{yz})},
\end{equation}
\end{subequations}
where $\bm{v}_{m}^{(\mathrm{xy})}$, $\bm{v}_{m}^{(\mathrm{xz})}$, and $\bm{v}_{m}^{(\mathrm{yz})}$ denote the \ac{2D} vectors obtained by projecting $\bm{v}_m$ onto the $(\mathrm{x},\mathrm{y})$, $(\mathrm{x},\mathrm{z})$, and $(\mathrm{y},\mathrm{z})$ planes, respectively.

{ For clarity, the notation for the real and quaternion vectors, together with their corresponding projections, is summarized in Table~\ref{tab:position-notation}}.

\begin{table}[t]
\caption{ Edge Vector Notation used in Real and Quaternion Domains.}
\label{tab:position-notation}
\centering
\renewcommand{\arraystretch}{1.1}
{
\begin{tabular}{ll}
\toprule
Symbol & Meaning / domain \\
\toprule
$\bm{v}_m$ & \makecell[l]{$m$-th edge vector in $\mathbb{R}^3$ \\ (\textit{e.g.}, $\bm{v}_m=[\grave{a}_m,\,\grave{b}_m,\,\grave{c}_m]^\mathsf{T}$)} \\
\midrule
$\nu_m$ & \makecell[l]{Quaternion representation of $\bm{v}_m$; \\ $\nu_m = \grave{a}_m + \mathbf{i} \grave{b}_m + \mathbf{j} \grave{c}_m + \mathbf{k}\cdot 0$} \\
\midrule
$\bm{v}_m^{(xy)}$ & \makecell[l]{$xy$-projection of $\bm{v}_m$ in $\mathbb{R}^2$ \\ (\textit{e.g.}, $\bm{v}_m^{(xy)}=[\grave{a}_m,\,\grave{b}_m]^\mathsf{T}$)} \\
\midrule
$\bm{v}_m^{(xz)}$ & $xz$-projection of $\bm{v}_m$ in $\mathbb{R}^2$ \\
\midrule
$\bm{v}_m^{(yz)}$ & $yz$-projection of $\bm{v}_m$ in $\mathbb{R}^2$ \\
\bottomrule
\end{tabular}
}
\end{table}

The corresponding Euclidean norms (distances) are given by $d_{m}^{(\mathrm{xy})}\triangleq\|\bm{v}_{m}^{(\mathrm{xy})}\|$, $d_{m}^{(\mathrm{xz})}\triangleq\|\bm{v}_{m}^{(\mathrm{xz})}\|$, and $d_{m}^{(\mathrm{yz})}\triangleq\|\bm{v}_{m}^{(\mathrm{yz})}\|$. 
Similarly, $\alpha_{mp}^{(\mathrm{xy})}$, $\alpha_{mp}^{(\mathrm{xz})}$, and $\alpha_{mp}^{(\mathrm{yz})}$ represent the \acp{ADoA} between two \ac{2D} projected vectors on the $(\mathrm{x},\mathrm{y})$, $(\mathrm{x},\mathrm{z})$, and $(\mathrm{y},\mathrm{z})$ planes, respectively.
For clarity of these parameters, please refer to Fig.\ref{fig:param_q} at the top of this page.

Based on \eqref{eq:inner_prod}--\eqref{eq:outer_prod}, the product of the quaternion edges $\nu_{m}$ and $\nu_{p}^*$, with $m\neq p$, can be expressed as
\begin{eqnarray}
\label{eq:q_prod}
\hspace{-5ex}\nu_{m}\nu_{p}^*
\hspace{-4ex} && =
\underbrace{
\left(\grave{a}_m\grave{a}_p+\grave{b}_m\grave{b}_p+\grave{c}_m\grave{c}_p\right)
}_{\left<\bm{v}_m,\bm{v}_p\right>}
+
\mathbf{i}
\underbrace{
\left(\grave{a}_p\grave{b}_m-\grave{a}_m\grave{b}_p\right)
}_{-\left|\bm{v}_{m}^{(\mathrm{xy})}\times \bm{v}_{p}^{(\mathrm{xy})}\right|}
\nonumber\\
&&\quad + \mathbf{j}
\underbrace{
\left(\grave{a}_p\grave{c}_m-\grave{a}_m\grave{c}_p\right)
}_{-\left|\bm{v}_m^{(\mathrm{xz})}\times \bm{v}_p^{(\mathrm{xz})}\right|}
+
\mathbf{k}
\underbrace{
\left(\grave{b}_p\grave{c}_m-\grave{b}_m\grave{c}_p\right)
}_{-\left|\bm{v}_{m}^{(\mathrm{yz})}\times \bm{v}_p^{(\mathrm{yz})}\right|}
\nonumber\\
&&  = d_{m}d_{p}\cos\alpha_{mp}
-
\mathbf{i}
d_m^{(\mathrm{xy})}d_p^{(\mathrm{xy})}\sin\alpha_{mp}^{(\mathrm{xy})}\nonumber \\
&&\quad -
\mathbf{j}
d_m^{(\mathrm{xz})}d_p^{(\mathrm{xz})}\sin\alpha_{mp}^{(\mathrm{xz})} -\mathbf{k}
d_m^{(\mathrm{yz})}d_p^{(\mathrm{yz})}\sin\alpha_{mp}^{(\mathrm{yz})}\!.
\end{eqnarray}

%
%

%
\begin{figure*}[t]
\begin{eqnarray}
\label{eq:q_kernel}
\hspace{-3em}&&
\bm{K}_\mathrm{q}
\triangleq
\bm{\nu}\bm{\nu}^\mathsf{H} 
=
\mathrm{diag}\left[
\bm{d}
\right]
\begin{bmatrix}
\cos\alpha_{11} & \cdots & \cos\alpha_{1M} \\
\vdots & \ddots & \vdots \\
\cos\alpha_{M1} & \cdots & \cos\alpha_{MM}
\end{bmatrix}
\mathrm{diag}\left[
\bm{d}
\right]
-
\mathbf{i}
\;
\mathrm{diag}\left[
\bm{d}^{(\mathrm{xy})}
\right]
\begin{bmatrix}
\sin\alpha_{11}^{(\mathrm{xy})} \!\!\!&\!\!\! \cdots \!\!\!&\!\!\! \sin\alpha_{1M}^{(\mathrm{xy})} \\
\vdots & \ddots & \vdots \\
\sin\alpha_{M1}^{(\mathrm{xy})} \!\!\!&\!\!\! \cdots \!\!\!&\!\!\! \sin\alpha_{MM}^{(\mathrm{xy})}
\end{bmatrix}
\mathrm{diag}\left[
\bm{d}^{(\mathrm{xy})}
\right]
\nonumber \\
\hspace{-3em}&&\quad
-
\mathbf{j}
\;
\mathrm{diag}
\left[
\bm{d}^{(\mathrm{xz})}
\right]
\begin{bmatrix}
\sin\alpha_{11}^{(\mathrm{xz})} \!\!\!&\!\!\! \cdots \!\!\!&\!\!\! \sin\alpha_{1M}^{(\mathrm{xz})} \\
\vdots & \ddots & \vdots \\
\sin\alpha_{M1}^{(\mathrm{xz})} \!\!\!&\!\!\! \cdots \!\!\!&\!\!\! \sin\alpha_{MM}^{(\mathrm{xz})}
\end{bmatrix}
\mathrm{diag}
\left[
\bm{d}^{(\mathrm{xz})}
\right]
-
\mathbf{k}
\;
\mathrm{diag}
\left[
\bm{d}^{(\mathrm{yz})}
\right]
\begin{bmatrix}
\sin\alpha_{11}^{(\mathrm{yz})} \!\!\!&\!\!\! \cdots \!\!\!&\!\!\! \sin\alpha_{1M}^{(\mathrm{yz})} \\
\vdots & \ddots & \vdots \\
\sin\alpha_{M1}^{(\mathrm{yz})} \!\!\!&\!\!\! \cdots \!\!\!&\!\!\! \sin\alpha_{MM}^{(\mathrm{yz})}
\end{bmatrix}
\mathrm{diag}
\left[
\bm{d}^{(\mathrm{yz})}
\right].
\end{eqnarray}
\hrule
\vspace{-3mm}
\end{figure*}

From \eqref{eq:q_prod}, the rank-$1$ quaternion-domain \ac{GEK} matrix, which incorporates all pairwise distances and \ac{ADoA} information among the nodes, is given in \eqref{eq:q_kernel}, shown at the top of the next page.
Assuming that measured values of all pairwise distance and \ac{ADoA} parameters appearing in \eqref{eq:q_prod} are available, the quaternion-domain \ac{GEK} matrix with measurement errors, denoted as $\tilde{\bm{K}}_\mathrm{q}$, can be constructed. 
Accordingly, the estimate of the quaternion edge vector $\bm{\nu}$ is given as
\begin{equation}
\label{eq:q_v_estimated}
    \hat{\bm{\nu}}=\sqrt{\lambda}\bm{u},
\end{equation}
where $(\lambda,\bm{u})$ denotes the pair consisting of the largest eigenvalue and its corresponding eigenvector of $\tilde{\bm{K}}_\mathrm{q}$.
%

The \ac{SVD} of $\tilde{\bm{K}}_\mathrm{q}$ is computed using the \ac{QSVD} method introduced in Section~\ref{subsec:QSVD}.
{Since $\tilde{\bm{K}}_\mathrm{q}$ has rank $1$ by construction, its equivalent complex representation has rank $2$ with identical eigenvalues~\cite{Zhang1997}}.
Furthermore, because the quaternion-domain \ac{GEK} $\tilde{\bm{K}}_{\mathrm{q}}$ is regular in our setting, the quaternion eigendecomposition is effectively equivalent to the \ac{QSVD}.
%

In the conventional \ac{SMDS} algorithm, the \ac{GEK} matrix is constructed based on the real-valued vectors described in \eqref{eq:r_v}. 
As a result, its rank is limited to $3$, as clearly shown in \eqref{eq:r_V_estimated}, which inherently restricts the noise reduction capability achievable through low-rank truncation via \ac{SVD}.
In contrast, the proposed \ac{QD-SMDS} algorithm utilizes a quaternion-domain representation, enabling the \ac{GEK} matrix to have rank-$1$.
This low-rank structure allows for maximal noise reduction, thereby significantly enhancing the robustness of localization performance against measurement errors.

Finally, we estimate the real-valued coordinate matrix $\bm{X}$ in \eqref{eq:r_X} from the estimated quaternion edge vector $\hat{\bm{\nu}}$.
First, the real, $\bm{\mathrm{i}}$-, and $\bm{\mathrm{j}}$-components of $\hat{\bm{\nu}}$ are extracted and rearranged according to the $(\mathrm{x},\mathrm{y},\mathrm{z})$ coordinates based on \eqref{eq:q_v}, yielding the estimated real-valued edge matrix as
\begin{equation}
\label{eq:V_rearrange}
    \hat{\bm{V}} \triangleq \left[\hat{\bm{v}}_1 , \cdots , \hat{\bm{v}}_M \right]^{\mathsf{T}} \in \mathbb{R}^{M \times 3}.
\end{equation}

Given the estimated edge vector matrix $\hat{\bm{V}}$, the same procedure as in \eqref{eq:MoorePenrose} can be applied to obtain the estimated coordinate matrix $\hat{\bm{X}}$.
Finally, a Procrustes transformation is applied to $\hat{\bm{X}}$ using the known positions of \acp{AN}.

{The computational complexity of the \ac{QD-SMDS} algorithm is dominated by the \ac{QSVD} operation, which corresponds to a truncated \ac{SVD} of the complex-valued representation of the quaternion kernel matrix in \eqref{eq:q_kernel}, having size $2M\times 2M$.
Therefore, the asymptotic complexity is $O\left(2(2M)^2\right)=O\left(M^2\right)=O\left(N_{\mathrm{A}}^2 (N_{\mathrm{A}}+N_\mathrm{T})^2\right)$, which is the same as that of the \ac{SMDS} algorithm.}

\subsubsection{Construction of the Quaternion-Domain GEK Matrix}
\label{subsec:QD-SMDS scenario}

To construct the quaternion-domain \ac{GEK} matrix in \eqref{eq:q_kernel}, additional phase difference information is required—beyond the typically measurable pairwise distance $d_m$ and \ac{ADoA} $\alpha_{mp}$—when the positional relationship between nodes is projected onto each plane.
Depending on the extent to which this additional information can be obtained from measurements, two practical scenarios are considered.

The first scenario, referred to as \textbf{Scenario I}, assumes that only the mutual distances $d_m$ and \acp{ADoA} $\alpha_{mp}$ between nodes are available, with no additional angular information.
Under this condition, only the real-domain \ac{GEK} matrix used in the conventional \ac{SMDS} algorithm, as defined in \eqref{eq:r_kernel}, can be directly constructed from measurements; the quaternion-domain \ac{GEK} cannot.
Therefore, it is necessary to first execute the conventional \ac{SMDS} to estimate the coordinates of \acp{TN}.
Based on these estimated positions, the angular information required for constructing the quaternion-domain \ac{GEK} matrix can then be computed.
With this information in place, the \ac{QD-SMDS} algorithm can subsequently be executed to refine the positioning accuracy. 

The second scenario, referred to as \textbf{Scenario II}, assumes that azimuth and elevation angles can be measured by using planar antennas, as proposed in~\cite{Zhang2018}, deployed at each \ac{AN}.
By appropriately orienting the planar antennas, it becomes possible to obtain the parameters illustrated in Fig. \ref{fig:azimuth,elevation}, where $\theta$ denotes the elevation angle and $\phi$ denotes the azimuth angle.
In the figure, red, blue, and green colors indicate the parameters obtained by projecting the edge vectors onto the $(\mathrm{x},\mathrm{y})$, $(\mathrm{x},\mathrm{z})$, and $(\mathrm{y},\mathrm{z})$ planes, respectively.

From the above, the quaternion-domain \ac{GEK} matrix $\bm{K}_{\mathrm{q}} $ can be constructed based on all measurable all pairwise distances and angular information:
\begin{equation}
\label{eq:mpara}
d_m,\theta_{m}^{(\mathrm{x})},\theta_{m}^{(\mathrm{y})},\theta_{m}^{(\mathrm{z})},\phi_{m}^{(\mathrm{xy})},\phi_{m}^{(\mathrm{yz})},\phi_{m}^{(\mathrm{xz})},\alpha_{mp},
\end{equation}
where the following parameters must be computed in advance from the quantities listed in \eqref{eq:mpara} as
\begin{subequations}
\begin{eqnarray}
\label{eq:d2}
   d_{m}^{(\mathrm{xy})}&=&d_m  \sin \theta_{m}^{(\mathrm{z})}, \\
   d_{m}^{(\mathrm{xz})}&=&d_m  \sin \theta_{m}^{(\mathrm{y})},\\
   d_{m}^{(\mathrm{yz})}&=&d_m  \sin \theta_{m}^{(\mathrm{x})},\\
\label{eq:t2}
   \alpha_{mp}^{(\mathrm{xy})}&=&\phi_{p}^{(\mathrm{xy})}-\phi_{m}^{(\mathrm{xy})},\\
   \alpha_{mp}^{(\mathrm{xz})}&=&\phi_{p}^{(\mathrm{xz})}-\phi_{m}^{(\mathrm{xz})}, \\
   \alpha_{mp}^{(\mathrm{yz})}&=&\phi_{p}^{(\mathrm{yz})}-\phi_{m}^{(\mathrm{yz})}.
\end{eqnarray}
\end{subequations}
%
\begin{figure}[!t]
\begin{center}
	\includegraphics[width=0.80\columnwidth,keepaspectratio=true]{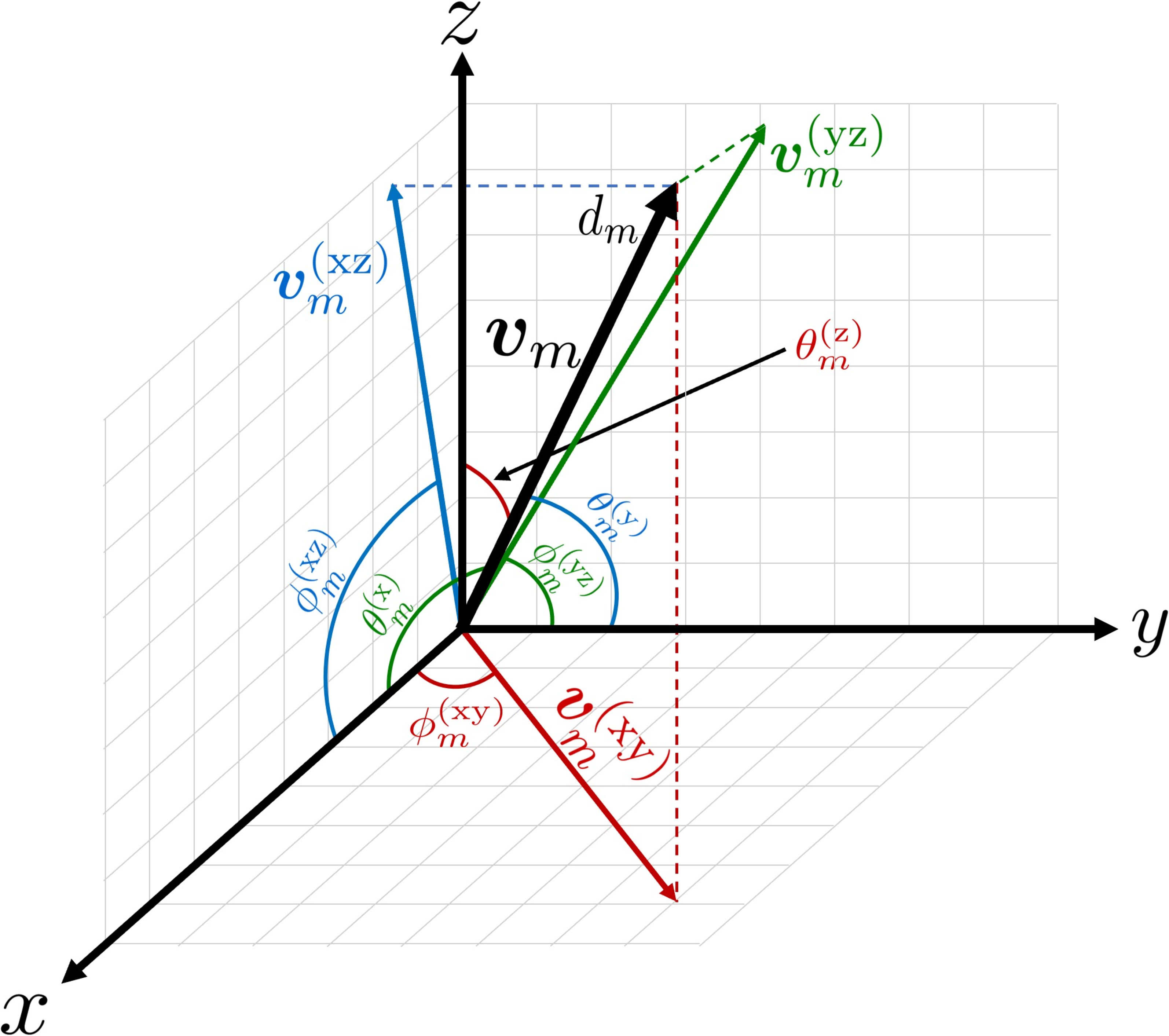}
	\caption{Parameters that can be obtained using a planar antenna.}
	\label{fig:azimuth,elevation}
	\vspace{-3ex}
\end{center}
\end{figure}

Finally, this subsection concludes with a summary of the \ac{QD-SMDS} algorithm in the form of pseudo-code, presented in Algorithm \ref{alg:QD-SMDS}, {together with a high-level architectural overview of the processing pipeline shown in Fig.~\ref{fig:qd_pipeline}}.

\begin{figure*}[t]
\centering
\begin{tikzpicture}[
    font=\small,
    block/.style={draw, rounded corners, align=center, minimum height=9mm, minimum width=28mm},
    line/.style={-{Stealth[length=2mm]}, thick}
]

\node[block] (in) {\textbf{Inputs}\\
Distances and \acp{ADoA},\\
coordinates of at least 4 \acp{AN}};

\node[block, right=6mm of in] (gek) {\textbf{Quaternion \ac{GEK}}\\\textbf{construction}\\
$\tilde{\bm{K}}_{\mathrm{q}}$ \eqref{eq:q_kernel}};

\node[block, right=6mm of gek] (lr) {\textbf{Low-rank}\\\textbf{processing}\\
(QSVD)};

\node[block, right=6mm of lr] (rec) {\textbf{Recovery}\\
$\hat{\bm{V}}\ \rightarrow\ \hat{\bm{X}}$};

\node[block, right=6mm of rec] (corr) {\textbf{Alignment}\\
(Procrustes)};

\draw[line] (in) -- (gek);
\draw[line] (gek) -- (lr);
\draw[line] (lr) -- (rec);
\draw[line] (rec) -- (corr);

\end{tikzpicture}
\caption{High-level architecture of the proposed quaternion-domain localization pipeline.}
\label{fig:qd_pipeline}

\vspace{-2ex}
\end{figure*}

\subsection{Handling Incomplete Data}
\label{sec:Imcomplete Data}

In the previous sections, it was assumed that all measurements between \acp{AN} and \acp{TN} were fully available, \textit{i.e.}, obtained without any missing data.
However, in practical scenarios, partial information loss may occur due to environmental factors such as \ac{NLOS} conditions.
In such cases, it becomes necessary to apply a matrix completion algorithm prior to executing the \ac{SMDS} or \ac{QD-SMDS} algorithms.

A typical example is the scenario in which some of the pairwise distance measurements in the network are missing.
In such case, the \ac{EDM} completion algorithm proposed in ~\cite{Dokmanic2015} can be employed to recover the missing entries.
Consequently, most studies assume that the pairwise distances are either directly measured or completed in advance.

Another practical scenario arises when the angle measurements in the network are partially unavailable, resulting in a partially observed \ac{GEK} matrix.
In this case, two different completion approaches can be considered.
The first approach leverages the fact that the \ac{EDM} is fully observed.
Classical \ac{MDS} algorithm can be applied to estimate the coordinates of \acp{TN}, from which the missing angle information can be inferred.
The second approach exploits the sparsity of the \ac{GEK} matrix itself.
A matrix completion method based on low-rank approximation can be directly applied to the sparse \ac{GEK} matrix.
For the real-domain \ac{GEK} matrix used in the \ac{SMDS} algorithm, the method proposed in~\cite{shabat2012} is sufficient.

\begin{algorithm}[!t]
\caption{Quaternion-Domain SMDS}\label{alg:QD-SMDS}
\begin{algorithmic}[1]
\Statex {\bf{Input:}}
\State \textit{Measured and estimated pairwise distances and \acp{ADoA}:
$d_{m},d_m^{(\mathrm{xy})},d_m^{(\mathrm{xz})},d_m^{(\mathrm{yz})},\alpha_{mp},\alpha_{mp}^{(\mathrm{xy})},\alpha_{mp}^{(\mathrm{xz})},\alpha_{mp}^{(\mathrm{yz})}$}
\State \textit{The coordinates of at least 4 ANs.}
\Statex {\bf{Steps:}}
\State \textit{Construct the quaternion-domain \ac{GEK} matrix $\tilde{\bm{K}}$ in \eqref{eq:q_kernel} using the input parameters.}
\State \textit{Perform \ac{QSVD} of the constructed \ac{GEK} matrix $\tilde{\bm{K}}$} \text{(see Section II-\ref{subsec:QSVD})}.
\State \textit{Obtain the edge vector $\hat{\bm{\nu}}$ using Eq. \eqref{eq:q_v_estimated}}.
\State \textit{Convert the estimated quaternion edge vector $\hat{\bm{\nu}}$ to the estimated real-valued edge matrix $\hat{\bm{V}}$.}
\State \textit{Compute $\hat{\bm{X}}$ from $\hat{\bm{V}}$ using Eq. \eqref{eq:MoorePenrose}}.
\State \textit{Apply the Procrustes transform to $\hat{\bm{X}}$ if needed} \text{(\textit{e.g.}, see \cite{Fiore2001}).}
\end{algorithmic}
\end{algorithm}


However, this algorithm cannot be directly applied to the quaternion-domain \ac{GEK} matrix in \ac{QD-SMDS}.
To address this, advanced quaternion matrix completion algorithms, such as those proposed in~\cite{Miao2019}, can be employed to recover the missing entries.
As demonstrated in the simulation results presented later, the \ac{QD-SMDS} algorithm tends to outperform the \ac{SMDS} algorithm in scenarios where angle measurements are partially unavailable.
This advantage arises from the fact that low-rank matrix completion methods generally yield better results when applied to large-scale matrices with lower rank.
While both \ac{SMDS} and \ac{QD-SMDS} construct \ac{GEK} of size $M \times M$, the rank of the matrix in \ac{SMDS} is $3$, whereas in \ac{QD-SMDS} it is only $1$.
This lower rank facilitates more accurate completion, thereby enhancing localization performance.
Even when using classical matrix completion methods based on the nuclear norm, such as the one proposed in~\cite{shabat2012}, the \ac{QD-SMDS} algorithm is still expected to outperform the \ac{SMDS} algorithm.
Although these methods cannot be directly applied to quaternion matrices, the quaternion-domain \ac{GEK} matrix can be decomposed into two complex-valued matrices, $\bm{Q}_{\mathrm{a}}$ and $\bm{Q}_{\mathrm{b}}$, via the Cayley-Dickson construction in~\eqref{eq:Cayley-Dickson}.
Matrix completion is then performed separately on these two complex matrices.
Since each has rank $2$, the completion tends to be more effective than applying the same algorithm directly to the rank-$3$ \ac{GEK} matrix in the \ac{SMDS} algorithm.
In other words, \ac{QD-SMDS} inherently exhibits greater robustness to missing data due to its structural properties, as shown in the following section.


\vspace{-1ex}
\subsection{Performance Assessment}
\label{sec:Simulation}

\subsubsection{Simulation Conditions}
\label{sec:Evironment}

Computer simulations were conducted to evaluate the performance of the proposed \ac{QD-SMDS} algorithm.
The simulation environment assumes a room with a dimensions of $30$m (length) $\times$ $30$m (width) $\times$ $10$m (height). 
\acp{AN} were placed at five locations: the four upper corners of the room, specifically at $(x,y,z) = (0,0,10),(30,0,10),(30,30,10),$ and $(0,30,10)$, as well as the origin $(x,y,z)=(0,0,0)$.
\acp{TN} were randomly placed at 15 locations within the interior, with their $\mathrm{x},\mathrm{y}$, and $\mathrm{z}$ coordinates independently drawn from a uniform distribution.

Distance measurements are modeled as Gamma-distributed random variables~\cite{Papoulis2002} with the mean equal to the true distance $d$ and a standard deviation of $\sigma_d$.
The \ac{PDF} of the measured distances $\tilde{d}$ corresponding to the true distance $d$ is given by
\begin{equation}
\label{eq:pdf_d}
    p_{\mathrm{D}}(d;\alpha.\beta)
    =
    \left(\beta^\alpha
    \Gamma(\alpha)
    \right)^{-1}
    \tilde{d}^{\left(\alpha-1\right)}
    e^{\frac{\tilde{d}}{\beta}}.\,
\end{equation}
where $\alpha\triangleq d^2/\sigma^2_d$ and $\beta\triangleq \sigma^2_d/d$.

In turn, angle measurement errors $\delta_{\theta}$ are assumed to follow a Tikhonov-distribution~\cite{Viterbi1966,Abreu2008}{, with the} \ac{PDF} of the measured angle $\tilde{\theta} = \theta + \delta_{\theta}$ corresponding to the true angle $\theta$ is given by
\begin{equation}
\label{eq:pdf_theta}
p_{\mathrm{\Theta}}\left(\tilde{\theta};\theta,\rho \right)
=
\frac{1}{2\pi I_0(\rho)}\exp\left[\rho\cos(\theta-\tilde{\theta})\right].
\end{equation}
 
The range of the angular error is determined by the angular parameter $\epsilon$, which represents the bounding angle of the central $90^{\mathrm{th}}$ percentile and is expressed as
\begin{equation}
\label{eq:angle_error}
\epsilon\triangleq\theta_{\mathrm{B}} \Big| \int_{-\theta_{\mathrm{B}}}^{\theta_{\mathrm{B}}} p_{\Theta}(\phi ; 0, \rho) d\phi=0.9.
\end{equation}

{The following performance metrics are considered the evaluations hereafter:
\begin{enumerate}
    \item \textbf{Average estimation error}, defined as
    \begin{equation}
        \label{eq:mse}
        \xi \triangleq \frac{1}{L} \sum_{l=1}^{L} \frac{1}{N_\mathrm{T}}\left\|\hat{\bm{X}}_{\mathrm{T}}^{(l)} - \bm{X}_{\mathrm{T}}^{(l)}\right\|_{\mathrm{F}}
        =
        \frac{1}{L}\sum_{l=1}^{L} \xi^{(l)},
    \end{equation}
    where $\hat{\bm{X}}_{\mathrm{T}}^{(l)}$ and $\bm{X}_{\mathrm{T}}^{(l)}$ denote the estimated and true coordinate matrices of the TNs, respectively, in the $l$-th Monte Carlo trial; $L$ is the total number of Monte Carlo trials; and $\xi^{(l)} \triangleq \frac{1}{N_\mathrm{T}}\big\|\hat{\bm{X}}_{\mathrm{T}}^{(l)} - \bm{X}_{\mathrm{T}}^{(l)}\big\|_{\mathrm{F}}$ represents the normalized Frobenius norm of the estimation error in the $l$-th trial.
    \item \textbf{Empirical \ac{CDF} of the localization error}, defined as
    \begin{equation}
        F_E(r) = \frac{1}{L} \sum_{l=1}^{L} \mathbf{1}\{\xi^{(l)} \leq r\}, \quad r \geq 0,
    \end{equation}
    where $\mathbf{1}\{\cdot\}$ is the indicator function.
\end{enumerate}
}

\subsubsection{Simulation Results Without Missing Data}
\label{sec:Result}

\begin{figure}[!t]
\centering
    \subfigure{\includegraphics[width=0.95\columnwidth,keepaspectratio=true]{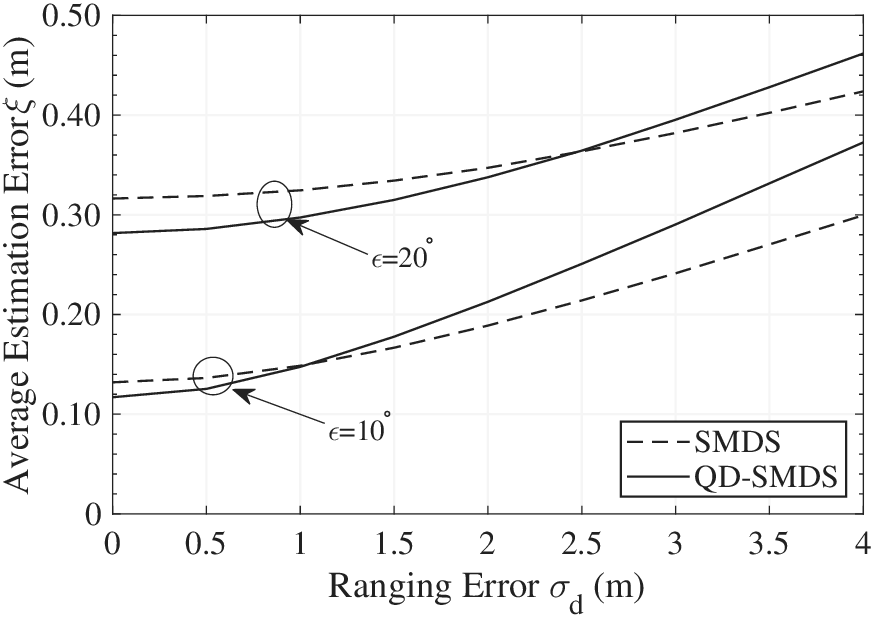}
    }   
    \vspace{1mm}   \subfigure{\includegraphics[width=0.95\columnwidth,keepaspectratio=true]{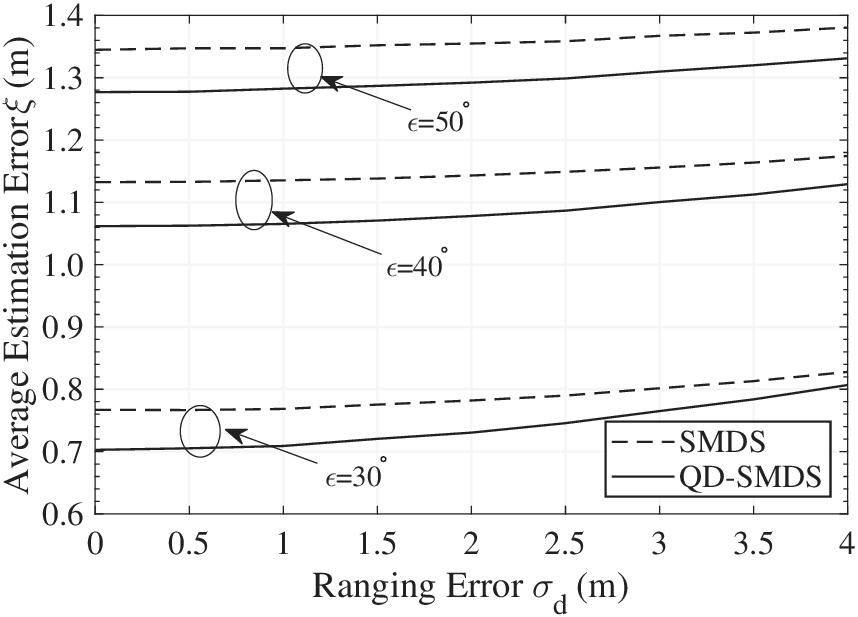}
    }
    \vspace{-2ex}
    \caption{Comparison of { average estimation error} between the \ac{SMDS} and \ac{QD-SMDS} algorithms in \textbf{Scenario I}.}
    \label{fig:SMDSvsQDSMDS_scenarioI}
\end{figure}

\begin{figure}[!t]
\centering
    \subfigure{
    \includegraphics[width=0.95\columnwidth,keepaspectratio=true]{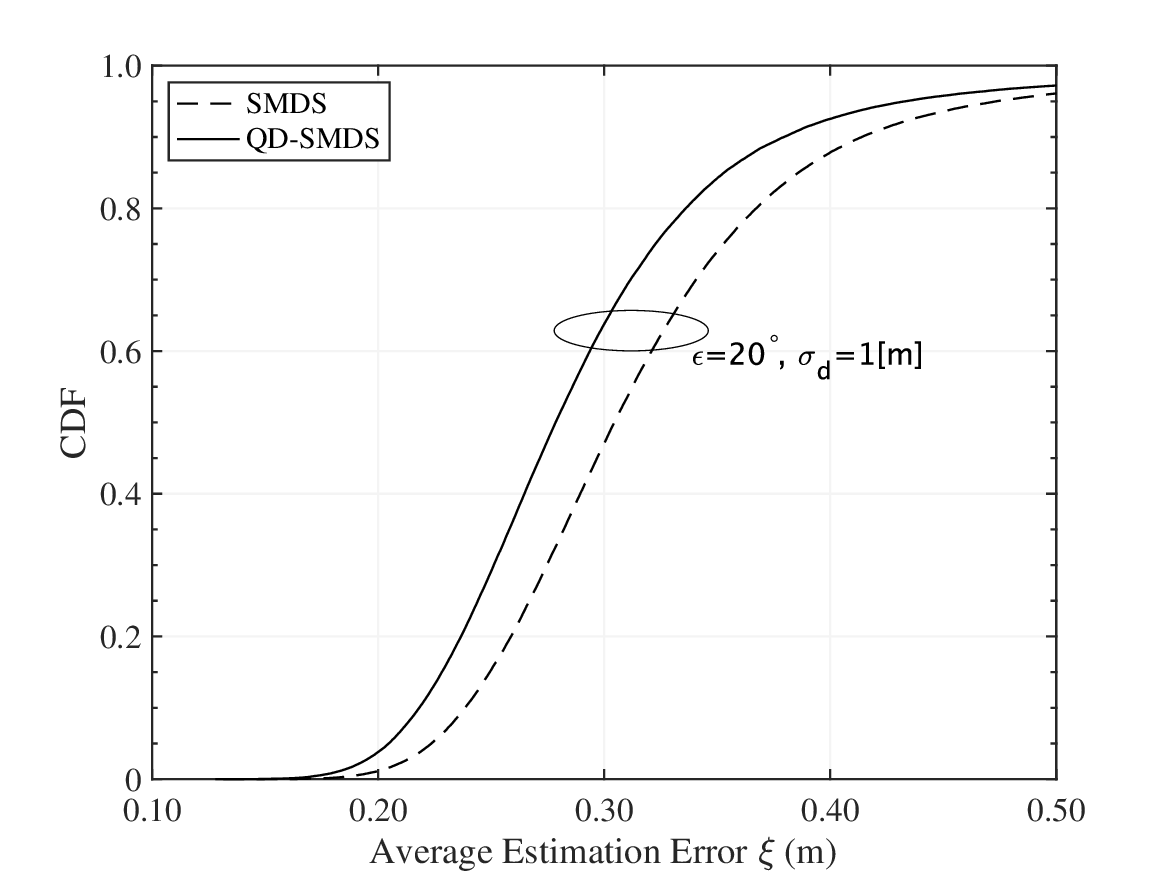}
    }   
    \vspace{1mm}   
    \subfigure{
    \includegraphics[width=0.95\columnwidth,keepaspectratio=true]{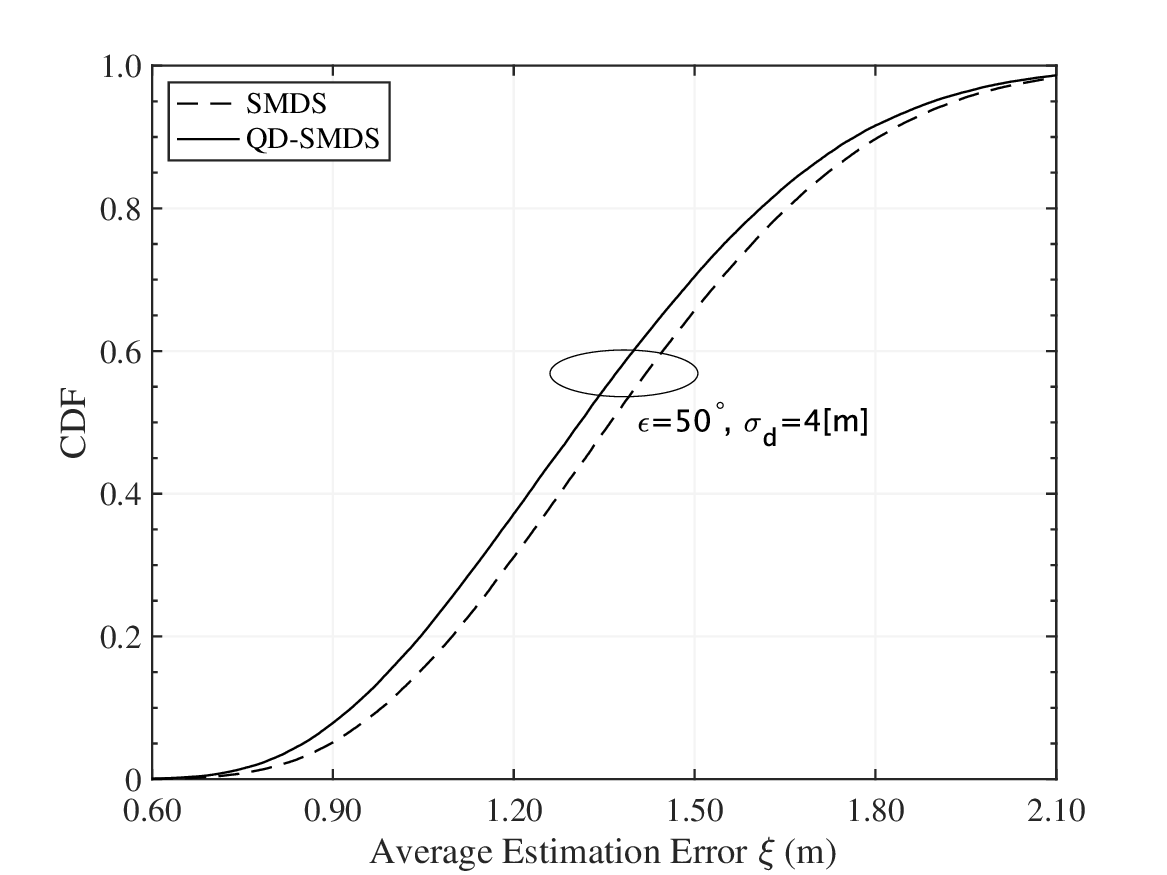}
    }
    \vspace{-2ex}
    \caption{Comparison of the empirical \ac{CDF} between the \ac{SMDS} and \ac{QD-SMDS} algorithms in \textbf{Scenario I}.}
    \label{fig:SMDSvsQDSMDS_scenarioI_CDF}
\end{figure}

Based on the above simulation conditions, we have compared the localization accuracy of the conventional \ac{SMDS} and the proposed \ac{QD-SMDS} algorithms under both \textbf{Scenario I} and \textbf{Scenario II} described in Section IV-A.
Fig. \ref{fig:SMDSvsQDSMDS_scenarioI} and { \ref{fig:SMDSvsQDSMDS_scenarioI_CDF}} show the localization accuracy comparison between \ac{SMDS} and \ac{QD-SMDS} in \textbf{Scenario I}.
{In Fig.~\ref{fig:SMDSvsQDSMDS_scenarioI}}, the horizontal axis represents the standard deviation of the Gamma-distributed distance measurements, as defined in~\eqref{eq:pdf_d}, while the vertical axis indicates the averaged estimation error, as defined in~\eqref{eq:mse}.
The results are plotted for various angular measurement errors $\epsilon\in \left\{10^\circ,20^\circ,30^\circ,40^\circ,50^\circ\right\}$.
{In Fig.~\ref{fig:SMDSvsQDSMDS_scenarioI_CDF}, the horizontal axis represents the average estimation error, whereas the vertical axis shows the empirical \ac{CDF}, \textit{i.e.}, the fraction of errors that do not exceed $\xi$ meters.
The results are shown for two combinations of angular and distance measurement errors, $(\sigma_d,\epsilon)=\{(1,20^\circ),(4,50^\circ)\}$. }
%

When the angle error is small ($\epsilon = 10^\circ$ and $20^\circ$), we can observe {in Fig.~\ref{fig:SMDSvsQDSMDS_scenarioI} } that the relative performance of the two methods { in terms of the average estimation error} varies depending on the distance error.
The \ac{QD-SMDS} algorithm outperforms the \ac{SMDS} algorithm up to $\sigma_d = 1.0$ m for $\epsilon=10^\circ$, and up to $\sigma_d = 1.8$ m for $\epsilon=20^\circ$.
However, as the distance error increases further, \ac{SMDS} begins to achieve higher accuracy.
This behavior can be attributed to the fact that when angle errors are minimal, the \ac{GEK} matrix can be constructed with high precision, making aggressive noise suppression via low-rank truncation less critical. 
Furthermore, \ac{SMDS}, which requires fewer (and noisy) parameters to construct the \ac{GEK} matrix, becomes advantageous under such conditions.
{This limitation of \ac{QD-SMDS} defines its operating regime, in which conventional \ac{SMDS} alone may be sufficient to achieve high accuracy when the angular measurement errors are small.}

In contrast, when the angle error exceeds $30^\circ$, the \ac{QD-SMDS} algorithm consistently outperforms \ac{SMDS}, with the performance gap widening as the angle error increases.
These results indicate that as the accuracy of the \ac{GEK} matrix deteriorates, the role of \ac{SVD}-based noise suppression through low-rank truncation becomes increasingly important.

\begin{figure}[!t]
\centering
	\subfigure{
	\includegraphics[width=0.95\columnwidth,keepaspectratio=true]{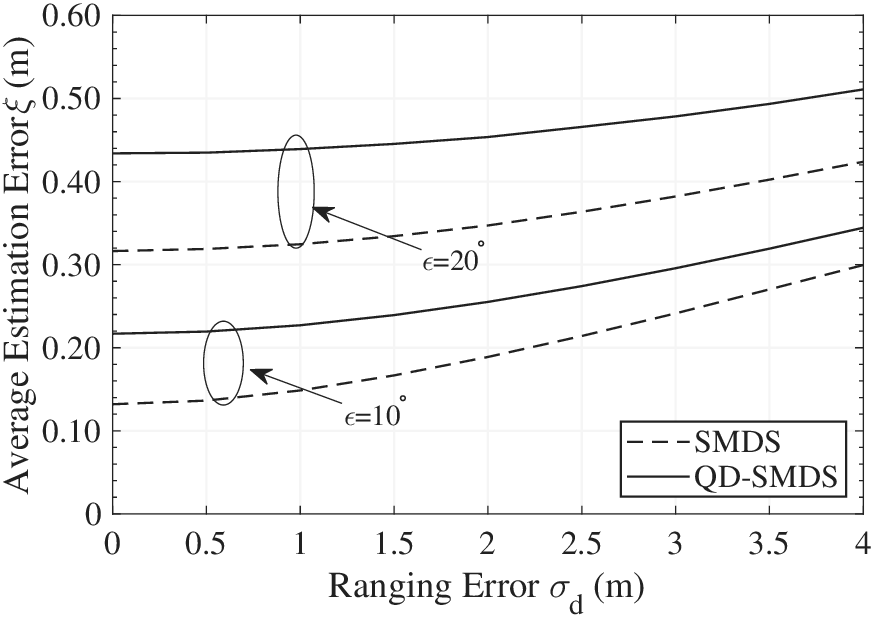}
	}
    \vspace{2mm}
	\subfigure{
	\includegraphics[width=0.95\columnwidth,keepaspectratio=true]{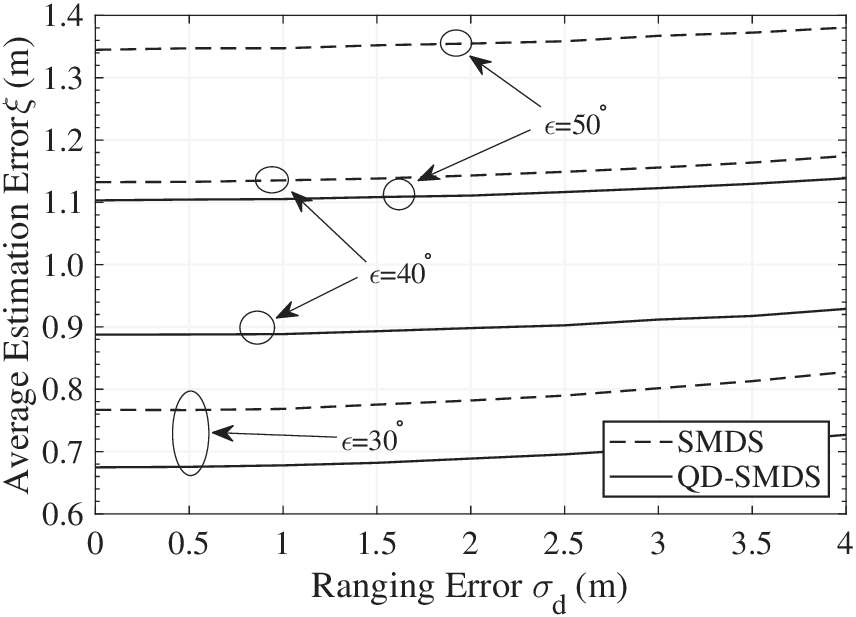}
	}
    \vspace{-2ex}
	\caption{Comparison of { average estimation error} between the \ac{SMDS} and \ac{QD-SMDS} algorithms in \textbf{Scenario II}.}
	\label{fig:SMDSvsQDSMDS_scenarioII}
	\vspace{-2ex}
\end{figure}

\begin{figure}[!t]
\centering

	\subfigure{
	\includegraphics[width=0.95\columnwidth,keepaspectratio=true]{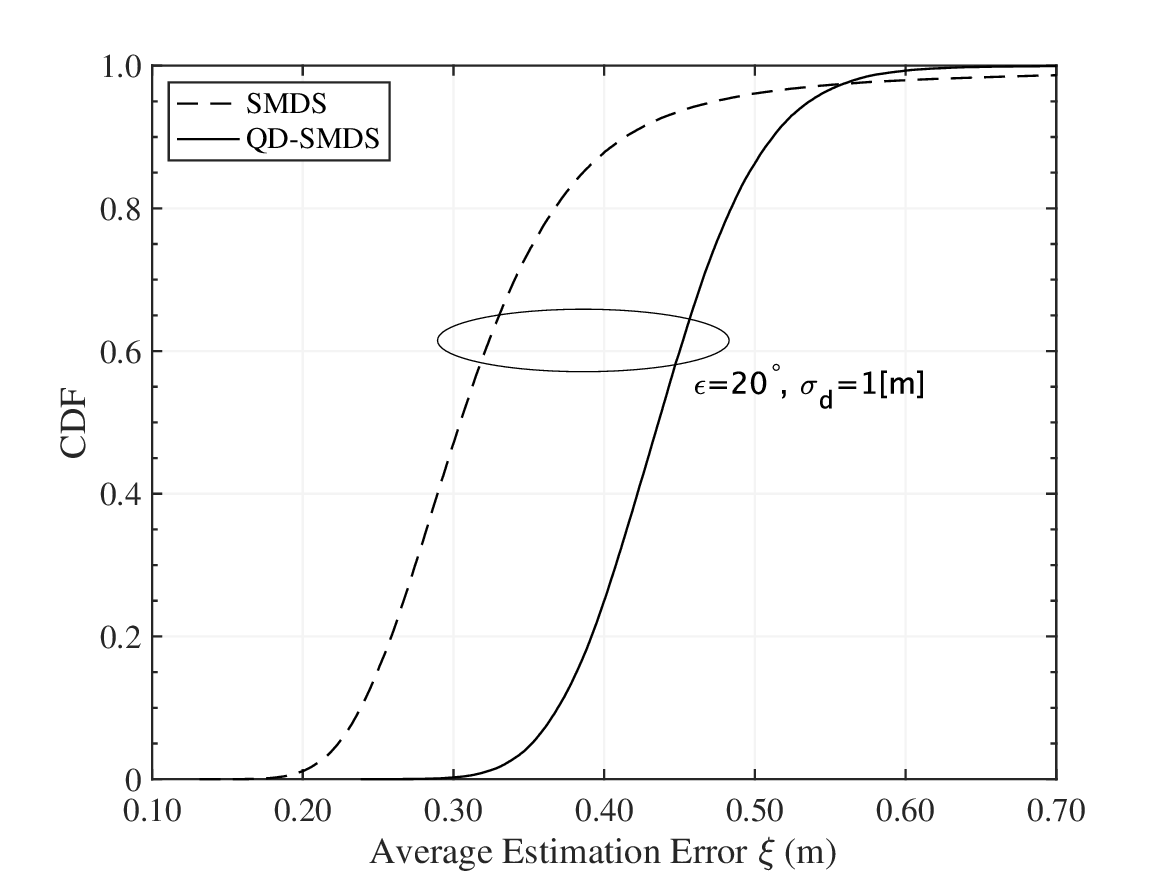}
	}
    \vspace{2mm}
	\subfigure{
	\includegraphics[width=0.95\columnwidth,keepaspectratio=true]{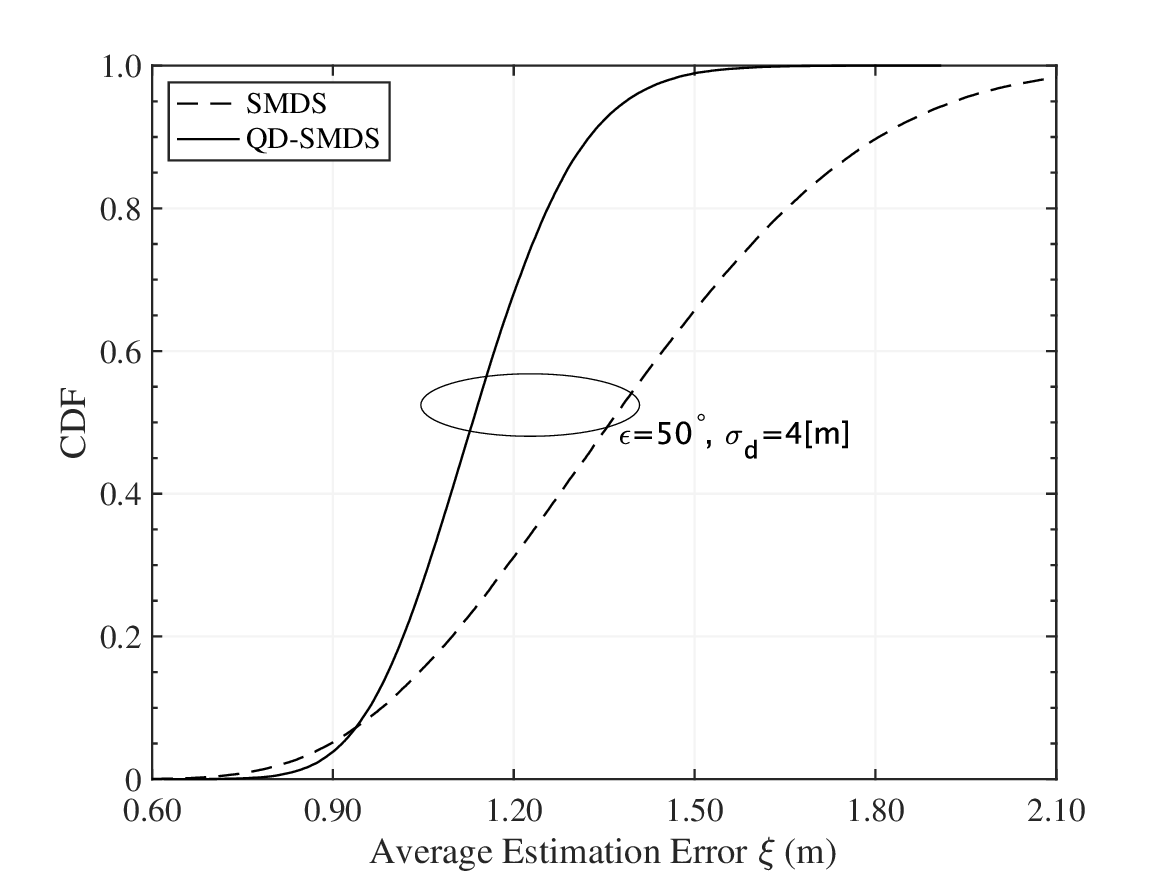}
	}
	\caption{ Comparison of empirical \ac{CDF} between the \ac{SMDS} and \ac{QD-SMDS} algorithms in \textbf{Scenario II}.}
	\label{fig:SMDSvsQDSMDS_scenarioII_CDF}
	\vspace{-2ex}

\end{figure}

As such, \ac{QD-SMDS}—which constructs a rank-$1$ \ac{GEK} matrix—offers improved robustness to angle errors compared to \ac{SMDS}, which employs a rank-$3$ \ac{GEK} matrix{; in particular, although errors from the first-stage \ac{SMDS} propagate into the construction of the quaternion-domain \ac{GEK} matrix, its rank-$1$ nature enables more effective noise suppression in the second stage}.
{On the other hand, Fig.~\ref{fig:SMDSvsQDSMDS_scenarioI_CDF} shows a clear advantage of \ac{QD-SMDS} in terms of the empirical \ac{CDF}: its distribution rises earlier than that of \ac{SMDS}, indicating consistently higher accuracy across the entire error range for both combinations of angle and distance errors.}

{{We conclude the discussion of \textbf{Scenario I} by noting that, although the practicality of this setup may appear limited due to the two-step procedure, similar approaches—where rough position estimates are first obtained to infer angular information and subsequently refined in a second step—have been considered in prior work~\cite{Abreu2007, Macagnano2013, Ghods2018TWC}.
\Ac{FLOP} counts for each proposed algorithm are reported later in Table~\ref{tab:FLOP_count}.
These results can assist users in determining whether the performance improvement achieved by this two-step procedure justifies the additional computational cost, given their available computational resources.}}

Fig.~\ref{fig:SMDSvsQDSMDS_scenarioII} compares the localization accuracy of \ac{SMDS} and \ac{QD-SMDS} in \textbf{Scenario II} {in terms of the average estimation error}. 
As also observed in Fig.~\ref{fig:SMDSvsQDSMDS_scenarioI}, \ac{SMDS} achieves higher localization accuracy when the angle error is up to $\epsilon = 20^\circ$.

However, beyond this threshold, \ac{QD-SMDS} significantly outperforms \ac{SMDS}, and the performance gap becomes more pronounced than in \textbf{Scenario I}. 
This improvement is attributed to the enhanced robustness of \ac{QD-SMDS} against measurement errors, enabled by the additional angular information that can be measured or estimated using planar antennas.

{A similar trend can be observed in Fig.~\ref{fig:SMDSvsQDSMDS_scenarioII_CDF}.
For small angular and distance errors, the \ac{SMDS} curve rises earlier.
In the second subfigure, corresponding to larger angular and distance errors, however, the pronounced leftward shift of the \ac{QD-SMDS} \ac{CDF} confirms a substantial reduction in localization error, demonstrating that the proposed method outperforms \ac{SMDS} by markedly reducing large-error estimates.}


\subsubsection{ Simulation Results With Random Missing Entries in \ac{GEK}}
\label{sec:Result2}

%
%
%
%

We now turn our attention to a practical scenario in which missing data results in a sparse \ac{GEK} matrix.
In \textbf{Scenario I}, since the \ac{SMDS} algorithm is initially employed for data supplementation, an effective performance comparison involving matrix completion is not feasible; therefore, this subsection focuses on \textbf{Scenario II}.

\begin{figure}[!t]
\begin{center}
	\subfigure{
	\includegraphics[width=0.95\columnwidth,keepaspectratio=true]{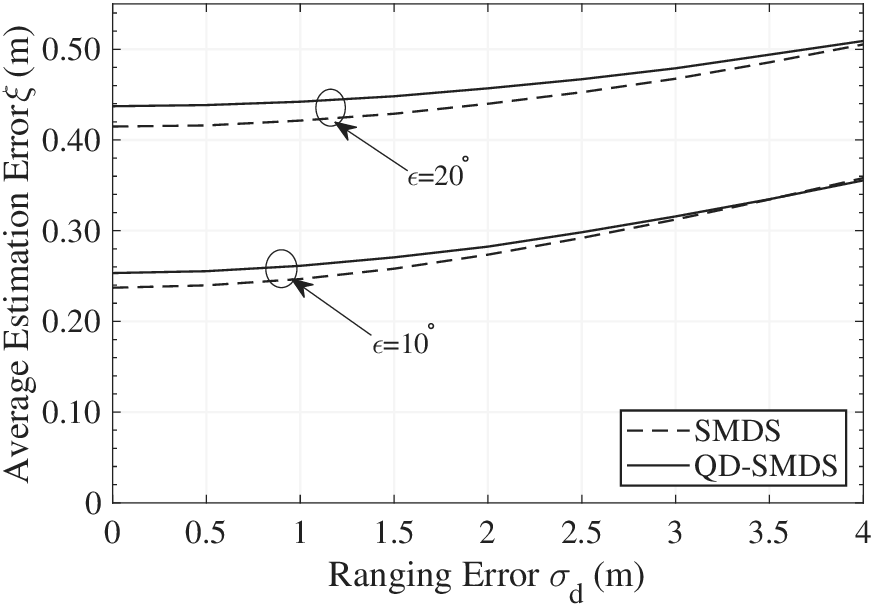}
	}
    \vspace{1mm}
	\subfigure{
	\includegraphics[width=0.95\columnwidth,keepaspectratio=true]{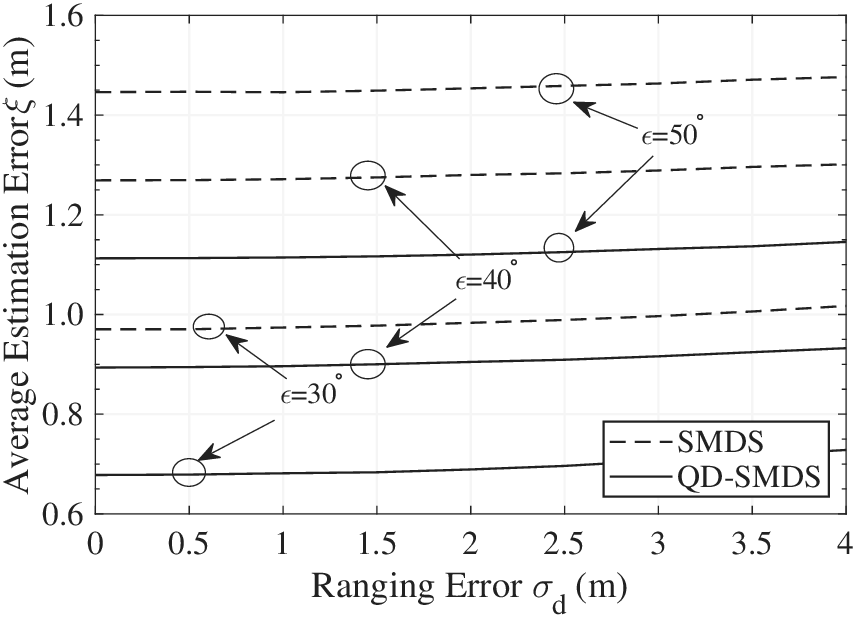}
	}
    \vspace{-2ex}
	\caption{Comparison of localization accuracy between the \ac{SMDS} and \ac{QD-SMDS} algorithms with the missing \ac{GEK} matrix in \textbf{Scenario II}.}
	\label{fig:ImcompleteData_scnearioII}
\end{center}
\end{figure}

In our setup, $30\%$ of the entries in both the real-domain \ac{GEK} matrix $\bm{K}_{\mathrm{r}}$ and the quaternion-domain \ac{GEK} matrix $\bm{K}_{\mathrm{q}}$ are randomly removed, while preserving matrix symmetry. 
For \ac{SMDS}, the incomplete real-domain \ac{GEK} matrix is first completed using {the \ac{IZMA}} algorithm described in~\cite{shabat2012}, followed by execution of the \ac{SMDS} algorithm.
{ The completion algorithm is terminated when both the reconstruction error on observed entries, and the update of the nuclear-norm threshold parameter $\lambda$ become smaller than 0.01. }
For \ac{QD-SMDS}, the incomplete quaternion-domain \ac{GEK} matrix is first decomposed into two complex matrices, each of which is individually completed using the same low-rank matrix completion method.
These are then recombined into a single quaternion matrix, after which the \ac{QD-SMDS} algorithm is executed.

Fig. \ref{fig:ImcompleteData_scnearioII} shows the localization accuracy comparison { in terms of the average estimation error} between \ac{SMDS} and \ac{QD-SMDS} in \textbf{Scenario II} under missing data conditions.
As observed, when the angle error is small (\textit{e.g.}, $\epsilon = 10^\circ$ and $20^\circ$), \ac{SMDS} still yields the best performance, although the gap between \ac{SMDS} and \ac{QD-SMDS} is negligible. 
In contrast, as the angle error increases, \ac{QD-SMDS} significantly outperforms \ac{SMDS}, with the performance gap becoming even more pronounced than in the case without missing data.

This improvement is attributed to the lower rank of the quaternion-domain \ac{GEK} matrix compared to the real-domain one. 
Since low-rank structures are more favorable for matrix completion, \ac{QD-SMDS} benefits from improved recovery accuracy, which directly enhances localization performance under larger angle errors.

Based on the numerical results and the computational effort involved, the \ac{SMDS} algorithm is preferable in scenarios with relatively small angle errors. 
In contrast, \ac{QD-SMDS} is more suitable when angle errors are large.
Moreover, the supplementary azimuth and elevation angle information obtained from planar antenna arrays becomes increasingly critical as the angle error escalates. 
Additionally, in the presence of partial angle information loss, \ac{QD-SMDS} remains effective regardless of the angle error magnitude, demonstrating its robustness to missing data.

{
\subsubsection{Simulation Results with Missing Distances}

\begin{figure}[!t]
\begin{center}

	\subfigure{
	\includegraphics[width=0.95\columnwidth,keepaspectratio=true]{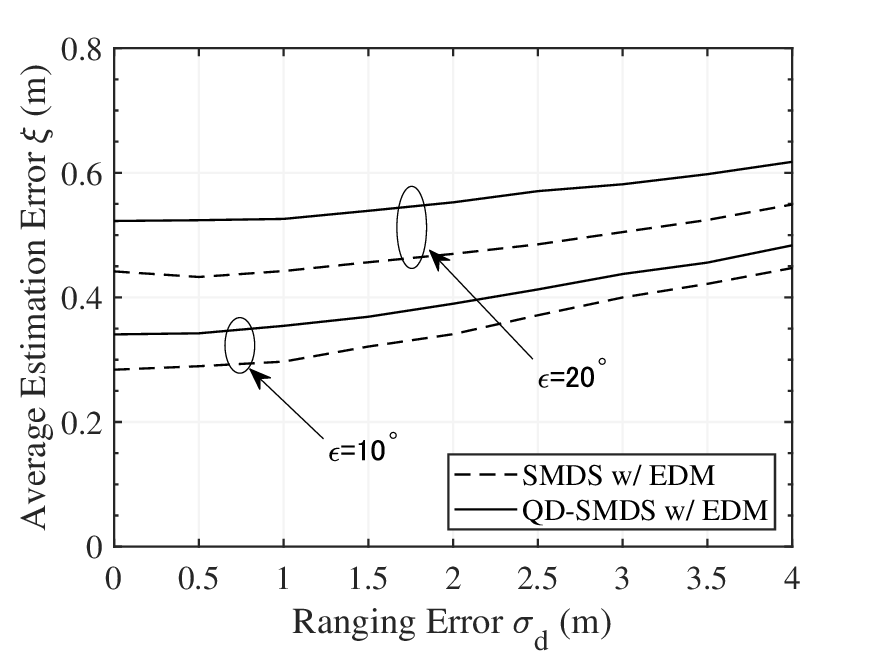}
	}
	\subfigure{
	\includegraphics[width=0.95\columnwidth,keepaspectratio=true]{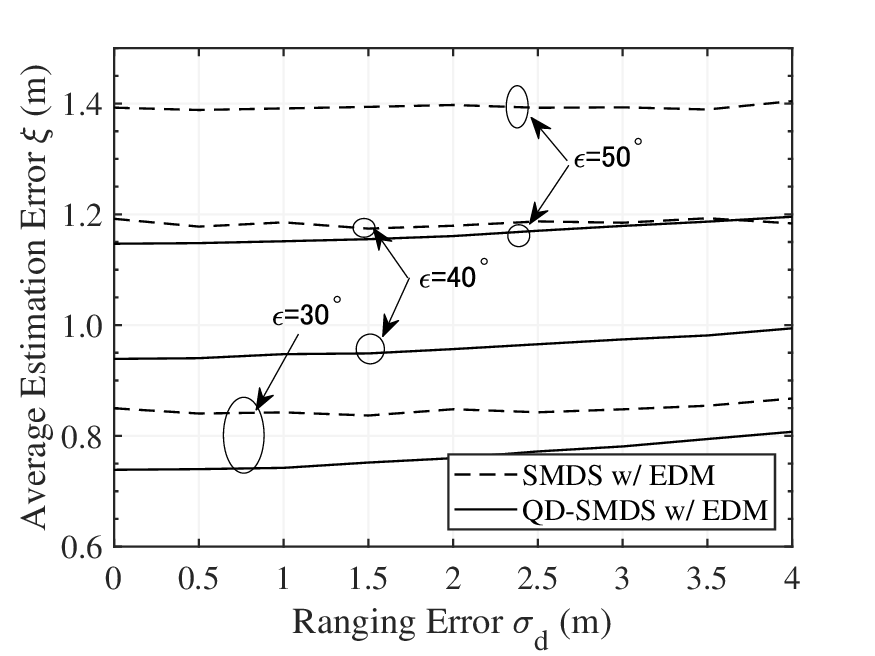}
	}
    \vspace{-2ex}
	\caption{{Comparison of localization accuracy between \ac{SMDS} and \ac{QD-SMDS} with 10\% missing distances and full angle information. The \ac{EDM} matrix is completed using the algorithm proposed in \cite{Dokmanic2015}.}}
	\label{fig:distances_missing}
\end{center}
\end{figure}

%
A more challenging (and relevant) impairment occurs when distance measurements are unavailable due to the loss of line-of-sight between an anchor and a target.
In this case, the degradation is more severe because a single missing distance induces a structured pattern of missing data in the \ac{GEK} matrix: the corresponding row and column become entirely unobserved, resulting in $2M-1$ missing \ac{GEK} entries.

Under such conditions, directly applying \ac{IZMA} to complete the \ac{GEK} matrix is not viable.
The completion method estimates missing entries by fitting coefficients (linear weights) from the observed elements.
If an entire row (or column) is unobserved, the associated coefficients remain unconstrained and the recovery becomes ill-posed.
Consequently, low-rank completion methods cannot, in general, uniquely recover fully unobserved rows/columns.

Instead, we first complete the \ac{EDM} (\textit{i.e.}, the missing distances) and subsequently construct a complete \ac{GEK} matrix to run \ac{SMDS} and \ac{QD-SMDS}.
Specifically, we consider a setting similar to \textbf{Scenario II}, in which full angle information is available but 10\% of the distances are missing.
The missing distances are recovered using the \ac{SDR}-based \ac{EDM} completion algorithm in~\cite{Dokmanic2015} with $\lambda = 1000$.
The results in Fig.~\ref{fig:distances_missing} exhibit the same qualitative trend as the case with randomly missing \ac{GEK} entries.
In particular, \ac{QD-SMDS} begins to outperform \ac{SMDS} only when the angular errors exceed $\epsilon = 20^\circ$, for the same reasons discussed above.}

%% file: TXT/5_QD-MRC-SMDS.tex

In \ac{QD-SMDS}, the primary source of computational complexity is the \ac{QSVD} process.
As described in Section II-B, \ac{QSVD} requires performing an \ac{SVD} on a complex-valued matrix whose size is twice that of the original quaternion-domain \ac{GEK}
matrix.
To reduce this computational overhead, this section revisits the structural properties of the quaternion-domain \ac{GEK} matrix and introduces low-complexity variants of the \ac{QD-SMDS} algorithm that eliminate the need for \ac{QSVD}.

\subsection{\ac{QD-MRC-SMDS}}

The quaternion edge vector defined in \eqref{eq:q_V} is partitioned into two sub-vectors: $\bm{\nu}_{\mathrm{AA}}$, corresponding to the edges between \acp{AN}, and $\bm{\nu}_{\mathrm{AT}}$, corresponding to the edges between \acp{AN} and \acp{TN}, as follows:
\begin{equation}
    \bm{\nu}=\left[
\bm{\nu}_{\mathrm{AA}}^{\mathsf{T}},\bm{\nu}_{\mathrm{AT}}^{\mathsf{T}}
\right]^{\mathsf{T}},
\label{eq:q_v_MRC}
\end{equation}
where $\bm{\nu}_{\mathrm{AA}}$ is known since it is fully determined by the coordinates of \acp{AN}, and $\bm{\nu}_{\mathrm{AT}}$ can be expressed by
\begin{equation}
    \bm{\nu}_{\mathrm{AT}}
    =
    \bm{B}_\mathrm{AA}\bm{\chi}_\mathrm{A} - \bm{B}_\mathrm{AT}\bm{\chi}_\mathrm{T},
    \label{eq:q_v_AT}
\end{equation}
where $\bm{\chi}_{\mathrm{A}}$ and $\bm{\chi}_{\mathrm{T}}$ denotes the quaternion coordinate vectors corresponding to the real-valued coordinate matrices $\bm{X}_\mathrm{A}$ and $\bm{X}_{\mathrm{T}}$ for \acp{AN} and \acp{TN}, respectively, as defined in \eqref{eq:r_Xa} and \eqref{eq:r_Xt}.
The corresponding structure matrices are given by
\begin{equation}
    \bm{B}_\mathrm{AA}
    \triangleq
    \bm{I}_{N_{\mathrm{A}}}
    \otimes
    \bm{1}_{N_{\mathrm{T}\times1}}
    \quad
    \text{and}
    \quad
    \bm{B}_\mathrm{AT}
    \triangleq
    \bm{1}_{N_{\mathrm{A}\times1}}
    \otimes
    \bm{I}_{N_{\mathrm{T}}}.
    \label{eq:anchor-target _matrix}
\end{equation}
 
From \eqref{eq:q_v_AT}, it is evident that the estimation of the \ac{TN} coordinates can be achieved by accurately estimating the quaternion edge vector  $\bm{\nu}_\mathrm{AT}$.
Furthermore, based on \eqref{eq:q_v_MRC}, the quaternion-domain \ac{GEK} matrix in \eqref{eq:q_kernel} can be expressed as
\begin{equation}
    \bm{K}_{\mathrm{q}}
    =
    \bm{\nu}\bm{\nu}^\mathsf{H}
    =\!
\left[
\begin{array}{c|c}
\bm{\nu}_{\mathrm{AA}}\bm{\nu}_{\mathrm{AA}}^{\mathsf{H}}
&
\bm{\nu}_{\mathrm{AA}}\bm{\nu}_{\mathrm{AT}}^{\mathsf{H}}
\\ \hline
\bm{\nu}_{\mathrm{AT}}\bm{\nu}_{\mathrm{AA}}^{\mathsf{H}}
&
\bm{\nu}_{\mathrm{AT}}\bm{\nu}_{\mathrm{AT}}^{\mathsf{H}}
\end{array}
\right]
\!=\!
\left[
\begin{array}{c|c}
 \bm{K}_1 & \bm{K}_2 \\ \hline
\bm{K}_2^\mathsf{H} & \bm{K}_3
\end{array}
\right].
\label{eq:q_kernel_MRC}
\end{equation}

%
\begin{algorithm}[!t]
\caption{QD-MRC-SMDS}
\label{alg:MRC-SMDS}
\begin{algorithmic}[1]
%
\Statex {\bf{Input:}}
\State  \textit{Measured and estimated pairwise distances and \acp{ADoA}: $\tilde{d}_m$, $\tilde{\alpha}_{mp}$, $\tilde{\phi}_{m}^{(\mathrm{xy})}$, $\tilde{\phi }_{m}^{(\mathrm{xz})}$, $\tilde{\phi}_{m}^{(\mathrm{yz})}$, $\tilde{\theta}_{m}^{(\mathrm{x})}$, $\tilde{\theta}_{m}^{(\mathrm{y})}$, $\tilde{\theta}_{m}^{(\mathrm{z})}$}
\State\textit{The coordinates of all \acp{AN}.}
\Statex {\bf{Steps:}}
\State \textit{Construct $\bm{\nu}_\mathrm{AA}$ and $\bm{K}_2$ in Eqs. \eqref{eq:q_V} and \eqref{eq:q_kernel}.}
\State \textit{Construct $\bm{B}_\mathrm{AA}$ and $\bm{B}_\mathrm{AT}$ in Eqs. \eqref{eq:anchor-target _matrix}.}
\State \textit{Compute the estimated quaternion coordinate vector $\hat{\bm{\chi}}_{\mathrm{T}}$ using Eq. $\eqref{eq:q_X_MRC-SMDS}$}
\State \textit{Convert $\hat{\bm{\chi}}_{\mathrm{T}}$ to the estimated real-valued coordinates matrix $\hat{\bm{X}}_{\mathrm{T}}$.}
\end{algorithmic}
\end{algorithm}


By exploiting the structure observed in \eqref{eq:q_kernel_MRC}, the edge vector can be estimated using simple linear filtering, without the need for more complex operations.

%

First, using \eqref{eq:q_v_AT}, $\bm{K}_{2}$ can be rewritten as
\begin{equation}
    \bm{K}_2 = 
\bm{\nu}_{\mathrm{AA}}\bm{\nu}^\mathsf{H}_{\mathrm{AT}}
=\bm{\nu}_{\mathrm{AA}}
\left(
\bm{B}_\mathrm{AA}\bm{\chi}_{\mathrm{A}}
 - 
\bm{B}_\mathrm{AT}\bm{\chi}_{\mathrm{T}}
\right)^\mathsf{H}.
\label{eq:K2}
\end{equation}
 
Next, \ac{MRC} is applied to suppress the influence of $\bm{\nu}_{\mathrm{AA}}$ on the right-hand side, resulting in
\begin{equation}
\label{eq:MRC}
    \frac{\bm{\nu}_{\mathrm{AA}}^\mathsf{H}}
{\|\bm{\nu}_{\mathrm{AA}}\|^2}
\bm{K}_2
=\left(
\bm{B}_\mathrm{AA}\bm{\chi}_{\mathrm{A}}
 - 
\bm{B}_\mathrm{AT}\bm{\chi}_{\mathrm{T}}
\right)^\mathsf{H}.
\end{equation}

Finally, taking into account the non-commutativity of both matrix and quaternion multiplications, \eqref{eq:MRC} is solved for $\bm{\chi}_{\mathrm{T}}$, thereby yielding a closed-form expression for the estimated \ac{TN} coordinates{\footnote{\setlength{\baselineskip}{10pt}{Since the condition $\bm{B}_{\mathrm{AT}}^\mathsf{T}\bm{B}_{\mathrm{AT}} = N_{\mathrm{A}}\bm{I}_{N_{\mathrm{T}}}$ is satisfied, the calculation of the Moore-Penrose pseudo-inverse does not require explicit matrix inversion; instead, it reduces to a scaled transpose operation.}}}:
\begin{equation}
\bm{\hat{\chi}}_{\mathrm{T}}
=
\frac{\bm{B}_\mathrm{AT}^\mathsf{T}}{N_{\mathrm{A}}}
\left(\bm{B}_\mathrm{AA}\bm{\chi}_{\mathrm{A}}
- 
\frac{1}{\|\bm{\nu}_{\mathrm{AA}}\|^2}
\bm{K}_2^\mathsf{H}
\bm{\nu}_{\mathrm{AA}}
\right).
\label{eq:q_X_MRC-SMDS}
\end{equation}

This low-complexity variant of the \ac{QD-SMDS} algorithm is hereafter referred to as \textit{\ac{QD-MRC-SMDS}}, and its pseudo-code is provided in Algorithm \ref{alg:MRC-SMDS}. 
Compared with the original \ac{QD-SMDS}, \ac{QD-MRC-SMDS} significantly reduces { the overall} complexity by eliminating expensive operations such as \ac{QSVD}, matrix inversion, and Procrustes transformation.
Instead, it enables direct coordinate estimation via simple quaternion-domain multiplications by utilizing a specific subset of the quaternion-domain \ac{GEK} matrix. 

{The complexity reduction is also evident in the asymptotic sense.
%
%
The most computationally intensive step is the matrix–vector multiplication $\bm{K}_2^\mathsf{H}\bm{\nu}_{\mathrm{AA}}$.
Its asymptotic complexity scales as $O\left(N_{\mathrm{A}}^2 N_\mathrm{T}(N_{\mathrm{A}}-1)/2\right)=O\left(N_{\mathrm{A}}^3 N_\mathrm{T}\right)$, \textit{i.e.}, linear in the number of targets $N_\mathrm{T}$.}
%

%
\subsection{Iterative QD-MRC-SMDS}
\label{sec:Iterative_MRC_SMDS}

While the \ac{QD-MRC-SMDS} algorithm offers extremely low computational complexity, it leverages only a subset of the information encoded in the \ac{GEK} matrix and therefore does not fully exploit all available data (\textit{i.e.}, $\bm{K}_3$).
To overcome this limitation, we extend \ac{QD-MRC-SMDS} into an iterative estimation algorithm that incorporates the full information contained in the \ac{GEK} matrix, with only a modest increase in computational cost.

\begin{algorithm}[!t]
\caption{Iterative QD-MRC-SMDS}
\label{alg:Iterative MRC-SMDS}
\begin{algorithmic}[1]
%
\Statex {\bf{Input:}}
\State \textit{Measured and estimated pairwise distances and \acp{ADoA}: $\tilde{d}_m$, $\tilde{\alpha}_{mp}$, $\tilde{\phi }_{m}^{(\mathrm{xy})}$, $\tilde{\phi}_{m}^{(\mathrm{xz})}$, $\tilde{\phi}_{m}^{(\mathrm{yz})}$, $\tilde{\theta}_{m}^{(\mathrm{x})}$, $\tilde{\theta}_{m}^{(\mathrm{y})}$, $\tilde{\theta}_{m}^{(\mathrm{z})}$}
\State\textit{The coordinates of all \acp{AN}.}
\Statex {\bf{Steps:}}
\State \textit{Construct $\bm{\nu}_\mathrm{AA}$, $\bm{K}_2$ and $\bm{K}_3$ via equations in Eqs. \eqref{eq:q_V} and \eqref{eq:q_kernel}.}
\State \textit{Initialize $\bm{\nu}_{\mathrm{AT}}$ using Eq. \eqref{eq:q_v_AT ini}.}
\Statex {\bf{for} $\tau = 1,\ldots,\tau_{\mathrm{max}}$}
\State \textit{Update $\bm{\nu}_{\mathrm{AT}}$ using Eq. \eqref{eq:q_v_AT update}.}
\Statex {\bf{end for}}
\State \textit{Obtain $\hat{\bm{\chi}}_{T}$ using Eq.\eqref{eq:q_X_ite_MRC}.}
\State \textit{Convert $\hat{\bm{\chi}}_{\mathrm{T}}$ to the estimated real-valued coordinates matrix $\hat{\bm{X}}_{\mathrm{T}}$.}
\end{algorithmic}
\end{algorithm}

First, from \eqref{eq:q_kernel_MRC}, we have
\begin{equation}
    \begin{bmatrix}
\bm{K}_2\\ \bm{K}_3
\end{bmatrix}
=
\begin{bmatrix}
\bm{\nu}_{\mathrm{AA}}\\ \bm{\nu}_{\mathrm{AT}}
\end{bmatrix}
\bm{\nu}_{\mathrm{AT}}^\mathsf{H}.
\label{eq:q_Ite_MRC}
\end{equation}

Since $\bm{\nu}_{\mathrm{AT}}$ appears twice in \eqref{eq:q_Ite_MRC}, it cannot be uniquely determined in a single step.
To address this, we reformulate the problem as an iterative process, from which the following update equation can be derived as 
\begin{equation}
    \bm{\nu}_{\mathrm{AT}}^{(\tau+1)}=
\frac{
\begin{bmatrix}
\bm{K}_2^\mathsf{H}&\bm{K}_3^\mathsf{H}
\end{bmatrix}
}
{
\left\|
\begin{bmatrix}
\bm{\nu}_{\mathrm{AA}}^\mathsf{H}& (\bm{\nu}_{\mathrm{AT}}^{(\tau)})^\mathsf{H}
\end{bmatrix}
\right\|^2
}
\begin{bmatrix}
\bm{\nu}_{\mathrm{AA}}\\ \bm{\nu}_{\mathrm{AT}}^{(\tau)}
\end{bmatrix},
\label{eq:q_v_AT update}
\end{equation}
where $\bm{\nu}_{\mathrm{AT}}^{(\tau)}$ denotes the estimate at the $\tau$-th iteration step ($0\le\tau\le \tau_\mathrm{max}$).

As an initial estimate for $\bm{\nu}_{\mathrm{AT}}$ we can take the result from the \ac{QD-MRC-SMDS} algorithm using \eqref{eq:MRC}, i.e., 
\begin{equation}
    \bm{\nu}_{\mathrm{AT}}^{(0)}=
\frac{
\bm{K}_2^\mathsf{H}
}
{\|
\bm{\nu}_{\mathrm{AA}}
\|^2
}
\bm{\nu}_{\mathrm{AA}},
\label{eq:q_v_AT ini}
\end{equation}

Finally, given the final estimate $\bm{\nu}_{\mathrm{AT}}^{(\tau_{\mathrm{max}})}$, the quaternion coordinate vector corresponding to \acp{TN} can be obtained from \eqref{eq:q_v_AT} as
\begin{equation}
    \bm{\hat{\chi}}_{\mathrm{T}}
=
\frac
{\bm{B}_\mathrm{AT}^\mathsf{T}}
{N_{\mathrm{A}}}
\left(
\bm{B}_\mathrm{AA}\bm{\chi}_{\mathrm{A}}
 - 
\bm{\nu}_{\mathrm{AT}}^{(\tau_{\mathrm{max}})}
\right).
\label{eq:q_X_ite_MRC}
\end{equation}

{ The computational complexity of iterative \ac{QD-MRC-SMDS} is dominated by the multiplication of
\begin{equation}
\begin{bmatrix}
\bm{K}_2^\mathsf{H} & \bm{K}_3^\mathsf{H}
\end{bmatrix} \in \mathbb{H}^{N_{\mathrm{A}} N_\mathrm{T} \times M},
\end{equation}
with
\begin{equation}
\begin{bmatrix}
\bm{\nu}_{\mathrm{AA}} \\ \bm{\nu}_{\mathrm{AT}}^{(\tau)}
\end{bmatrix} \in \mathbb{H}^{M \times 1},
\end{equation}
yielding a complexity of $O\bigl(N_{\mathrm{A}} N_\mathrm{T}\bigl(N_{\mathrm{A}}(N_{\mathrm{A}}-1)/2 + N_{\mathrm{A}} N_\mathrm{T}\bigr)\bigr)=O\left(N_{\mathrm{A}}^3 N_\mathrm{T}+N_{\mathrm{A}}^2 N_\mathrm{T}^2\right)$, which adds an $O\left(N_{\mathrm{A}}^2 N_\mathrm{T}^2\right)$ term compared with standard \ac{QD-MRC-SMDS}.}

This iterative variant of the \ac{QD-MRC-SMDS} algorithm is hereafter referred to as \textit{iterative \ac{QD-MRC-SMDS}}, and its pseudo-code is provided in Algorithm \ref{alg:Iterative MRC-SMDS}. 

\subsection{Performance Assessment}
\subsubsection{Fixed Point and One Step Convergence Analysis}
%
\begin{figure}[!t]
\begin{center}
\includegraphics[width=0.91\columnwidth,keepaspectratio=true]{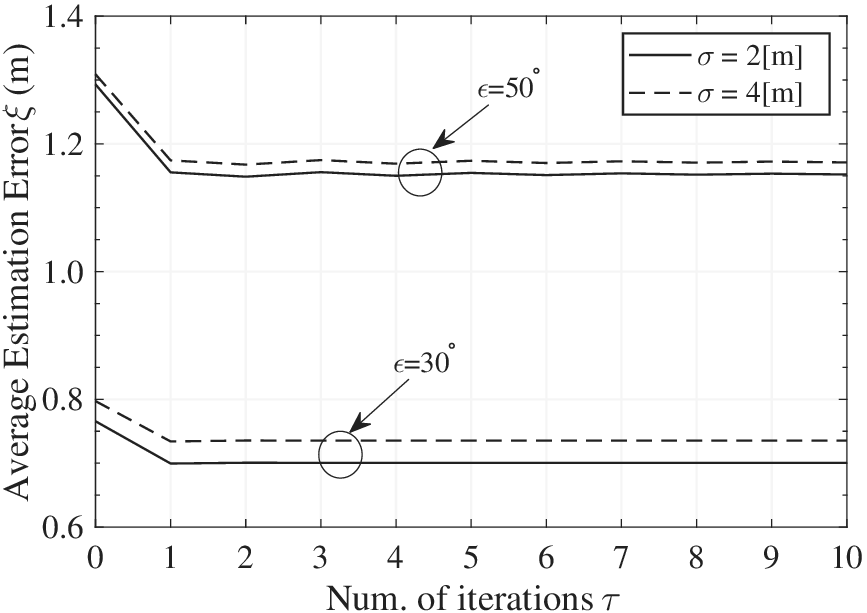}
	\caption{{Convergence of iterative \ac{QD-MRC-SMDS} under various ranging
and angle measurement errors.}}
	\label{fig:converge}
\end{center}
\end{figure}

{
\label{sec:iterative_convergence}

{
Before comparing the performance of the proposed algorithms, we analyze in this section the fixed-point and convergence properties of iterative \ac{QD-MRC-SMDS}, by exploiting the structure of the quaternion-domain \ac{GEK} matrix.
We begin by characterizing the fixed points of the iterative update in \eqref{eq:q_v_AT update}.
A quaternion vector $\bm{\nu}_{\mathrm{AT}}^{\star}$ is a fixed point if $\bm{\nu}_{\mathrm{AT}}^{(\tau+1)} = \bm{\nu}_{\mathrm{AT}}^{(\tau)} = \bm{\nu}_{\mathrm{AT}}^{\star}$, which yields
\begin{equation}
\label{eq:fixed_point}
\bm{\nu}_{\mathrm{AT}}^{\star}
=
\frac{
\bm{K}_2^\mathsf{H}\bm{\nu}_{\mathrm{AA}}
+
\bm{K}_3^\mathsf{H}\bm{\nu}_{\mathrm{AT}}^{\star}
}{
\|\bm{\nu}_{\mathrm{AA}}\|^2 + \|\bm{\nu}_{\mathrm{AT}}^{\star}\|^2
}.
\end{equation}

Expanding $\bm{K}_2^\mathsf{H}=\bm{\nu}_{\mathrm{AT}}\bm{\nu}_{\mathrm{AA}}^\mathsf{H}$ and $\bm{K}_3^\mathsf{H}=\bm{\nu}_{\mathrm{AT}}\bm{\nu}_{\mathrm{AT}}^\mathsf{H}$ from \eqref{eq:q_kernel_MRC} and noting that $\bm{\nu}_{\mathrm{AA}}^\mathsf{H}\bm{\nu}_{\mathrm{AA}}=\|\bm{\nu}_{\mathrm{AA}}\|^2\in\mathbb{R}$ commutes with any quaternion, \eqref{eq:fixed_point} becomes
\begin{equation}
\label{eq:fixed_point_expanded}
\bm{\nu}_{\mathrm{AT}}^{\star}
=
\frac{
\bm{\nu}_{\mathrm{AT}}
\bigl(
\|\bm{\nu}_{\mathrm{AA}}\|^2
+
\bm{\nu}_{\mathrm{AT}}^\mathsf{H}\bm{\nu}_{\mathrm{AT}}^{\star}
\bigr)
}{
\|\bm{\nu}_{\mathrm{AA}}\|^2 + \|\bm{\nu}_{\mathrm{AT}}^{\star}\|^2
}.
\end{equation}
Setting $\bm{\nu}_{\mathrm{AT}}^{\star}=\bm{\nu}_{\mathrm{AT}}$ yields $\bm{\nu}_{\mathrm{AT}}^\mathsf{H}\bm{\nu}_{\mathrm{AT}}=\|\bm{\nu}_{\mathrm{AT}}\|^2\in\mathbb{R}$, so that \eqref{eq:fixed_point_expanded} reduces to $\bm{\nu}_{\mathrm{AT}}$, confirming that the true edge vector is a fixed point of the iterative update.

We now analyze the one-step behavior of the update.
Substituting $\bm{K}_2^\mathsf{H}=\bm{\nu}_{\mathrm{AT}}\bm{\nu}_{\mathrm{AA}}^\mathsf{H}$ and $\bm{K}_3^\mathsf{H}=\bm{\nu}_{\mathrm{AT}}\bm{\nu}_{\mathrm{AT}}^\mathsf{H}$ into \eqref{eq:q_v_AT update}, the first iterate can be written as
\begin{equation}
\label{eq:one_step_direction}
\bm{\nu}_{\mathrm{AT}}^{(1)}
=
\bm{\nu}_{\mathrm{AT}}\,\eta,
\quad
\eta\triangleq\frac{
\|\bm{\nu}_{\mathrm{AA}}\|^2
+
\bm{\nu}_{\mathrm{AT}}^\mathsf{H}\bm{\nu}_{\mathrm{AT}}^{(0)}
}{
\|\bm{\nu}_{\mathrm{AA}}\|^2 + \|\bm{\nu}_{\mathrm{AT}}^{(0)}\|^2
}\in\mathbb{H},
\end{equation}
where $\eta$ is in general a quaternion because the inner product $\bm{\nu}_{\mathrm{AT}}^\mathsf{H}\bm{\nu}_{\mathrm{AT}}^{(0)}\in\mathbb{H}$.

Then, when $\bm{\nu}_{\mathrm{AT}}^{(0)}$ is given as in \eqref{eq:q_v_AT ini}, we get
\begin{equation}
\label{eq:eta_one}
\bm{\nu}_{\mathrm{AT}}^{(0)}
=
\frac{
\bm{K}_2^\mathsf{H}
}{
\|\bm{\nu}_{\mathrm{AA}}\|^2
}
\bm{\nu}_{\mathrm{AA}}
=
\bm{\nu}_{\mathrm{AT}}\,
\frac{
\|\bm{\nu}_{\mathrm{AA}}\|^2
}{
\|\bm{\nu}_{\mathrm{AA}}\|^2
}
=
\bm{\nu}_{\mathrm{AT}},
\end{equation}
which substituted into \eqref{eq:one_step_direction} yields $\bm{\nu}_{\mathrm{AT}}^{(1)}=\bm{\nu}_{\mathrm{AT}}$.

It is important to note that $\bm{\nu}_{\mathrm{AT}}^{(0)}$ and $\bm{\nu}_{\mathrm{AT}}^{(1)}$ are obtained from different subsets of the \ac{GEK} matrix: the former relies solely on $\bm{K}_2$ via \eqref{eq:q_v_AT ini}, whereas the latter additionally incorporates $\bm{K}_3$ through the iterative update \eqref{eq:q_v_AT update}.
In the noiseless case, the rank-$1$ structure of the \ac{GEK} matrix guarantees that $\bm{\nu}_{\mathrm{AT}}^{(0)}$ and $\bm{\nu}_{\mathrm{AT}}^{(1)}$ both recover $\bm{\nu}_{\mathrm{AT}}$ exactly, so that a single iteration of \eqref{eq:q_v_AT update} is already exact, and any further iteration leaves the estimate unchanged.
In the noisy case, we only have access to a perturbed version of the \ac{GEK} matrix $\bm{\tilde K}_\mathrm{q}$ which is no longer exactly rank-$1$ but remains near-rank-$1$.
Owing to this near-rank-$1$ structure, the first iteration of \eqref{eq:q_v_AT update} still captures the dominant reduction of the estimation error, as confirmed by the simulation results.

}

{This result is corroborated by} Fig.~\ref{fig:converge}, which shows the average estimation error of the iterative \ac{QD-MRC-SMDS} algorithm as a function of the number of iterations under distance error conditions of $\sigma_\mathrm{d}=2$ m and $4$ m.
All other simulation parameters are identical to those used in Section~IV-C.
As observed from the figure, the performance converges within a single iteration under both conditions, { consistent with the one-step convergence established above.}
This confirms that setting $\tau_\mathrm{max} = 1$ is sufficient, and the additional computational cost introduced by the iterative procedure corresponds to evaluating \eqref{eq:q_v_AT update} only once.
}

\subsubsection{Simulation Results}

%
\begin{figure}[!t]
\centering
	\subfigure{
	\includegraphics[width=0.95\columnwidth,keepaspectratio=true]{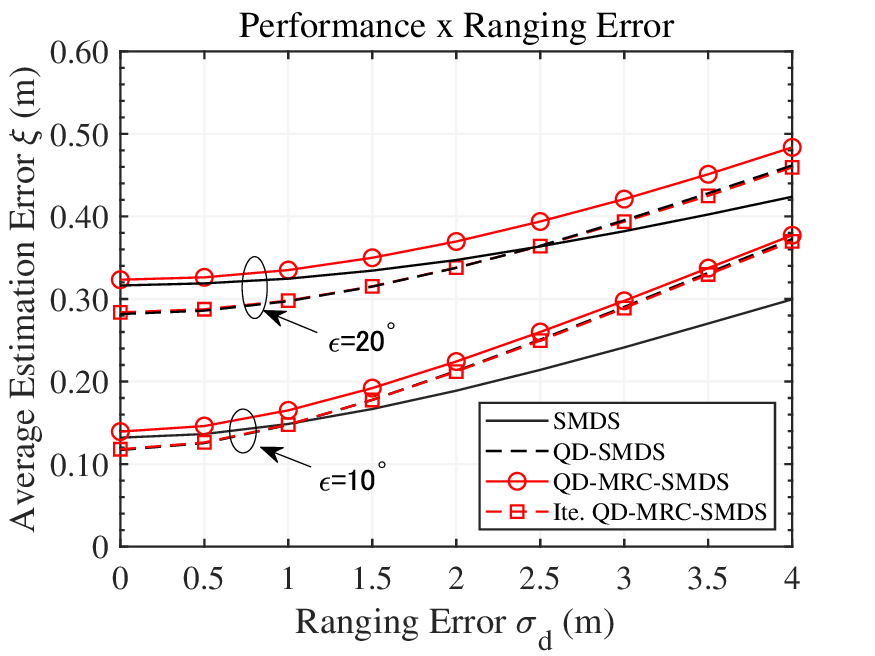}
	}
	\subfigure{
	\includegraphics[width=0.95\columnwidth,keepaspectratio=true]{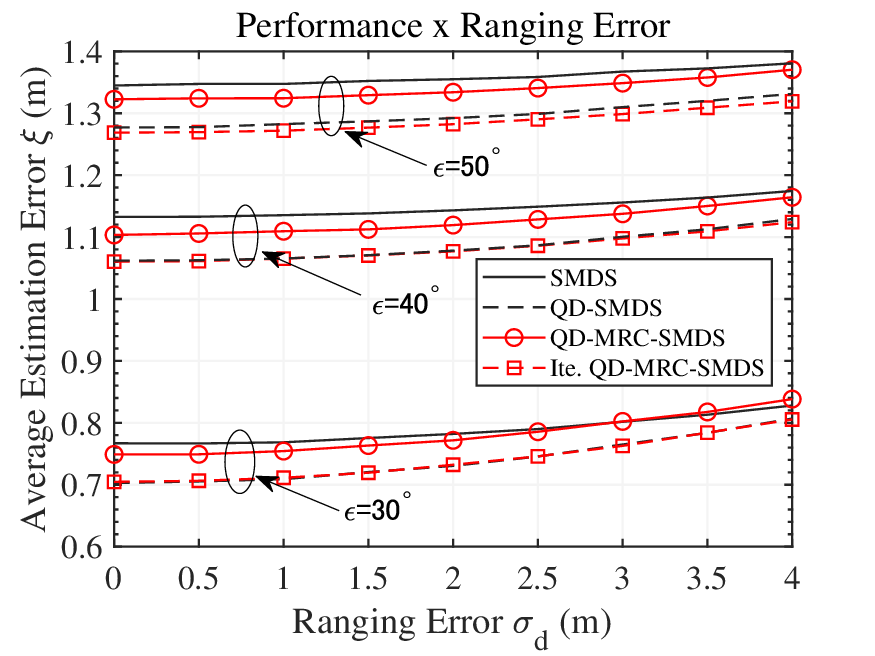}
	}
    \vspace{-2ex}
	\caption{Comparison of localization accuracy between the \ac{QD-SMDS}, \ac{QD-MRC-SMDS}, and iterative \ac{QD-MRC-SMDS} algorithms in \textbf{Scenario I}.}
	\label{fig:SMDSvsQDSMDS_w/o_SVD_scenarioI}
\end{figure}

\begin{figure}[!t]
\centering
	\subfigure{
	\includegraphics[width=0.95\columnwidth,keepaspectratio=true]{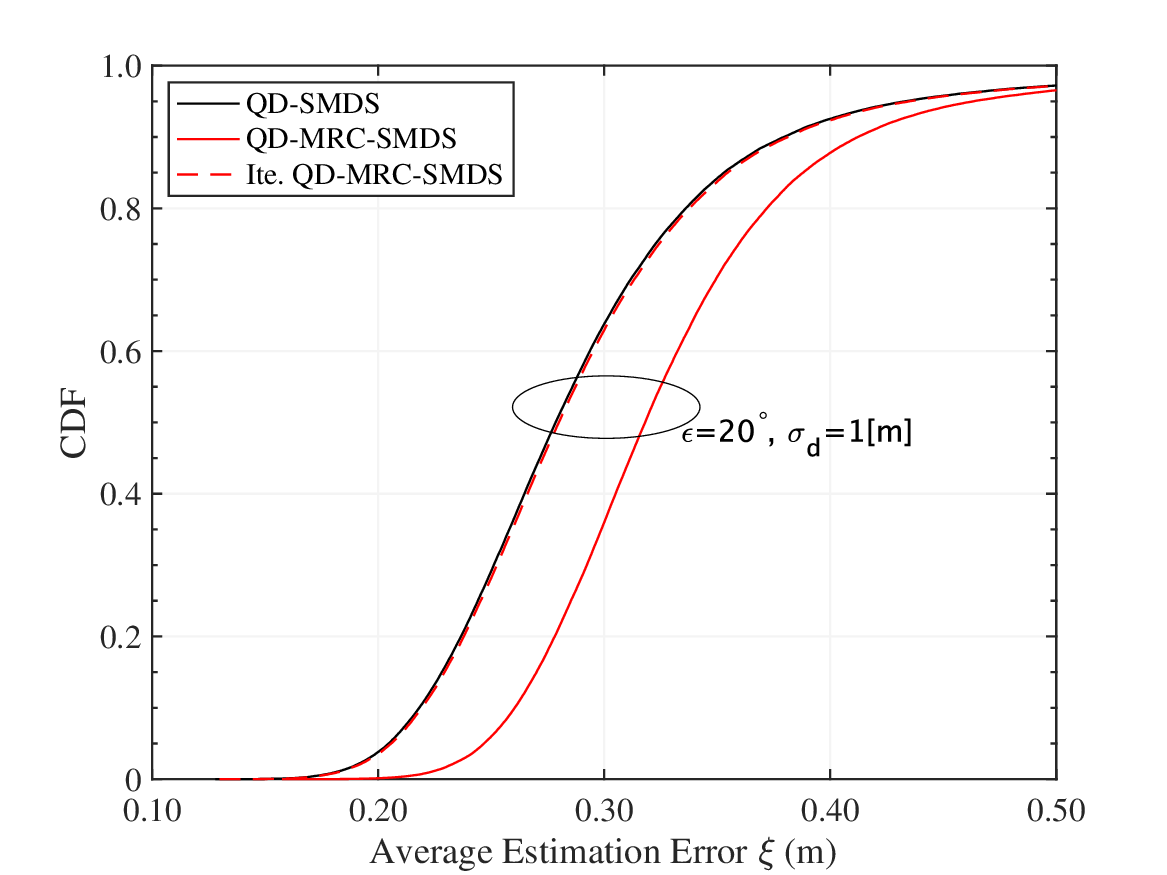}
	}
    \vspace{2mm}
	\subfigure{
	\includegraphics[width=0.95\columnwidth,keepaspectratio=true]{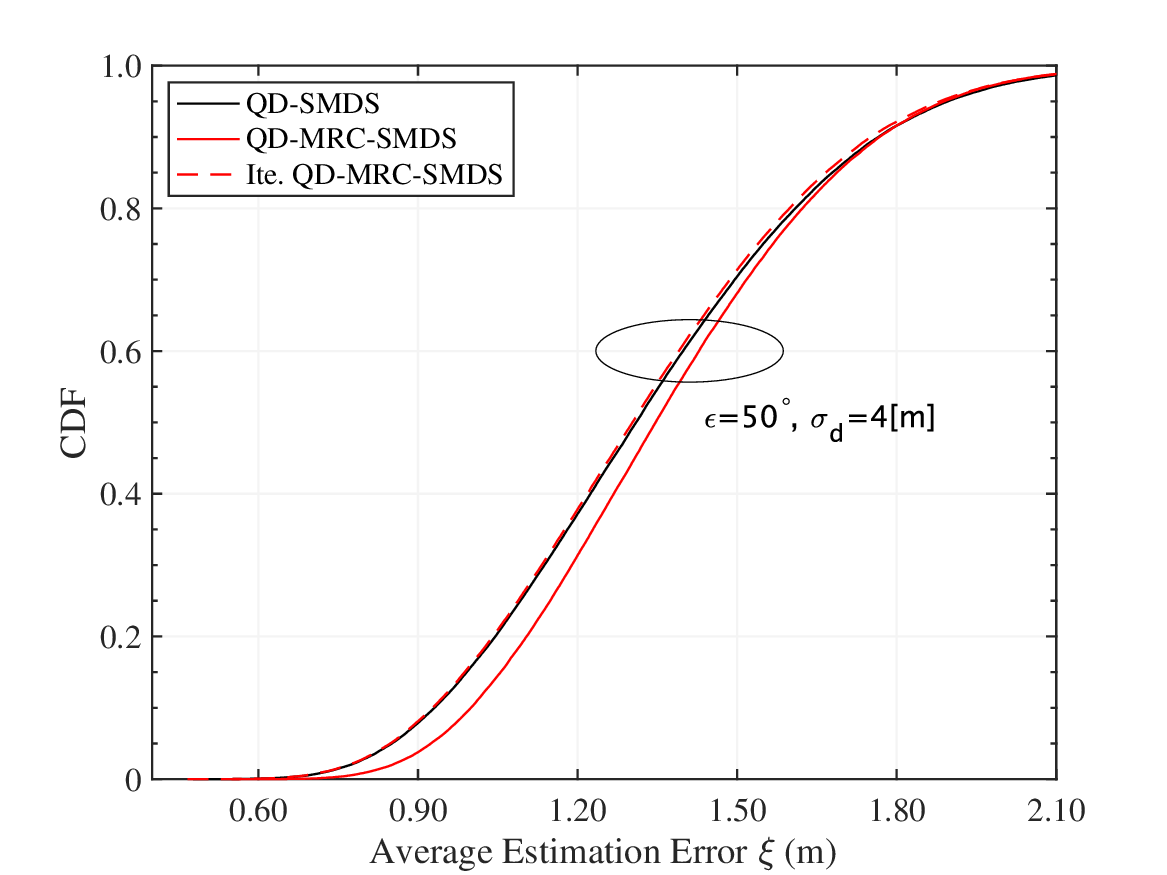}
	}
	\caption{ Empirical CDFs of \ac{QD-SMDS}, \ac{QD-MRC-SMDS}, and iterative \ac{QD-MRC-SMDS} algorithms in \textbf{Scenario I}.}
	\label{fig:SMDSvsQDSMDS_w/o_SVD_scenarioI_CDF}
\end{figure}

%
%

{ In this subsection, we compare the localization accuracies of the \ac{QD-MRC-SMDS} and iterative \ac{QD-MRC-SMDS} algorithms} under the same simulation conditions as in Figs. \ref{fig:SMDSvsQDSMDS_scenarioI} and \ref{fig:SMDSvsQDSMDS_scenarioII} and { under the two distinct scenarios.}

{ \noindent \emph{Comparisons under Scenario 1:}} Figs. \ref{fig:SMDSvsQDSMDS_w/o_SVD_scenarioI} { and \ref{fig:SMDSvsQDSMDS_w/o_SVD_scenarioI_CDF} show the average estimation error and the empirical \ac{CDF} of both methods in \textbf{Scenario I}}, with the \ac{SMDS} algorithm included as reference (black curve).
Notably, both \ac{QD-MRC-SMDS} and its iterative variant achieve high localization accuracy without relying on \ac{SVD}-based noise suppression.
Even the lowest-complexity \ac{QD-MRC-SMDS} incurs only a modest performance degradation compared to \ac{QD-SMDS}.
Moreover, its ability to maintain high accuracy under large angle errors indicates that \ac{QD-MRC-SMDS} inherits the robustness to angular uncertainties observed in \ac{QD-SMDS}.
When compared to the \ac{SMDS} performance shown in Fig. \ref{fig:SMDSvsQDSMDS_scenarioI}, \ac{QD-MRC-SMDS} demonstrates superior localization accuracy, especially severe angular error conditions (\textit{i.e.}, $\epsilon\ge 30^\circ$), underscoring its robustness and computational efficiency.
Furthermore, the iterative \ac{QD-MRC-SMDS} algorithm achieves performance nearly equivalent to that of \ac{QD-SMDS}, suggesting that the full potential of the \ac{GEK} matrix can be exploited without requiring low-rank truncation via \ac{SVD}.

%

%
{ \noindent \emph{Comparisons under Scenario 2:} Next,} Figs. \ref{fig:SMDSvsQDSMDS_w/o_SVD_scenarioII} { and \ref{fig:SMDSvsQDSMDS_w/o_SVD_scenarioII_CDF} show the average estimation error and the empirical \ac{CDF} of each method in \textbf{Scenario II}, respectively}.
While the overall trend is similar to that observed in \textbf{Scenario I}, the performance gap between the iterative \ac{QD-MRC-SMDS} and \ac{QD-SMDS} increases with larger angular errors.

This can be attributed primarily to the greater degradation in the accuracy of the \ac{GEK} matrix caused by increased measurement errors in the additional angular parameters of the azimuth and elevation angles. In this case, it appears that noise suppression in the SVD-based approach works slightly better than in the iterative one.
Nevertheless, the performance gap remains minimal—even at the largest angular error ($\epsilon=50^\circ$), the { average} estimation error difference is within $0.05$ m.

%% file: TXT/6_Discussions.tex
\subsection{\Ac{FLOP} Count of \ac{SMDS}-based Approaches}

To enable a fair comparison of the computational effort among the \ac{SMDS} variants presented in this paper, the total number of \acp{FLOP} required by each algorithm is reported.

%
\begin{figure}[!t]
\centering
	\subfigure{
	\includegraphics[width=0.95\columnwidth,keepaspectratio=true]{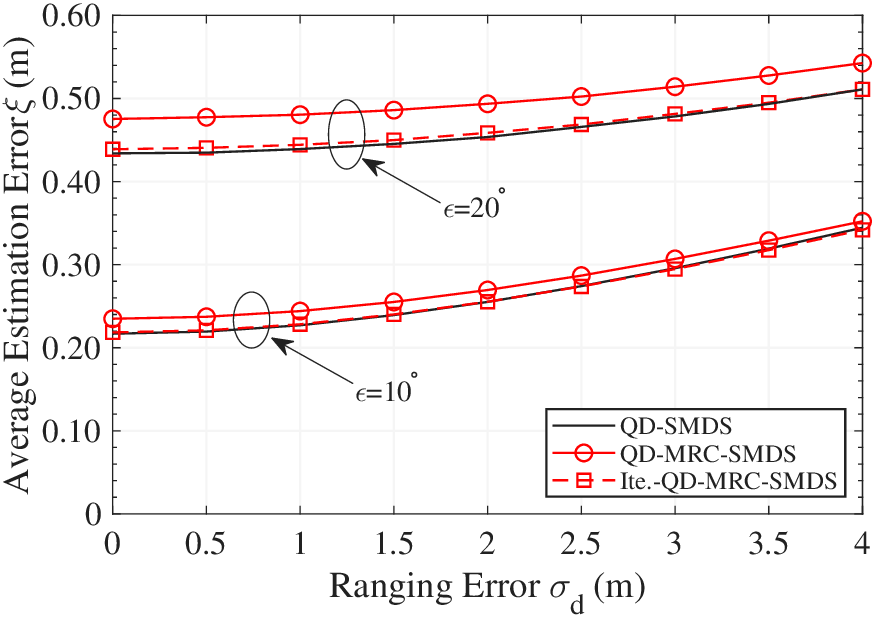}
	}
    \vspace{2mm}
	\subfigure{
	\includegraphics[width=0.95\columnwidth,keepaspectratio=true]{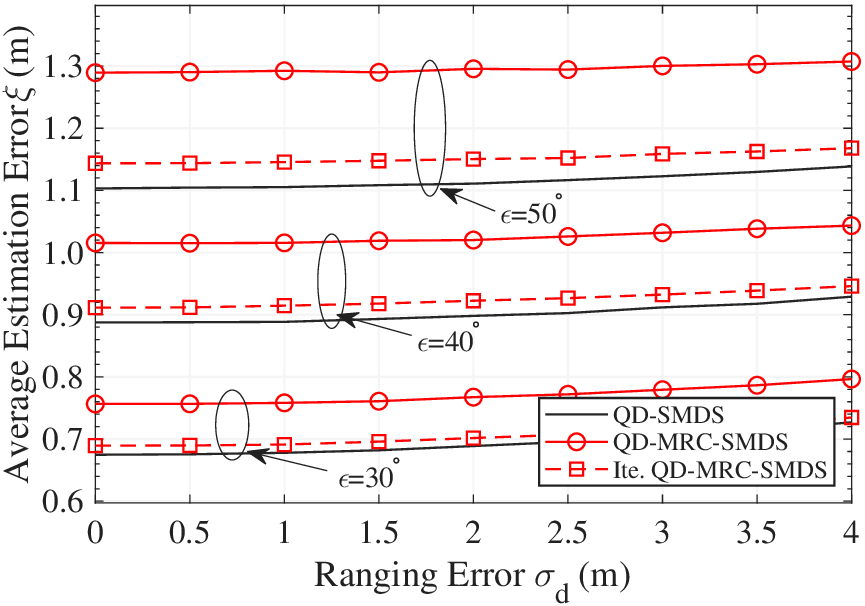}
	}
	\caption{Comparison of localization accuracy between the \ac{QD-SMDS}, \ac{QD-MRC-SMDS}, and iterative \ac{QD-MRC-SMDS}  algorithms in \textbf{Scenario II}.}
	\label{fig:SMDSvsQDSMDS_w/o_SVD_scenarioII}
	\vspace{-2ex}
\end{figure}
\begin{figure}[!t]
\centering

	\subfigure{
	\includegraphics[width=0.95\columnwidth,keepaspectratio=true]{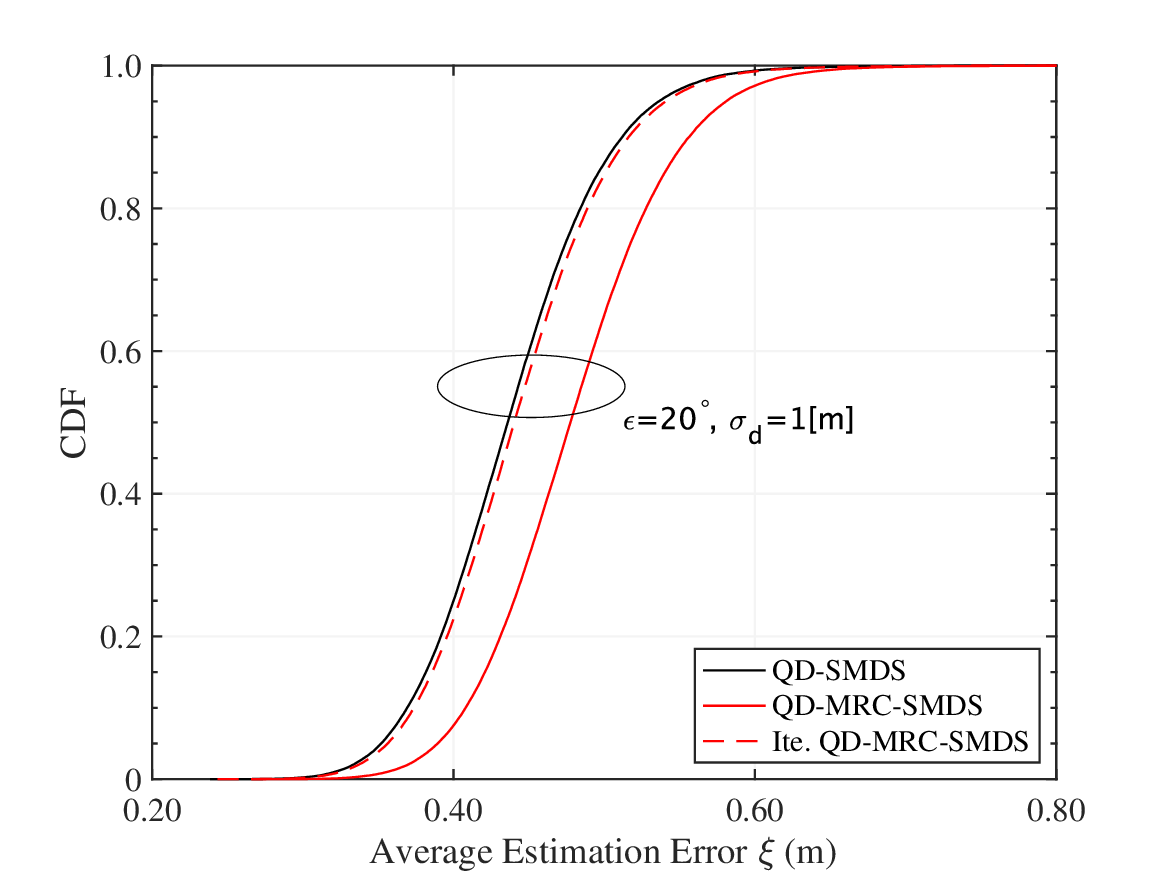}
	}
    \vspace{2mm}
	\subfigure{
	\includegraphics[width=0.95\columnwidth,keepaspectratio=true]{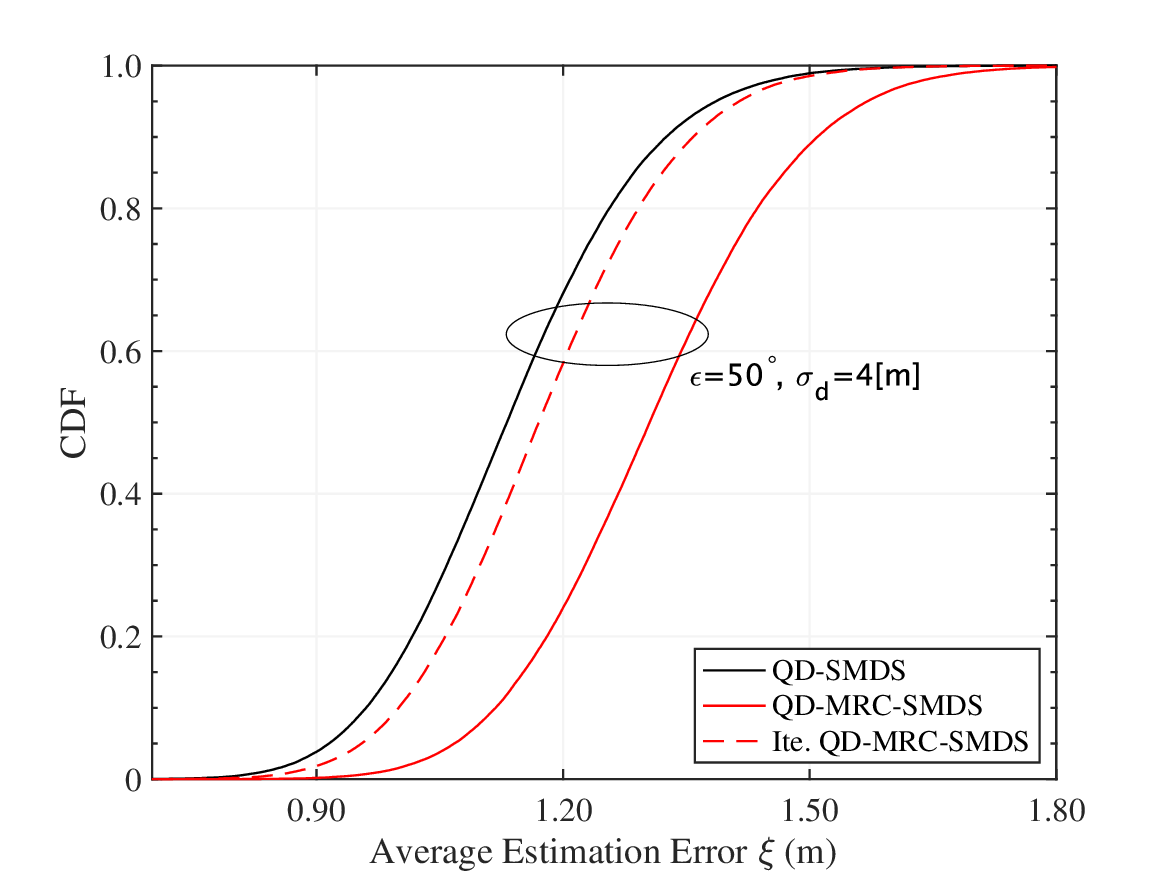}
	}
	\caption{ Comparison of empirical CDF between the \ac{QD-SMDS}, \ac{QD-MRC-SMDS}, and iterative \ac{QD-MRC-SMDS}  algorithms in \textbf{Scenario II}.}
	\label{fig:SMDSvsQDSMDS_w/o_SVD_scenarioII_CDF}
\end{figure}

A detailed derivation of the \ac{FLOP} count for each algorithm is provided in { Appendix~A} and summarized in Table~\ref{tab:FLOP_count}, which also includes the \ac{FLOP} count for the configuration used in the previous simulations.

{
While both SMDS and QD-SMDS rely on SVD-based low-rank truncation, the QSVD required by QD-SMDS operates on the $2M\times2M$ complex-valued equivalent of the quaternion GEK matrix, making it $\sim 32 \times$ more expensive in the dominant cubic term compared to the real SVD in SMDS. 
Table \ref{tab:FLOP_count} quantifies this: for $N_{\mathrm{A}}=5$ and $N_{\mathrm{T}}=15$, QD-SMDS already costs $\sim 31 \times$ more FLOPs than SMDS. However, this overhead is entirely avoided by QD-MRC-SMDS, which replaces QSVD with a single matrix–vector multiplication scaling as $O(N_{\mathrm{A}}^3 N_{\mathrm{T}})$, i.e., linearly in $N_\mathrm{T}$. 
As a result, QD-MRC-SMDS is actually cheaper than SMDS even in the quaternion domain, while the iterative variant remains within the same order of magnitude. For large networks, the FLOP count formulas in Table \ref{tab:FLOP_count} allow a system designer to evaluate the costs of these alternatives for any $N_{\mathrm{A}}$ and $N_{\mathrm{T}}$.

To further illustrate this, even for a large network with $N_{\mathrm{T}} = 1000$ target nodes, the \ac{FLOP} counts of \ac{QD-MRC-SMDS} and its iterative variant are on the order of $10^8$ and $10^9$, respectively (exact FLOP count can be obtained using the formulas in Table \ref{tab:FLOP_count}).
Assuming a single core CPU operating at $10$~GFLOPS (a throughput comparable to the single-core floating-point performance of some Raspberry Pi models)---with no parallelisation, multi-threading, or hardware acceleration---these algorithms can be executed in approximately $30$~ms and $110$~ms, respectively, confirming their suitability for real-time deployment even in large-scale networks.
Note that the \ac{FLOP} counts for SMDS and QD-SMDS would be on the order of $10^{11}$ and $10^{13}$, respectively; at this scale, additional computational resources such as multi-core parallelisation or specialised hardware accelerators would be required to make them viable for real-world deployment.
}

{  The advantages of switching from \ac{SMDS} to its computationally efficient quaternion-domain variants is evident when combining the analysis of the overall localization performance with the computational effort in Table~\ref{tab:FLOP_count}.
These variants provide improved localization performance for larger angular errors while requiring lower computational effort.}

Based on these results, system designers can select the most appropriate algorithm by balancing the computational effort against localization accuracy.
For latency-critical applications, the low-complexity \ac{QD-MRC-SMDS} represents the most practical choice.

Conversely, when computational resources are sufficient and maximum accuracy is required—even under large angular errors—\ac{QD-SMDS} becomes the preferred option.
Finally, iterative \ac{QD-MRC-SMDS} achieves an excellent trade-off between accuracy and complexity, making it a strong candidate for a wide range of practical deployment scenarios.

\begin{table}[!t]
 \caption{{\ac{FLOP} count formulas for all \ac{SMDS} variations. 
 The total \ac{FLOP} count for the configuration used in the reported simulations is also provided.}}
 \label{tab:FLOP_count}
 \centering
 \scriptsize
 {
 \resizebox{\columnwidth}{!}{%
 \begin{tabular}{lcc}
     \toprule
     Algorithm & \ac{FLOP} count formula & \makecell{\ac{FLOP} count for \\ $N_{\mathrm{A}} = 5, N_\mathrm{T} = 15$} \\
     \midrule 
     \ac{SMDS} & \makecell[l]{$3M^2 + \frac{20}{3}M^3 + 3(M+1)$ \\
     $+ 3(N_\mathrm{A}+N_\mathrm{T})(2(N_\mathrm{A}+M)-1)$ \\ 
     $+ 54 N_\mathrm{A} + 21 N_\mathrm{T} + 397.$} &  $\approx 4.13 \times 10^6$\\
     \midrule
     \ac{QD-SMDS} & \makecell[l]{$12M^2 + \frac{640}{3}M^3 + 4M + 1$ \\
    $+ 3(N_{\mathrm{A}}+N_\mathrm{T})(2(M+N_{\mathrm{A}})-1)$  \\ 
    $+ 54 N_\mathrm{A} + 21 N_\mathrm{T} + 397.$} & $\approx 1.31 \times 10^8$  \\
     \midrule
     \ac{QD-MRC-SMDS} & \makecell[l]{$12M^2 + 16 N_{\mathrm{A}} N_\mathrm{T} \left(N_{\mathrm{A}}(N_{\mathrm{A}}-1) - 1 \right)$  \\ $+ 8(N_{\mathrm{A}}(N_{\mathrm{A}}-1)/2) - 1 + 8 N_{\mathrm{A}} N_\mathrm{T}$ \\ $+ N_\mathrm{T} (8N_{\mathrm{A}} - 4)$} & $\approx 1.11 \times 10^5$\\
     \midrule
     \makecell[l]{Iterative \\ \ac{QD-MRC-SMDS}} &  \makecell[l]{$12M^2 + 16 N_{\mathrm{A}} N_\mathrm{T} \left(N_{\mathrm{A}}(N_{\mathrm{A}}-1) - 1 \right)$ \\ $+ 8(N_{\mathrm{A}}(N_{\mathrm{A}}-1)/2) - 1 + 8N_{\mathrm{A}} N_\mathrm{T}$ \\ $+ N_\mathrm{T} (8N_{\mathrm{A}}  - 4) + 16N_{\mathrm{A}} N_\mathrm{T} (2M - 1)$\\  $+ 8M - 1 + 4N_{\mathrm{A}} N_\mathrm{T}$} & $\approx 3.14 \times 10^5$ \\ 
     \bottomrule
 \end{tabular}
 }
 }
\end{table}

{

\subsection{Complexity Comparisons with Other Localization Methods}

To place our \ac{QD-SMDS} approaches in context, we compare the computational complexity of our methods with other localization algorithms. 
Among convex optimization approaches, most \ac{SDP}-based methods tie the complexity primarily to the number of targets $N_\mathrm{T}$.
For fixed dimension and typical \ac{SDP} solvers, the complexity grows polynomially but very steeply with $N_\mathrm{T}$, with a worst case bound of $O(N_\mathrm{T}^6)$~\cite{Biswas2006}. 
In contrast, the proposed \ac{QD-SMDS} approaches scale quadratically in $N_\mathrm{T}$, which makes them more suitable than \ac{SDP}-based solvers for large-scale networks.

A comparison with Bayesian approaches is more nuanced, since their complexity typically depends on additional framework-specific parameters.
In the method proposed in \cite{Naseri2019}, the hybrid Bayesian message-passing algorithm scales linearly with the number of targets, with a per-iteration cost on the order of $O(N_{\mathrm{A}} N_\mathrm{T} M N)$, where $M$ and $N$ denote the transmitted and drawn sample sizes, respectively.
Similarly, for other Bayesian methods~\cite{Ihler2005,Li2022}, the per-iteration cost is linear in $N_\mathrm{T}$. 
However, unlike Bayesian or message-passing localization algorithms–which require multiple iterations and repeated processing of all anchor–target measurements–\ac{QD-SMDS} provides a single-pass, closed-form computation.
In particular, the \ac{QD-MRC-SMDS} variant scales linearly in $N_\mathrm{T}$ without requiring iterative refinement.

A third class of competing techniques consists of machine-learning-based localization methods~\cite{Wang2015, Tsuchida2019, Tasaki2020, Ammad2025, Shahbazian2023}, which learn a mapping from measured features to positions. 
This line of indoor positioning research employs deep neural networks using \ac{CSI} fingerprints and explicitly separates a computationally intensive offline training phase from a lightweight online localization phase consisting of a single forward pass through the network.
Thus, while learning-based methods trade substantial iterative offline complexity and data-collection effort for very low online cost, \ac{QD-SMDS} offers a non-iterative, training-free alternative with explicit and predictable complexity for any given $N_{\mathrm{A}}$ and $N_\mathrm{T}$.

{
A fourth class of localization methods exploits tensor decomposition by arranging measurement data into higher-order arrays and applying low-rank factorizations to recover position information~\cite{LiuDSP2025,ChenICASSP2018}.
These methods operate on raw received signals---such as wideband array signals, RSS, or ToA---and are designed for passive emitter localization, which differs from the cooperative WSN setting addressed here.
Moreover, they rely on iterative solvers with variable convergence and require a grid search over the target area, whereas \ac{QD-SMDS} is non-iterative, grid-free, and operates directly on pre-extracted pairwise distance and angle measurements.
}

\subsection{Implementation Feasibility and Potential Deployment Challenges}

Although the evaluation of QD-SMDS in this paper is conducted through controlled simulations, we emphasize that real-life implementation of the method can be easily carried out since the technique is an \ac{MDS}-based approach \cite{saeed2019state}, for which explicitly experimental results can be found in related literature \cite{patwari2003relative, costa2006, shang2004improved, wu2009nmdsmle, vo2008weighted}.

To elaborate further, the method is compatible with practical range- and angle-based localization systems. 
The distance measurements required by QD-SMDS can be obtained from widely used technologies such as \ac{UWB} \ac{ToA}/\ac{TDoA} devices or \ac{RSSI}-based ranging modules. 
Prior work using these technologies has demonstrated that such systems can provide sufficiently accurate range estimates under realistic conditions~\cite{Alarifi2016,Silva2014}.

In addition to range measurements, QD-SMDS utilizes azimuth and elevation angle estimates to form the \ac{GEK} matrix.
Such angle information can be obtained using planar antenna arrays or coprime planar arrays capable of two-dimensional \ac{DoA} estimation. 
Practical array-processing techniques—including polynomial-rooting approaches and subspace-based methods such as MUSIC and ESPRIT—enable the extraction of both azimuth and elevation angles from compact antenna configurations~\cite{Zhang2018,Xiaofei2011,Xu2018}.

However, the deployment of distance- and angle-capable antenna arrays introduces additional considerations.
Accurate \ac{DoA} estimation requires careful antenna calibration, adequate signal-to-noise ratio, and mitigation of multipath propagation, all of which may otherwise bias angle measurements.
In addition, clock synchronization for \ac{ToA}/\ac{TDoA} systems, antenna calibration for \ac{UWB} and array-based platforms, and anchor placement geometry are known to influence the quality of the collected range measurements~\cite{Maran2010}. 

Consistent with most of the localization literature, the present work focuses on the algorithmic contribution and its performance under simulated situations. 
A dedicated real-world deployment and measurement campaign therefore represents a promising direction for future work.
}

%% file: TXT/7_Conclusion.tex
In this paper, we proposed a novel \ac{QD-SMDS} algorithm, developed by reformulating the classical \ac{SMDS} algorithm within the quaternion domain to enable low-complexity, simultaneous localization of multiple targets using data aggregated from a large number of wireless sensor nodes. 
By constructing the \ac{GEK} matrix in the quaternion domain, the matrix rank can be reduced to one, even in \ac{3D} Euclidean space, thereby maximizing the noise suppression effect via \ac{QSVD}.
Moreover, the proposed method is inherently compatible with low-rank matrix completion techniques, which further enhances its robustness against missing data.
Simulation results demonstrate that \ac{QD-SMDS} consistently outperforms the conventional \ac{SMDS}, particularly in scenarios with significant angle measurement errors.
Its advantage becomes more pronounced when both azimuth and elevation information are available.
However, the use of \ac{QSVD}—along with matrix inversion and the Procrustes transformation required for coordinate recovery—introduces substantial computational complexity. 
To address this limitation, we further developed a computationally efficient variants of \ac{QD-SMDS} by exploiting the structural properties of the quaternion-domain \ac{GEK} matrix. 
These variants enable coordinate estimation through simple quaternion matrix multiplications, achieving localization performance comparable to that of the original \ac{QD-SMDS}, while significantly reducing computational cost.

%% file: TXT/appendix.tex
{
\subsection{\Ac{FLOP} Count Calculations}
\subsubsection{\ac{SMDS}}
The \ac{SMDS} algorithm starts by building the \ac{GEK} matrix in \eqref{eq:r_kernel}.
This computation yields a matrix with entries $\cos(\alpha_{mn}) d_m d_n$ at row $m$ and column $n$, requiring two multiplications per matrix entry.
Additionally, we also need to evaluate the cosines of the measured angles, and the number of \acp{FLOP} for such an operation is hardware-dependent. 
For simplicity, we will assume that one cosine evaluation is equivalent to 1 \ac{FLOP}, yielding
\begin{equation}
    \text{\acp{FLOP} \eqref{eq:r_kernel}} = 2M^2 + M^2 = 3M^2
\end{equation}

In principle, the \ac{SVD} can be computed using a truncated method that returns only the three dominant eigenpairs~\cite{Calvetti1994}.
However, the corresponding \ac{FLOP} count depends on the number of iterations required for convergence.
Therefore, to avoid iteration-dependent \ac{FLOP} counts, we assume that a full \ac{SVD} is used, for which the number of \ac{FLOP}s is approximately~\cite{Golub2013}
\begin{equation}
    \text{\ac{FLOP}s for \ac{SVD}} \approx \frac{20}{3}M^3.
\end{equation}

The next operation in \eqref{eq:r_V_estimated} involves three square-root operations and multiplying each column by the corresponding singular value.
For simplicity, we assume that a square-root operation is equivalent to one \ac{FLOP}, yielding
\begin{equation}
    \text{\ac{FLOP}s for \eqref{eq:r_V_estimated}} = 3 + 3M = 3(M+1).
\end{equation}

The next operation is \eqref{eq:MoorePenrose}, where the pseudoinverse
\begin{equation}
\label{eq:pseudoinverse}
\left[
\begin{array}{c|c}
  \bm{I}_{N_{\mathrm{A}}} & \bm{0}_{N_{\mathrm{A}}\times N_{\mathrm{T}}}\\\hline
  \multicolumn{2}{c}{\bm{C}}
\end{array}
\right]^{-1} \in \mathbb{R}^{(N_{\mathrm{A}} + N_\mathrm{T}) \times (N_{\mathrm{A}}+M)}
\end{equation}
is computed.
Since this pseudoinverse needs to be computed only once and reused across simulations, we assume it is already available and do not include its computation in the total \ac{FLOP} count.
Thus, computing \eqref{eq:MoorePenrose} amounts to
\begin{equation}
    \text{\ac{FLOP}s for \eqref{eq:MoorePenrose}} = 3(N_{\mathrm{A}}+N_\mathrm{T})(2(N_{\mathrm{A}}+M)-1).
\end{equation}

In the final step, the estimated coordinate system is aligned with the true coordinate system using the Procrustes transformation \cite{Fiore2001}. The alignment consists of three operations: translation, scaling, and rotation.

First, both coordinate systems are translated such that their centroids coincide with the origin. For the true anchor-node coordinate matrix $\bm{X}_{\mathrm{A}} \in \mathbb{R}^{N_\mathrm{A} \times 3}$, the centroid is given by
\begin{equation}
\bar{\bm{X}}_{\mathrm{A}} 
= 
\frac{1}{N_{\mathrm{A}}} 
\sum_{k=1}^{N_{\mathrm{A}}} 
\bm{X}_{\mathrm{A},k},
\end{equation}
where $\bm{X}_{\mathrm{A},k}$ denotes the coordinates of the $k$-th anchor node. The centered coordinates are then obtained as
\begin{equation}
\bm{X}_{\mathrm{A},k}^{(\mathrm{O})} 
= 
\bm{X}_{\mathrm{A},k} - \bar{\bm{X}}_{\mathrm{A}}.
\end{equation}
The same centering procedure is applied to the estimated anchor coordinates.

Next, the centered matrix $\bm{X}_{\mathrm{A}}^{(\mathrm{O})}$ is normalized using its Frobenius norm,
\begin{equation}
\bm{X}_{\mathrm{A}}^{(\mathrm{norm})}
=
\frac{\bm{X}_{\mathrm{A}}^{(\mathrm{O})}}
{\|\bm{X}_{\mathrm{A}}^{(\mathrm{O})}\|_{\mathrm{F}}},
\end{equation}
and analogously for the estimated coordinate matrix.

The optimal rotation is obtained by first computing the cross-covariance matrix
\begin{equation}
\bm{S}
=
\left(
\bm{X}_{\mathrm{A}}^{(\mathrm{norm})}
\right)^{\mathsf{T}}
\hat{\bm{X}}_{\mathrm{A}}^{(\mathrm{norm})},
\end{equation}
followed by its singular value decomposition
\begin{equation}
\bm{S} = \bm{U} \bm{\Sigma} \bm{V}^{\mathsf{T}}.
\end{equation}
The rotation matrix is then given by
\begin{equation}
\bm{R} = \bm{V}\bm{U}^{\mathsf{T}}.
\end{equation}

Finally, the aligned estimate is obtained as
\begin{equation}
\hat{\bm{X}}_{\mathrm{aligned}}
=
s\, \hat{\bm{X}} \bm{R}
+
\bar{\bm{X}}_{\mathrm{A}},
\end{equation}
where the scaling factor is
\begin{equation}
s
=
\frac{\|\bm{X}_{\mathrm{A}}^{(\mathrm{O})}\|_{\mathrm{F}}}
{\|\hat{\bm{X}}_{\mathrm{A}}^{(\mathrm{O})}\|_{\mathrm{F}}}.
\end{equation}

Since the true anchor-node coordinates are static throughout the simulations, quantities depending solely on $\bm{X}_{\mathrm{A}}$ are computed once and are therefore excluded from the reported floating-point operation (FLOP) counts.

The computational complexity of the Procrustes alignment is summarized as follows:

\begin{itemize}

\item \textbf{Normalization of $\hat{\bm{X}}_{\mathrm{A}}$:}
\begin{equation}
15 N_\mathrm{A}.
\end{equation}

\item \textbf{Computation of the rotation matrix:}
\begin{equation}
18N_{\mathrm{A}} + 396.
\end{equation}

\item \textbf{Final alignment transformation:}
\begin{equation}
21 N_\mathrm{A} + 21 N_\mathrm{T} + 1.
\end{equation}

\end{itemize}

Combining all contributions yields
\begin{equation}
\text{\acp{FLOP} Procrustes}
=
54 N_\mathrm{A} + 21 N_\mathrm{T} + 397.
\end{equation}

Adding everything together yields
\begin{align}
    \text{\acp{FLOP} \ac{SMDS}} & \approx 3M^2 + \frac{20}{3}M^3 + 3(M+1) \nonumber \\
    & + 3(N_\mathrm{A}+N_\mathrm{T})(2(N_\mathrm{A}+M)-1) \nonumber \\ 
    &+ 54 N_\mathrm{A} + 21 N_\mathrm{T} + 397.
\end{align}
where $M = N_{\mathrm{A}}(N_{\mathrm{A}}-1)/2 + N_{\mathrm{A}} N_{\mathrm{T}}$.

\subsubsection{\ac{QD-SMDS}}
For \ac{QD-SMDS} (and the other two quaternion-domain variants), the quaternion \ac{GEK} matrix in \eqref{eq:q_kernel} must first be constructed.
By inspection, this costs four times as much as constructing the real \ac{GEK} matrix in \eqref{eq:r_kernel}, yielding
\begin{equation}
    \text{\ac{FLOP}s for \eqref{eq:q_kernel}} = 12M^2.
\end{equation}

The next operation is \ac{QSVD}, which amounts to performing an \ac{SVD} on a $2M\times 2M$ complex matrix.
On average, a complex arithmetic operation is four times more expensive than a real one.
Thus, we approximate the \ac{FLOP} count for \ac{QSVD} as
\begin{equation}
    \text{\ac{FLOP}s for \ac{QSVD}} \approx 4 \frac{20}{3}(2M)^3 = \frac{640}{3}M^3.
\end{equation}

Next, the quaternion edge vector is estimated via \eqref{eq:q_v_estimated}, which involves one square-root operation and $M$ quaternion-scalar multiplications, i.e.,
\begin{equation}
    \text{\ac{FLOP}s for \eqref{eq:q_v_estimated}} = 1 + 4M.
\end{equation}

Finally, the quaternion edge vector is converted to the real domain, after which the Procrustes transformation and \eqref{eq:MoorePenrose} are applied as in the \ac{SMDS} case.
This yields a total \ac{FLOP} count of
\begin{align}
    \text{\ac{FLOP}s for \ac{QD-SMDS}} &= 12M^2 + \frac{640}{3}M^3 + 4M + 1 \nonumber \\
    &+ 3(N_{\mathrm{A}}+N_\mathrm{T})(2(M+N_{\mathrm{A}})-1) \nonumber \\ 
    &+ 54 N_\mathrm{A} + 21 N_\mathrm{T} + 397.
\end{align}

\subsubsection{\ac{QD-MRC-SMDS}}
\label{app:Flop_QD-MRC-SMDS}
After constructing the quaternion \ac{GEK} matrix, \ac{QD-MRC-SMDS} reduces to computing \eqref{eq:MRC}.
First, note that, due to the structure of
\begin{align}
    \bm{B}_\mathrm{AA}
    &\triangleq
    \bm{I}_{N_{\mathrm{A}}}
    \otimes
    \bm{1}_{N_{\mathrm{T}}\times 1}
    \nonumber\\
    &=
    \begin{bmatrix}
    \bm{1}_{N_{\mathrm{T}}\times 1}&\bm{0}_{N_{\mathrm{T}}\times 1}&\cdots&\bm{0}_{N_{\mathrm{T}}\times 1}\\
    \bm{0}_{N_{\mathrm{T}}\times 1}&\bm{1}_{N_{\mathrm{T}}\times 1}&\cdots&\bm{0}_{N_{\mathrm{T}}\times 1}\\
    \vdots&\vdots&\ddots&\vdots\\
    \bm{0}_{N_{\mathrm{T}}\times 1}&\bm{0}_{N_{\mathrm{T}}\times 1}&\cdots&\bm{1}_{N_{\mathrm{T}}\times 1}
    \end{bmatrix}
    \in\mathbb{R}^{N_{\mathrm{A}}N_{\mathrm{T}}\times N_{\mathrm{A}}},
\end{align}
\begin{equation}
    \bm{B}_\mathrm{AT}
    \triangleq
    \bm{1}_{N_{\mathrm{A}}\times 1}
    \otimes
    \bm{I}_{N_{\mathrm{T}}}
    =
    \begin{bmatrix}
    \bm{I}_{N_{\mathrm{T}}}\\
    \bm{I}_{N_{\mathrm{T}}}\\
    \vdots\\
    \bm{I}_{N_{\mathrm{T}}}
    \end{bmatrix}
    \in\mathbb{R}^{N_{\mathrm{A}}N_{\mathrm{T}}\times N_{\mathrm{T}}},
    \label{eq:anchor-target_matrix_1}
\end{equation}
the product $\bm{B}_\mathrm{AA}\bm{\chi}_{\mathrm{A}}$ simply repeats each entry of the quaternion vector $\bm{\chi}_{\mathrm{A}}\in\mathbb{H}^{N_{\mathrm{A}}\times 1}$, $N_\mathrm{T}$ times.
We therefore do not count this operation toward the total \ac{FLOP} count.

We also note that a quaternion operation is, on average, 16 times more costly than a real-valued operation.
Hence, computing $\bm{K}_2^\mathsf{H}\bm{\nu}_{\mathrm{AA}}$ requires $16 N_{\mathrm{A}} N_\mathrm{T}\left(N_{\mathrm{A}}(N_{\mathrm{A}}-1)-1\right)$ \acp{FLOP}.

For the vector quaternion norm $\|\bm{\nu}_{\mathrm{AA}}\|^2$, we require 7 \acp{FLOP} (4 multiplications and 3 additions) per quaternion norm.
Summing these values over all entries requires additional $N_{\mathrm{A}}(N_{\mathrm{A}}-1)/2-1$ scalar additions, for a total of $8\left(N_{\mathrm{A}}(N_{\mathrm{A}}-1)/2\right)-1$ \acp{FLOP}.
The ratio then requires $N_{\mathrm{A}} N_\mathrm{T}$ scalar--quaternion divisions, which cost $4 N_{\mathrm{A}} N_\mathrm{T}$ \acp{FLOP}.
The subtraction corresponds to $N_{\mathrm{A}} N_\mathrm{T}$ quaternion subtractions, which again cost $4 N_{\mathrm{A}} N_\mathrm{T}$ \acp{FLOP}.
Thus, the expression in parentheses costs
\begin{align}
    &\text{\ac{FLOP}s}\left(\bm{B}_\mathrm{AA}\bm{\chi}_{\mathrm{A}}
    -
    \frac{1}{\|\bm{\nu}_{\mathrm{AA}}\|^2}
    \bm{K}_2^\mathsf{H}
    \bm{\nu}_{\mathrm{AA}}
    \right)
    \nonumber\\
    &= 16 N_{\mathrm{A}} N_\mathrm{T}\left(N_{\mathrm{A}}(N_{\mathrm{A}}-1)-1\right)
    + 8\left(N_{\mathrm{A}}(N_{\mathrm{A}}-1)/2\right) - 1
    \nonumber\\
    &+ 8 N_{\mathrm{A}} N_\mathrm{T}.
\end{align}

The factor $\bm{B}_\mathrm{AT}^\mathsf{T}/N_{\mathrm{A}}$ can be implemented by substituting the ones in $\bm{B}_\mathrm{AT}^\mathsf{T}$ with $1/N_{\mathrm{A}}$, which we omit from the \ac{FLOP} calculation.
The remaining multiplication can be treated as a sparse multiplication of a real matrix with a quaternion vector, where each multiplication and addition costs 4 \acp{FLOP}.
If implemented efficiently, computing one entry of $\bm{\hat{\chi}}_{\mathrm{T}}$ requires $N_{\mathrm{A}}$ multiplications and $N_{\mathrm{A}}-1$ additions, for a total of $8N_{\mathrm{A}}-4$ \acp{FLOP}.
This brings the total \ac{FLOP} count of \ac{QD-MRC-SMDS} to
\begin{align}
    &\text{\ac{FLOP}s for \ac{QD-MRC-SMDS}} \\
    &=12M^2 + 16 N_{\mathrm{A}} N_\mathrm{T} \left(N_{\mathrm{A}}(N_{\mathrm{A}}-1) - 1 \right)
    \nonumber \\
    &+ 8\left(N_{\mathrm{A}}(N_{\mathrm{A}}-1)/2\right) - 1 + 8 N_{\mathrm{A}} N_\mathrm{T}
    \nonumber \\ 
    &+ N_\mathrm{T} (8N_{\mathrm{A}} - 4).
\end{align}

\subsubsection{Iterative \ac{QD-MRC-SMDS}}
We first note that if we substitute $\bm{\nu}_{\mathrm{AT}}^{(\tau_{\mathrm{max}})} = \bm{\nu}_{\mathrm{AT}}^{(0)}$ in \eqref{eq:q_X_ite_MRC}, where $\bm{\nu}_{\mathrm{AT}}^{(0)}$ is defined in \eqref{eq:q_v_AT ini}, then \eqref{eq:q_X_ite_MRC} becomes equivalent to \ac{QD-MRC-SMDS}, for which the \ac{FLOP} count was derived in Appendix~\ref{app:Flop_QD-MRC-SMDS}.

The additional computation is \eqref{eq:q_v_AT update}, which involves a quaternion matrix--vector multiplication costing $16N_{\mathrm{A}} N_\mathrm{T}(2M-1)$ \acp{FLOP}.
The denominator is a quaternion-vector norm, which (as noted above) costs $8M-1$ \acp{FLOP}.
Finally, the ratio between a vector and a scalar requires $4N_{\mathrm{A}} N_{\mathrm{T}}$ \acp{FLOP}.
This yields the total \ac{FLOP} count for iterative \ac{QD-MRC-SMDS}:
\begin{align}
    &\text{\ac{FLOP}s for iterative \ac{QD-MRC-SMDS}} \\
    &= 12M^2 + 16 N_{\mathrm{A}} N_\mathrm{T} \left(N_{\mathrm{A}}(N_{\mathrm{A}}-1) - 1 \right) \nonumber \\ 
    &+ 8(N_{\mathrm{A}}(N_{\mathrm{A}}-1)/2) - 1 + 8 N_{\mathrm{A}} N_\mathrm{T} \nonumber \\
    &+ N_\mathrm{T} (8N_{\mathrm{A}} - 4) \nonumber \\ 
    &+ 16N_{\mathrm{A}} N_{\mathrm{T}} (2M - 1) + 8M - 1 + 4N_{\mathrm{A}} N_{\mathrm{T}}.
\end{align}
 A single iteration is sufficient for convergence of iterative \ac{QD-MRC-SMDS}; therefore, we do not include $\tau$ in the total \ac{FLOP} count.

\subsection{Effect of Simulation Environment on Localization Accuracy}
\begin{figure}[!t]
\centering

	\subfigure{
	\includegraphics[width=0.95\columnwidth,keepaspectratio=true]{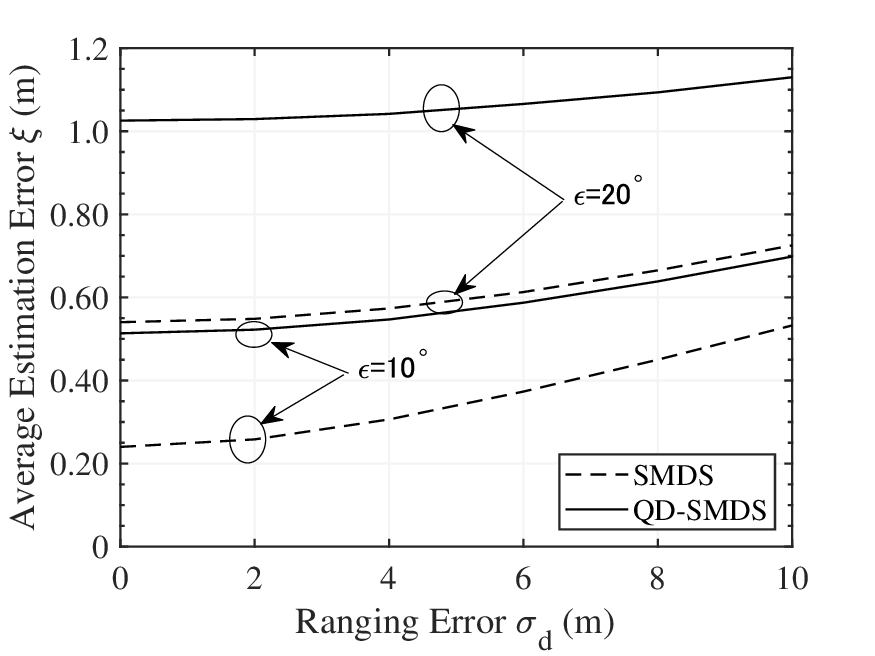}
	}
    \vspace{2mm}
	\subfigure{
	\includegraphics[width=0.95\columnwidth,keepaspectratio=true]{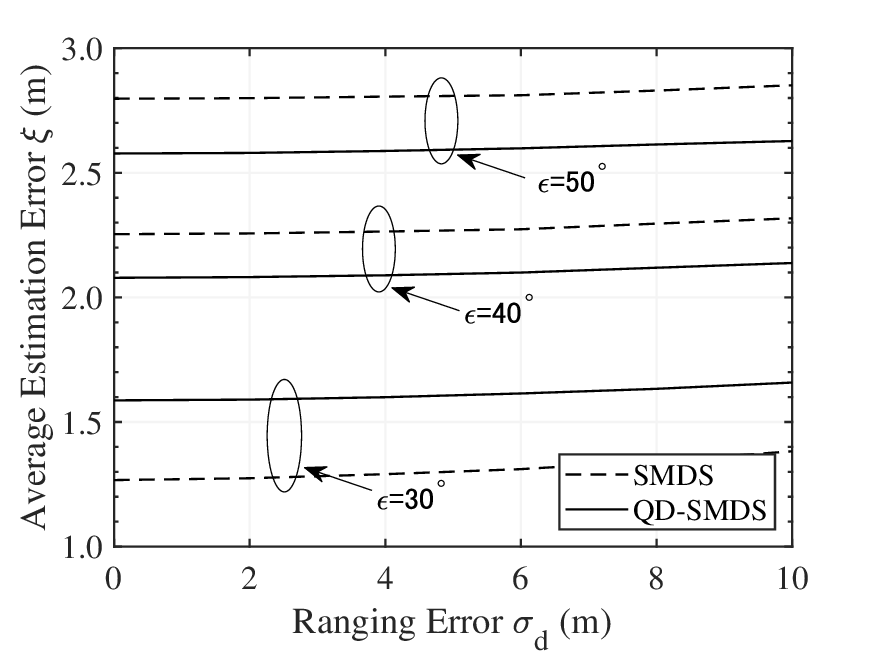}
	}
	\caption{ Reconsidering \textbf{Scenario II} with larger room dimensions and more \acp{TN}.}
	\label{fig:ScenarioII_new_topology}
\end{figure}

In the main part of the article, the impact of changes in network topology (\textit{e.g.}, room size and number of \acp{TN}) on localization performance was not explicitly investigated.
Additional experiments show that the main conclusion remains valid: \ac{QD-SMDS} begins to outperform \ac{SMDS} primarily when the angular measurements are highly unreliable.

To further support this claim, we revisit \textbf{Scenario II} in a new simulation environment  with room dimensions of $100$~m (length) $\times$ $100$~m (width) $\times$ $30$~m (height).
\Acp{AN} are placed at five locations: the four upper corners of the room, specifically at $(x,y,z)=(0,0,10)$, $(100,0,30)$, $(100,100,30)$, and $(0,100,30)$, as well as the origin $(x,y,z)=(0,0,0)$.
\Acp{TN} are randomly deployed at 30 locations within the room interior, with their $\mathrm{x}$, $\mathrm{y}$, and $\mathrm{z}$ coordinates independently drawn from a uniform distribution.
All other simulation settings, including the noise distributions, follow those in Section IV-C-1.

The results shown in Fig.~\ref{fig:ScenarioII_new_topology} demonstrate that the conclusions presented in the main text remain unchanged: \ac{QD-SMDS} outperforms \ac{SMDS} only under sufficiently large angular errors.
While the crossover point at which \ac{QD-SMDS} surpasses \ac{SMDS} may vary depending on the network topology, the qualitative performance advantage of \ac{QD-SMDS} over \ac{SMDS} remains consistent.
}